\documentclass[%
 aip,
 amsmath,amssymb,
 reprint,%
]{revtex4-1}	

\usepackage[utf8]{inputenc}
\usepackage[T1]{fontenc}
\usepackage{mathptmx}

\usepackage{amsmath}
\usepackage{amsfonts}
\usepackage{graphicx}
\usepackage{amssymb}
\usepackage{bbold}
\usepackage{scalerel}
\usepackage{xcolor}
\usepackage{ulem}

\begin{document}

\preprint{APS/123-QED}

\title[Primitive Model Electrolytes in the Near and Far Field]{Primitive Model Electrolytes in the Near and Far Field:\\ Decay Lengths from DFT and Simulations}

\author{P. Cats}
\email{p.cats@uu.nl}
 \affiliation{Institute for Theoretical Physics, Center for Extreme Matter and Emergent Phenomena, Utrecht University, Princetonplein 5, 3584 CC Utrecht, the Netherlands}

\author{R. Evans}%
\affiliation{HH Wills Physics Laboratory, University of Bristol, Bristol BS8 1TL, United Kingdom}%
 
 \author{A. H\"artel}%
\affiliation{Institute of Physics, University of Freiburg, Hermann-Herder-Stra\ss{}e 3, Freiburg 79104, Germany}%

 \author{R. van Roij}
  \affiliation{Institute for Theoretical Physics, Center for Extreme Matter and Emergent Phenomena, Utrecht University, Princetonplein 5, 3584 CC Utrecht, the Netherlands}

\date{\today}

\begin{abstract}
Inspired by recent experimental observations of anomalously large decay lengths in concentrated electrolytes, we revisit
the Restricted Primitive Model (RPM) for an aqueous electrolyte. We investigate the asymptotic decay lengths of the
one-body ionic density profiles for the RPM in contact with a planar electrode using classical Density Functional
Theory (DFT), and compare these with the decay lengths of the corresponding two-body correlation functions in bulk
systems, obtained in previous Integral Equation Theory (IET) studies. Extensive Molecular Dynamics
(MD) simulations are employed to complement the DFT and IET predictions. Our DFT calculations incorporate electrostatic
interactions between the ions using three different (existing) approaches: one based on the simplest mean field
treatment of Coulomb interactions (MFC), whilst the other two employ the Mean Spherical Approximation (MSA).
The MSAc invokes only the MSA bulk direct correlation function whereas the MSAu also incorporates the MSA bulk
internal energy. Although MSAu yields profiles that agree best with MD simulations in the near field, in the far field we observe that the decay lengths are consistent between IET, MSAc, and MD simulations, whereas those from 
MFC and MSAu deviate significantly. Using DFT we calculated the solvation force, which relates directly to surface
force experiments. We find that its decay length is neither qualitatively nor quantitatively close to the large decay
lengths measured in experiments and conclude that the latter cannot be accounted for by the primitive model. The
anomalously large decay lengths found in surface force measurements require an explanation that lies \textit{beyond} primitive
models.

\end{abstract}

\maketitle

\section{Introduction}

Electrolytes are important in many physical and biological phenomena and are crucial in many technological applications. A basic topic that continues to attract enormous interest is the structure of the  Electric Double Layer (EDL), i.e. how ions are distributed  in a liquid electrolyte in contact with a charged surface. 
Models describing the EDL have progressed  from a simplistic double layer capacitor model, proposed by Helmholtz  \cite{Helmholtz} from which the name EDL originates, to the first Primitive Model (PM) description by  Gouy\cite{Gouy}, Chapman\cite{Chapman}, and Debye and H\"uckel\citep{DH} (DH), where the electrolyte is modelled explicitly  in terms of  discrete ions, the charge carriers, embedded in a uniform dielectric medium, to current all-atom models, where both the solvent molecules and the ions are treated explicitly.
In recent years, classical Density Functional Theory (DFT) and Integral Equation Theories (IET)  have been employed, alongside  Molecular Dynamics (MD) and Monte Carlo simulations, to treat the EDL. Given the rich collection of theories and simulation methods used to investigate various models, one might have expected a comprehensive description of the EDL to have emerged. Recent experiments suggest otherwise. Over the last few years, several experimental groups have measured anomalously large decay lengths of the force between two charged cylindrical surfaces immersed in concentrated electrolytes (e.g. concentrations larger than about  1 M of NaCl dissolved in water) or in ionic liquids \cite{Gebbie_etal_2015,Cheng_etal_2015,Espinosa_etal_2014,Smith_etal_2016}; see also the summary article Ref.~\onlinecite{Gebbie_2017}. We refer to these experiments as Surface Force Apparatus (SFA) studies. Although these measurements relate to \textit{confined} liquids, it is well-known that the solvation force, as  measured by SFA, is determined by the asymptotic decay of the one-body density profiles at an individual surface, i.e. by the structure of the EDL in the far-field region, well away from the surface/electrode. The tails of the density profiles at each surface 'talk' to each other thereby determining the asymptotics of the solvation force.
The key observation is that the decay lengths measured in SFA experiments  are \textit{very} much longer than the Debye length obtained from DH theory, the length scale that must pertain in the dilute limit where the ionic concentration vanishes.

Understanding fully the structure of EDLs at high ionic concentrations is clearly important for fundamental reasons. Moreover, this is also directly relevant for practical devices that hinge on mobile ions in a liquid. For instance, room temperature ionic liquids confined in the pores of supercapacitors find applications in energy storage \cite{Simon_2008,Merlet_2012,Limmer_2013} and heat-to-current conversion \cite{Hartel_2015}, and porous carbon electrodes immersed in aqueous electrolytes can be used for harvesting blue energy \cite{Brogioli_2009,Janssen_2017} or desalinating water \cite{Poroda_etal_2013}. These engineering applications are in addition to the important role of water-dissolved ions in, for instance, biology (the action potential, homeostasis, etc.) and geology (mineral stability, dissolution rates, etc.).

In this paper we investigate the structure of EDLs, focusing on the decay lengths of the one-body density profiles and how these are determined by the decay of two-body correlation functions in the bulk liquid. The former aspect is investigated using DFT and MD simulation while the latter is examined using IET and MD simulation. We specialize to the Restricted Primitive Model (RPM) where the ionic species have equal size and equal but opposite charge. This choice simplifies theoretical treatments: number and charge density profiles, and the corresponding two-body correlation functions, (essentially) decouple, allowing us to treat both pieces independently. Implementing  DFT, we consider three different treatments of the electrostatic interactions, while employing the same Fundamental Measure Theory (FMT) to describe the hard-sphere (HS) interactions that mimic the steric forces. 

The paper is arranged as follows: Section II  sets out the basic theory for homogeneous as well as for inhomogeneous electrolytes. In Section ~\ref{sec:Inhom_elec}, the three functionals for the electrostatic interactions are introduced: a mean-field Coulomb functional, and two functionals based on the Mean Spherical Approximation (MSA). Sec.III lays out the details of the model, its parameters and how we translate between DFT and simulation. Sec.IV describes the results of our DFT calculations and MD simulations and how these connect with results from previous IET studies \cite{Attard,Ennis,Evans} that examined the decay of bulk pair correlation  functions. Our MD simulations were designed to check predictions of DFT for the one-body density profiles  in the near field, Sec.~\ref{sec:near_field}, and to examine the asymptotic decay of both the one- and two-body profiles in Sec.~\ref{sec:far_field}. Our DFT results for the decay length of the solvation force, obtained from the grand potential of the RPM confined between two planar electrodes, are presented in Sec.~\ref{sec:solv_force} where they are compared with the decay lengths measured in IET and DFT studies of the structure of the RPM and with decay lengths measured in SFA experiments.
Sec.~V describes a summary and discussion of our results  whilst Sec.~VI  provides concluding remarks.


\section{DFT for the Primitive Model}
\begin{figure}
\includegraphics[width=0.9\columnwidth]{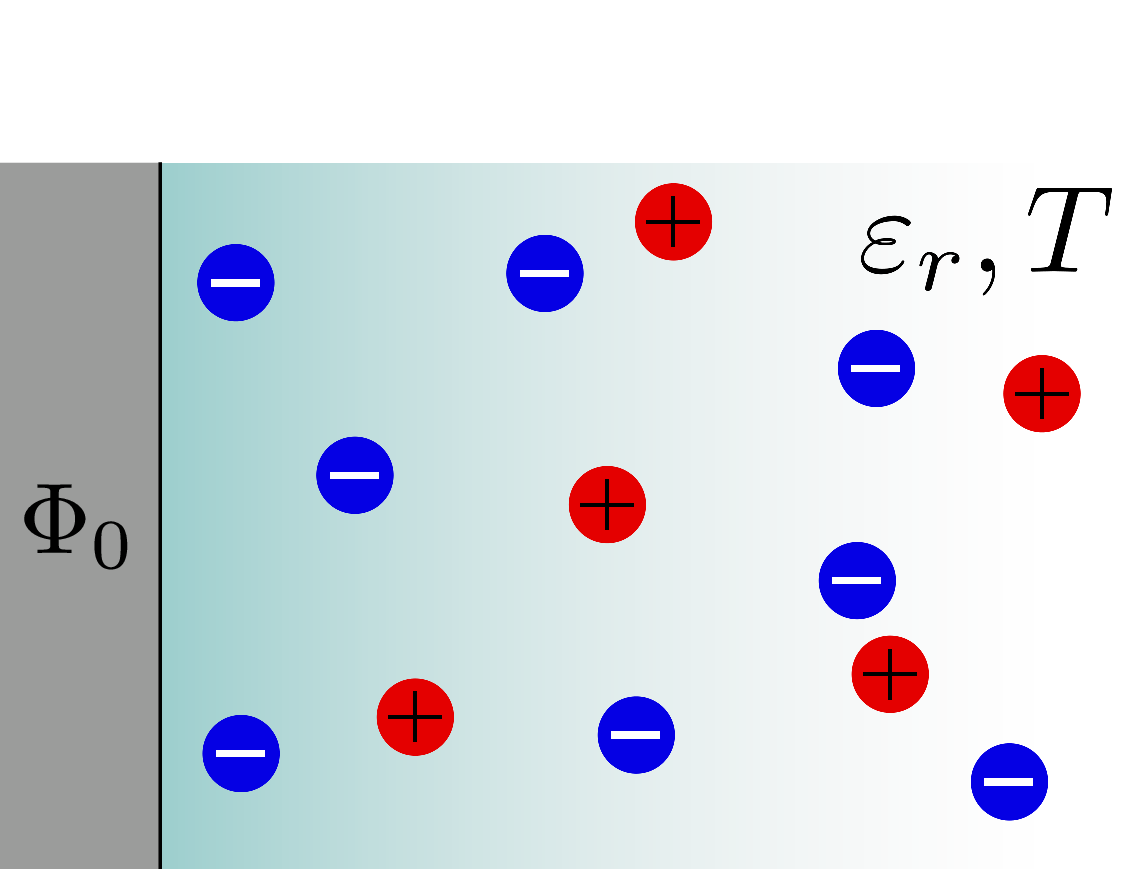}
\caption{\label{Fig:system} An illustration of the system considered throughout this study. The cations and anions (in red and blue, respectively) have equal size and valency and reside in a dielectric continuum at temperature $T$ and relative dielectric permittivity  $\varepsilon_r$.  The electrolyte is in contact with an electrode at surface potential $\Phi_0$, which causes a nonzero charge density profile near the electrode as depicted by the background shading. }
\end{figure}

We investigate the Primitive Model (PM) of an aqueous electrolyte, either in a homogeneous bulk state or in contact with a planar electrode at surface potential $\Phi_0$, as depicted in Fig.~\ref{Fig:system}. The PM is the model in which the solvent is treated as a dielectric medium with constant dielectric permittivity $\varepsilon_r\varepsilon_0$ and temperature $T$.  The ions are modeled as hard spheres with diameter $d_j$ and charge $ez_j$, where $j$ refers to the species and $e$ is the elementary charge. We consider mainly the Restricted Primitive Model (RPM), in which the electrolyte consists of two species (cat- and anions) that are characterized by $z_\pm=\pm 1$ and $d_\pm\equiv d$. The pair potentials of the RPM are defined by
\begin{align}\label{Eq:pair_pot}
\beta u_{ij}(r)=
\begin{cases}
\displaystyle \infty, &r<d;\\
\displaystyle z_iz_j\frac{\lambda_B}{r}, &r\geq d;
\end{cases}
\end{align}
with ${\lambda_B=\beta e^2/4\pi\varepsilon_r\varepsilon_0}$ the Bjerrum length; the distance between two point charges at which the electrostatic energy equals the thermal energy $\beta^{-1}=k_BT$. 
One way of tackling the inhomogeneous  PM is by applying DFT.
The starting point of classical DFT is the grand potential functional $\Omega$ of the density profiles $\rho_j(\mathbf{r})$, which reads \cite{Mermin,Evans1979,HansenMC}
\begin{align}
\Omega [ \{\rho\} ] =\mathcal{F}[\{\rho\}]-\sum_j \int \mathrm{d}\mathbf{r} \rho_j(\mathbf{r})\left[\mu_j-V^j_{ext}(\mathbf{r})\right],
\end{align}
with $\mathcal{F}$ the intrinsic Helmholtz free energy functional, $\mu_j$ the chemical potential and $V_{ext}^j$ the external potential, for each species $j$, and where $\{\rho\}=\{\rho_i|i=1,2\ldots \nu\}$ denotes the set of density profiles with $\nu$ being the number of species in the system. Here, $\mathcal{F}$ is an intrinsic property of the system which depends on the temperature and the interparticle interactions, but not on $\mu_j-V_{ext}^j$. The grand potential functional has the property that it is minimized for a given set $\mu_j-V^j_{ext}(\mathbf{r})$ by the equilibrium density profiles $\rho_{0,j}(\mathbf{r})$, i.e. $\delta\Omega/\delta\rho_j|_{\rho_{0,j}}=0$, resulting in the Euler-Lagrange equation
\begin{align}\label{Eq:EL}
\left.\frac{\delta \mathcal{F}[\{\rho\}]}{\delta \rho_j(\mathbf{r})}\right|_{\rho_{0,j}}=\mu_j-V^j_{ext}(\mathbf{r}).
\end{align}
Therefore, once an explicit form of $\mathcal{F}$ is constructed, one can find the equilibrium density profiles $\{\rho_0\}$ by solving the Euler-Lagrange equation. Then $\Omega[\{\rho_0\}]$ is the thermodynamic equilibrium grand potential. 

However, $\mathcal{F}$ is in general not  known exactly, so DFT hinges on approximations for $\mathcal{F}$. It is convenient to separate $\mathcal{F}$ into an ideal gas free-energy functional $\mathcal{F}_{id}$  obtained by turning off all the interparticle interactions:
\begin{align}
\beta\mathcal{F}_{id}[\{\rho\}]=\sum_j\int \mathrm{d}\mathbf{r}\rho_j(\mathbf{r})\left[ \log \left(\Lambda_j^3\rho_j(\mathbf{r})\right)-1\right],
\end{align}
with $\Lambda_j$ the thermal wavelength of species $j$, and the excess (over ideal) functional $\mathcal{F}_{ex}$ that accounts for the interactions, i.e.
\begin{align}\label{Eq:F_splitting}
\mathcal{F}[\{\rho\}]=\mathcal{F}_{id}[\{\rho\}]+\mathcal{F}_{ex}[\{\rho\}].
\end{align}
Importantly, $\mathcal{F}_{ex}$ is also the generator for the direct correlation functions, in particular
\begin{align}\label{Eq:c2F}
c^{(2)}_{ij}(\mathbf{r},\mathbf{r}')=-\beta\frac{\delta^2\mathcal{F}_{ex}[\{\rho\}]}{\delta\rho_{i}(\mathbf{r})\delta\rho_{j}(\mathbf{r}')}
\end{align}
is the pair (two-body) direct correlation function  which is related to the total (pair) correlation function via the Ornstein-Zernike (OZ) equation. For a uniform liquid with constant (bulk) densities $\{\rho_{b}\}$, the OZ equation \cite{OZ} reads: 
\begin{align}\label{Eq:OZ}
h_{ij}(r)=c_{ij}^{(2)}(r)+\sum_k\rho_{b,k}\int\mathrm{d}\mathbf{r'}c^{(2)}_{ik}(|\mathbf{r}-\mathbf{r'}|)h_{kj}(r'),
\end{align}
where the sum is over species $k$. Eqs.~\eqref{Eq:c2F} and~\eqref{Eq:OZ}  reveal an elegant relation between the total correlation functions $h_{ij}$ and the direct correlation functions $c_{ij}$ obtained from  free-energy functionals.

Constructing approximate DFTs that generate accurate one-body (density) profiles for fluids at substrates  and for two-body correlation functions in bulk is a challenge across liquid-state physics \cite{EvansJPCM}. For the PM this is especially demanding due to the long range character of the Coulomb potential. Tackling Coulombic interactions within DFT is non-trivial, as explained in detail in a recent review \cite{Andreas_2017}.

In subsequent sections we present theories for both homogeneous and inhomogeneous systems, i.e. without and with electrodes, respectively. We consider to what extent state of the art density functional theories incorporate correlation effects.

\subsection{Homogeneous Electrolytes}
Here, we review briefly the properties of bulk (homogeneous) fluid systems, considering results of  both the mean-field Debye-H\"uckel (DH) theory and the Mean Spherical Approximation (MSA) which is used as a closure to solve the (bulk) OZ equation. Subsequent subsections build upon these results.

\subsubsection{Debye H\"uckel Theory}
Electrolyte solutions were investigated in detail by Debye and H\"uckel \cite{DH} (DH) using the (linearized) Poisson-Boltzmann equations. For the RPM, DH determined the total electrostatic potential around a fixed ion and showed that each ion is screened by a cloud of ions of opposite charge over a typical distance $\kappa_D^{-1}$, where $\kappa_D=\sqrt{8\pi\lambda_B\rho_b}$. This means that the average potential surrounding that ion decays exponentially with the decay length $\kappa_D^{-1}$ for $r>d$. DH also calculated the electrostatic Helmholtz free energy $F^{ES}$ given by  \cite{DH}
\begin{align}
\frac{\beta F^{ES}}{V}&=-\frac{1}{8\pi d^3}\left[(d\kappa_D)^2-2d\kappa_D+2\log(d\kappa_D+1)\right],
\label{Eq:F_DH}
\end{align} 
where $V$ is the total volume. In the dilute limit $d\kappa_D\rightarrow0$ this reduces to the famous limiting law, exact in the dilute limit, 
\begin{align}\label{Eq:F_ll}
\frac{\beta F^{ES}}{V}= -\frac{\kappa_D^3}{12\pi},
\end{align}
which predicts that the electrostatic free energy density is negative (so has a cohesive  character) and vanishes with bulk concentration as $\rho_b^{3/2}$; note that in the RPM both species have the same bulk concentration $\rho_b$.

Although DH wrote down a generalized free energy expression, appropriate to models that encompass more species with different radii and valencies, the DH theory provides an accurate description  for  dilute systems only. In order to progress one requires more sophisticated extensions to DH theory that can tackle concentrated electrolytes.  An important extension is the mean spherical approximation (MSA), a closure to the OZ equation in which the hard core repulsion between ions is enforced from the outset by requiring the radial distribution functions  $g_{ij}(r)=h_{ij}(r)+1$ to vanish inside the hard core.

\subsubsection{Mean-Spherical Approximation}\label{sec:MSA_bulk}
The MSA is frequently employed to solve the OZ Eq.~\eqref{Eq:OZ}. 
For the PM the MSA imposes the conditions:
\begin{align}
g_{ij}(r)&=0, & r<d_{ij},\\
c_{ij}^{(2)}(r)&=-z_iz_j\frac{\lambda_B}{r}, & r\geq d_{ij},
\end{align} 
where $d_{ij}=(d_i+d_j)/2$ denotes the average hard sphere diameter of species\textit{ i} and \textit{j}. The first is an exact condition whereas the second constitutes the approximation. The full solution for the direct and total correlation functions, as well as the energy within the MSA, was found by Blum and others in the 70's \cite{Blum_1,Blum_2,Blum_3,Hiroike} building upon the pioneering work of Waisman and Lebowitz \cite{Waisman_1970,Waisman_1972_I,Waisman_1972_II}. The solution for the direct correlation functions can be written as 
\begin{align}\label{Eq:msa_full}
   c^{MSA}_{ij}(r)=c^{HS}_{ij}(r)+\Delta c^{MSA}_{ij}(r),  
\end{align}
where the first term is the well-known Percus-Yevick direct correlation function for hard spheres (HS), see for instance Ref.~\onlinecite{HansenMC}, and the second arises from the electrostatic interactions. For the RPM, $\Delta c^{MSA}_{ij}(r)$ takes the simple form:
\begin{align}\label{Eq:cMSA}
\Delta c_{ij}^{MSA}(r)&=
\begin{cases}
\displaystyle -z_iz_j\frac{\lambda_B}{r}\frac{2Dr-r^2}{D^2}, & r < d;\\
\displaystyle -z_iz_j\frac{\lambda_B}{r}, & r \geq d,\\
\end{cases}
\end{align}
where $D=d+1/\Gamma$, with $2\Gamma$ a parameter depending on $\rho_b$  and discussed below. The results for the electrostatic internal and free energy of the RPM are given by:
\begin{align}
\beta\frac{U^{ES}}{V}&=-\lambda_B\frac{2\rho_{b}\Gamma(\rho_b)+d q \eta(\rho_b)}{1+d\Gamma(\rho_b)},\\
\beta\frac{F^{ES}}{V}&=-\lambda_B\frac{2\rho_{b}\Gamma(\rho_b)+d q \eta(\rho_b)}{1+d\Gamma(\rho_b)}+\frac{\Gamma^3(\rho_b)}{3\pi}\label{Eq:MSAF},
\end{align} 
where $q=\rho_+z_++\rho_-z_-$ is the charge density. Of course, this vanishes in the bulk; for convenience, we retain $q$ for future reference.
However, for future reference, we retain this definition.
For every state point $\{\rho_bd^3, d/\lambda_B\}$, the parameters $\Gamma(\rho_b)$ and $\eta(\rho_b)$ must be determined self-consistently using the relations:
\begin{align}
\Gamma^2&=\pi \lambda_B \frac{2\rho_{b}-2d^2q\eta+2\rho_{b}d^4\eta^2}{\left(1+d\Gamma\right)^2} \label{Eq:MSA_gamma_bulk},\\
\eta&=\frac{1}{H(\Gamma)}\frac{dq}{1+d\Gamma}, \label{Eq:MSA_eta_bulk}
\end{align}
where it is understood  $H$ and $\Gamma$ are functions of $\rho_b$, and $H(\rho_b)$ is given by
\begin{align}\label{Eq:MSA_H_bulk}
H=\frac{d^3 2\rho_{b}}{1+d\Gamma}+\frac{2}{\pi}\left(1-\frac{\pi}{6}d^3 2\rho_{b} \right).
\end{align}
The parameter $2\Gamma$ reduces to the inverse Debye length $\kappa_D$ in the limit $d\kappa_D\rightarrow0$. However, whereas $\kappa_D^{-1}$ plays the role of a screening length in the dilute limit as we will see, $1/2\Gamma$ is merely an intermediate parameter of the theory and should not be regarded as a physical screening length. The parameter $\eta$ characterizes the symmetry of the electrolyte; it vanishes for the RPM and also for symmetric $z:z$ electrolytes with ion valencies $z$ provided the ionic radii are equal. In general, however, $\eta$  is non-zero for asymmetric electrolytes, see for instance Ref.~\onlinecite{Roth_shells}. The Helmholtz free  energy in the MSA shares some similarity with DH theory and in the $\lim_{d\rightarrow 0}$ this reduces to the limiting law in Eq.~\eqref{Eq:F_ll}.

We now turn our attention to inhomogeneous systems, for which DFT provides a powerful theoretical framework.

\subsection{Inhomogeneous Electrolytes}\label{sec:Inhom_elec}
DFT is designed to treat both the thermodynamic and structural equilibrium properties of inhomogeneous many-body systems. The key ingredient is the excess Helmholtz free energy functional $\mathcal{F}_{ex}$ defined by Eq.~\eqref{Eq:F_splitting}, which for our case should contain both the hard-core interactions and the Coulomb interactions of the ions as described by the pair potential Eq.~\eqref{Eq:pair_pot}. Those two types of interactions (hard-core and Coulombic) will be treated separately, and we split $\mathcal{F}_{ex}$ accordingly as
\begin{align}
\mathcal{F}_{ex}[\{\rho\}]=\mathcal{F}_{ex}^{HS}[\{\rho\}]+\mathcal{F}_{ex}^{ES}[\{\rho\}].
\end{align}
The first term on the right-hand side is the Helmholtz excess functional that accounts for the hard-core repulsion; this is well-described by White-Bear II (WBII) version of Fundamental Measure Theory (FMT) for hard spheres (HS), see e.g. Ref.~\onlinecite{Roth_FMT}. The second term accounts for the electrostatic interactions, which are inherently difficult to treat \cite{Andreas_2017}. In the next paragraphs we describe three functionals that treat the electrostatic (Coulombic)  interactions: a functional based on a mean-field approximation, one that uses the MSA direct correlation function Eq.~\eqref{Eq:cMSA} and one that uses both the MSA direct correlation function and the MSA expression for the Helmholtz free energy Eq.~\eqref{Eq:MSAF}. For simplicity, we focus on the RPM, but our treatment can be extended to more general cases.

\subsubsection{Mean-Field Coulomb Functional}
The easiest way to include electrostatics is within a mean-field approximation (that we call MFC), i.e. we set
\begin{align}\label{Eq:FC}
\beta\mathcal{F}_{ex}^{ES}[\{\rho\}]=\beta\mathcal{F}_{ex}^{\scaleto{ MFC}{4pt}}[\{\rho\}]&\equiv\frac{1}{2}\int\mathrm{d}\mathbf{r}Q(\mathbf{r})\phi(\mathbf{r}),
    \end{align}
where $\phi(\mathbf{r})$ denotes the dimensionless electrostatic potential, and $eQ(\mathbf{r})$ the total charge density ${Q(\mathbf{r})=Q_{ion}(\mathbf{r})+Q_{ext}(\mathbf{r})}$, with $Q_{ion}(\mathbf{r})=\sum_j z_j\rho_j(\mathbf{r})$ denoting the charge density of the ions and $Q_{ext}(\mathbf{r})$ the charge density of fixed charges, such as those on the electrode. The potential and charge density are related by the Poisson equation
\begin{align}\label{Eq:Poisson}
\nabla^2\phi(\mathbf{r})=-4\pi\lambda_B Q(\mathbf{r}).
\end{align}
Eq.~\eqref{Eq:FC} corresponds to treating Coulombic  contributions on a mean-field level; correlation effects are omitted. Note that the free energy vanishes for a homogeneous bulk system, where $Q(\mathbf{r})=0$ and $\phi(\mathbf{r})=0$. We have chosen to  include the fixed charges in $\mathcal{F}_{ex}^{MFC}$, whereas formally these should be included in the external potential. However, writing $\mathcal{F}_{ex}^{MFC}$ this way is convenient since it allows us to treat the full electrostatic potential that includes contributions from the external charges and the response of the ionic charges. With this choice, it is understood implicitly that $V_{ext}^j(\mathbf{r})$ contains only the non-electrostatic part of the external potential.

It is well-known that mean-field approaches remain reliable  if the density fluctuations are small at all positions. This implies that the accuracy of this  MFC functional is restricted to low values of the bulk ionic densities and of the fixed-charge densities. In order to describe systems with stronger electrostatic coupling, we must extend the theory. This can be achieved using results from MSA. In the following subsections we  borrow from the presentation of  Ref.~\onlinecite{Andreas_2017}. 

\subsubsection{Mean Spherical Approximation: Correlation Function}\label{sec:MSAc}
 Given the relation in Eq.~\eqref{Eq:c2F} between the pair direct correlation function and the excess Helmholtz free energy functional, a natural way to implement the explicit MSA result Eq.~\eqref{Eq:cMSA} is 
\begin{align}
\beta&\mathcal{F}_{ex}^{ES}[\{\rho\}]=\beta\mathcal{F}_{ex}^{MSAc}[\{\rho\}]\equiv\nonumber\\
&-\frac{1}{2}\sum_{ij}\int\mathrm{d}\mathbf{r}\int\mathrm{d}\mathbf{r'}\rho_i(\mathbf{r})\Delta c^{MSA}_{ij}(|\mathbf{r}-\mathbf{r'}|;\rho_{b})\rho_j(\mathbf{r'}).
\end{align}
This approximation, which has origins in Ref.~\onlinecite{MSAc}, inputs the MSA direct correlation functions evaluated at the \textit{bulk} densities $\rho_{b,\pm}=\rho_{b}$. Thus the functional is built around a certain \textit{bulk} reference system.
It is convenient to split this functional into a mean-field contribution MFC, as in Eq.~\eqref{Eq:FC}, plus corrections, i.e. 
\begin{align}
\Delta c^{MSA}_{ij}(r)=-z_iz_j\frac{\lambda_B}{r}+\Delta c^{MSAc}_{ij}(r)
\end{align}
where the first term is the MFC contribution and from Eq.~\eqref{Eq:cMSA} one finds for the RPM that
\begin{align}
\Delta c_{ij}^{MSAc}(r)=
\begin{cases}
\displaystyle z_iz_j\frac{\lambda_B}{r}\frac{\left(r-D\right)^2}{D^2} & r < d;\\
0 & r \geq d.\\
\end{cases}
\end{align}
The quantity $D$ was introduced previously just below Eq.~\eqref{Eq:cMSA}. 
It follows that, $\mathcal{F}_{ex}^{ES}=\mathcal{F}_{ex}^{MFC}+\mathcal{F}_{ex}^{MSAc}$, where  the first term is given by Eq.~\eqref{Eq:FC}, and
\begin{align}\label{Eq:FMSAc}
\beta&\mathcal{F}_{ex}^{MSAc}[\{\rho\}]=\\
&-\frac{1}{2}\sum_{ij}\underset{|\mathbf{r}-\mathbf{r}'|<d}{\int\mathrm{d}\mathbf{r}\int}\mathrm{d}\mathbf{r'}\rho_i(\mathbf{r})\Delta c^{MSAc}_{ij}(|\mathbf{r}-\mathbf{r'}|;\rho_{b})\rho_j(\mathbf{r'}).\nonumber
\end{align}
 Within the RPM, Eq.~\eqref{Eq:FMSAc} reduces to 
\begin{align}\label{Eq:FMSAcRPM}
\beta&\mathcal{F}_{ex}^{MSAc}[\{\rho\}]=\\
&-\frac{1}{2}\int\mathrm{d}\mathbf{r}\int\mathrm{d}\mathbf{r'}Q_{ion}(\mathbf{r})\Delta c^{MSAc}(|\mathbf{r}-\mathbf{r'}|;\rho_{b})Q_{ion}(\mathbf{r}'),\nonumber
\end{align}
where the $z_iz_j$ term in $c_{ij}^{MSAc}$ is used in defining the charge densities $Q_{ion}$. Hence, we have shown explicitly that this functional depends only on the charge density profiles $Q_{ion}(\mathbf{r})$ and not on the total number density profile $\rho_+(\mathbf{r})+\rho_-(\mathbf{r})$. The total \textit{bulk} density is manifest via the spatially-constant parameter ${D=d+1/\Gamma(\rho_b)}$ that enters direct correlation functions of the bulk reference system. We emphasize that the part of the MSA direct correlation function incorporating the short-range steric repulsions, i.e. $c^{HS}_{ij}$ in Eq.~\eqref{Eq:msa_full}, is treated by an accurate HS (FMT) functional; see Refs.~\onlinecite{Andreas_2017,Gillespie}. The review by Roth \cite{Roth_FMT} provides an excellent account of the FMT for HS.


\subsubsection{Mean Spherical Approximation: Free Energy}
In the previous sub-section structural information from the  MSA, i.e. the bulk direct correlation function, was used in constructing the approximate electrostatic DFT functional. However, we saw earlier that the MSA also provides the internal and free energy of the homogeneous bulk system. A natural way to incorporate the bulk free energy density from MSA into a functional is by replacing the bulk densities with local or weighted densities. Specifically, we replace the charge density $q$  and the total density $2\rho_b$, respectively, with the weighted densities \cite{Blum_c,Roth_shells}
\begin{align}
    \tilde{n}_Z(\mathbf{r})=&\int\mathrm{d}\mathbf{r'} \left(\rho_+(\mathbf{r}')-\rho_-(\mathbf{r}')\right)\omega(|\mathbf{r}-\mathbf{r}'|),\label{Eq:nZ}\\
    \tilde{n}_N(\mathbf{r})=&\int\mathrm{d}\mathbf{r'} \left(\rho_+(\mathbf{r}')+\rho_-(\mathbf{r}')\right)\omega(|\mathbf{r}-\mathbf{r}'|),\label{Eq:nN}
\end{align}
where the weight function ${\omega(r)=\delta(r-D/2)/\pi D^2}$ is chosen.  That is, ions are smeared out over a shell with diameter $D$, which is supposed to represent the range over which the charge is screened in bulk. However, as pointed out in Sec.~\ref{sec:MSA_bulk}, the parameter $\Gamma$ should not be regarded as an inverse screening length. Notwithstanding, we follow the methodology of Refs.~\onlinecite{Roth_shells,Blum_c}.
Replacing directly the densities results in the reduced free energy density, see Eq.~\eqref{Eq:MSAF}, used in Ref.~\onlinecite{Roth_shells}
\begin{widetext}
\begin{align}\label{Eq:MSA_phi_msa}
\Phi^{MSA}(\{\tilde{n}(\mathbf{r})\})=-\lambda_B \frac{\tilde{n}_{N}(\mathbf{r})\Gamma(\{\tilde{n}(\mathbf{r})\})+d\tilde{n}_Z(\mathbf{r})\eta(\{\tilde{n}(\mathbf{r})\})}{1+d\Gamma(\{\tilde{n}(\mathbf{r})\})}+\frac{\Gamma(\{\tilde{n}(\mathbf{r})\})^3}{3\pi}.
\end{align}
\end{widetext}
where $\{\tilde{n}(\mathbf{r})\}=\{\tilde{n}_N(\mathbf{r}),\tilde{n}_Z(\mathbf{r})\}$. Here, $\Gamma$ and $\eta$ are point-wise versions of Eqs.~\eqref{Eq:MSA_gamma_bulk}-\eqref{Eq:MSA_H_bulk}, i.e. they are determined in exactly the same way as for the bulk values but using $\tilde{n}_N(\mathbf{r})$ and $\tilde{n}_Z(\mathbf{r})$ at points $\mathbf{r}$ instead of $2\rho_b$ and $q$. Note that, although $\eta$ vanishes in the RPM in the bulk, the quantity $\eta(r)$ can be non-zero in the RPM when there is a non-zero fixed charge density, i.e. near charged surfaces. 
The additional functional that arises from this treatment reads
\begin{align}\label{Eq:FMSAu}
\beta\mathcal{F}_{ex}^{MSAu}[\{\rho\}]=\int\mathrm{d}\mathbf{r}\,\Phi^{MSA}(\{\tilde{n}(\mathbf{r})\}),
\end{align}
and the approximation becomes $\beta\mathcal{F}_{ex}^{ES}=\beta\mathcal{F}_{ex}^{MFC}+\beta\mathcal{F}_{ex}^{MSAc}+\beta\mathcal{F}_{ex}^{MSAu}$; see Eq.~(30) of Ref.~\onlinecite{Roth_shells}. We use the superscript $u$ to indicate the energy route.

This final addition to the electrostatic functional brings both advantages and disadvantages. By including this additional contribution  one obtains rather accurate results for density profiles  for a wide range of parameters, compared to simulations \cite{Roth_shells,Gillespie}. The contribution is also significant for the energetics, especially at lower concentrations where the electrostatic free energy scales with $\rho_b^{3/2}$. Moreover, when entering the realm of asymmetric electrolytes the $\eta$ term in Eq.~\eqref{Eq:MSAF} becomes important and can give a substantial contribution to the bulk free energy. On the downside, it turns out that this functional  breaches various requirements of consistency (see Appendix~\ref{App:ES_cons}). We shall show these considerations are important in determining  the asymptotic decay of bulk pair correlation functions and one-body density profiles. 

Three electrostatic functionals are employed in this paper. The simplest functional $\mathcal{F}_{ex}^{MFC}$, which uses only the Coulomb potential, is referred to as the mean-field Coulomb functional. The second functional, $\mathcal{F}_{ex}^{MFC}+\mathcal{F}_{ex}^{MSAc}$, which uses the bulk direct correlation function from the MSA, is referred to as the MSAc functional. And the third functional, $\mathcal{F}_{ex}^{MFC}+\mathcal{F}_{ex}^{MSAc}+\mathcal{F}_{ex}^{MSAu}$, which uses both the bulk direct correlation function and the free energy result from the MSA, is referred to as the MSAu functional. In Ref.~\onlinecite{Roth_shells} the authors use the acronym FMT/fMSA for the third functional.

\section{ DFT Calculations and MD Simulations for the RPM at a Planar Electrode}\label{sec:DFT_MD}
We  apply the density functionals of Sec.~\ref{sec:Inhom_elec} to 1:1 ionic solutions in an aqueous medium (the solvent is not treated explicitly) with a constant dielectric relative permittivity $\varepsilon_r=78$ and temperature $T=293.41$ K, corresponding to a Bjerrum length of $\lambda_B=0.73$ nm. The electrolyte consists of equal-sized cat- and anions, with hard-core diameters $d_+=d_-=d=0.5$ nm,  in contact with a planar electrode located at $z=0$ and at a fixed surface potential $\Phi_0$. Given the planar symmetry, and in the absence of any symmetry breaking transition,  the ionic density profiles $\rho_+(z)$ and $\rho_-(z)$ are a function of the distance $z$ from the wall. For the RPM we  define the dimensionless charge and \textit{ excess }  number densities as
\begin{align}\label{Eq:rho_Z}
\rho_Z(z)&=\frac{\rho_+(z)-\rho_-(z)}{\rho_b},\\
\rho_N(z)&=\frac{\rho_+(z)+\rho_-(z)}{\rho_b}-2,\label{Eq:rho_N}
\end{align}
where the bulk densities $\rho_{b,+}=\rho_{b,-}\equiv \rho_b$. As a measure for the concentrations we use the dimensionless quantity $d\kappa_D$, which scales as $\sqrt{\rho_b}$. In the electrolyte literature it is customary to introduce a reduced temperature $T^*=d/\lambda_B$. For the model we consider $T^*\approx 0.685$, which is far above the critical temperature $T^*_c\approx 0.05$ of the $1:1$ RPM \cite{RPMCP}. Thus, we avoid complications associated with liquid-gas phase separation. In practice we consider a planar slit geometry with two identical charged walls at $z=0$ and $z=H$ separated by a distance $H$, sufficiently large that the density profiles for  $z=H/2$ are very close to their bulk values at the specified chemical potential and temperature. 
In order to test the predictions of the various DFT approximations, we carried out extensive Molecular Dynamics (MD) simulations of the density profiles for the same range of parameters 
using the ESPResSo package \cite{weik_espresso_2019}.

In the simulation we measure energy in $k_{\textrm{B}}T$, 
length in $1$ nm, and time in $[\textrm{length}\sqrt{\textrm{mass}/\textrm{energy}}]$ which is set by a mass of $3\cdot 10^{-23}$ g, resulting in a time unit of $2.699$ ps. 
Whereas our DFT calculations are performed in a grand canonical ensemble with fixed chemical potentials $\mu_\pm$ and fixed surface potential $\Phi_0$, the MD simulations are naturally performed in the canonical ensemble with fixed numbers of ions $N_\pm$ and fixed surface charge densities $\pm eQ_\text{W}$. We employ two oppositely charged electrodes and fix $N=N_+=N_-$. Direct comparison between DFT and MD results is  possible because we focus on matching bulk behaviour in the center of the slit at $z=H/2$. 
The oppositely charged walls allow us to 
account for the surface charge density $eQ_\text{ext}=eQ_\text{W}(\delta(z)-\delta(z-H))$ at the walls by applying an additional 
constant force to all particles in the simulations. 
As for a parallel plate capacitor, this force on the ionic charges stems from the electric field $4\pi k_\text{B}T\lambda_\text{B}Q_\text{W}/e$. 

The electrostatic interactions between the ions are treated in ESPResSo using the P3M method 
\cite{weik_espresso_2019, hockney_computer_1988}, a sophisticated Ewald method. The hard core interactions between ions are modeled by the
Weeks-Chandler-Anderson potential \cite{andersen_relationship_1971, weeks_role_1971}
\begin{align}
 u_{\textrm{WCA}}(r) = \left\{ 
\begin{matrix} 4\epsilon \left( 
\left( \frac{\sigma_{\textrm{LJ}}}{r} \right)^{12} - 
\left( \frac{\sigma_{\textrm{LJ}}}{r} \right)^6 + 
\frac{1}{4}
\right) & r<d \\ 0 & r\geq d , \end{matrix} \right.
\end{align}
with $\epsilon=0.5\cdot 10^4 k_BT$ and $\sigma_\text{LJ}=d/2^{1/6}$ such that the potential is purely (and strongly) repulsive and its derivative is continuous at the diameter $d$. 
In order to model the effect of the hard walls we set the wall-ion interaction potential to $u_\text{WCA}(-\tfrac{d}{2}+z)$ for the wall at $z=0$ and 
$u_\text{WCA}(H+\tfrac{d}{2}-z)$ for the wall at $z=H$. 

Ion trajectories are calculated in a simulation box of volume $L_x\times L_y\times L_z$ with periodic boundary conditions and $L_z=H$. 
In order to restrict electrostatic interactions to the volume between the two walls (without contributions from periodic copies in the $z$-direction), we use an electric layer correction which is built into ESPResSo. This method allows one to use the aforementioned fast P3M method that assumes periodicity in all three dimensions and then efficiently corrects for the unwanted contribution from the periodicity in the $z$-direction\cite{elc-method-part1,*elc-method-part2}. The method requires an additional region of empty space in the form of an extension of the simulation box in the $z$-direction and we set its length to $0.15 L_z$.
Further, we choose $L_x=L_y$ such that the number of ions $ N$ is sufficiently large to fix the average densities of ions when we compare results between MD and DFT. 
For making comparisons, the starting MD values are obtained by preliminary DFT simulations in order to achieve approximately  the same $d\kappa_D$ and the potential $\Phi_0$ of interest. Then, $d\kappa_D$ and the surface potential $\Phi_0$ were deduced from the MD simulations, which are subsequently used as input for the DFT calculations. The input values for the MD simulations ($Q_W$, $N$ and $L_x$) and DFT calculations ($d\kappa_D$ and $\Phi_0$) for the following three sets were 
\begin{itemize}
\setlength{\itemsep}{0pt}
\setlength{\itemindent}{0.3cm}
    \item[Set 1] $Q_\text{W}=0.00427$ nm$^{-2}$, $N=1977$, $L_x=50$ nm \\
    \hspace*{0.3cm}$d\kappa_D=0.619$, $\Phi_0=1$ mV, 
    \item[Set 2] $Q_\text{W}=0.913$ nm$^{-2}$,  $N=1983$, $L_x= 22.5$  nm \\
    \hspace*{0.3cm}$d\kappa_D=1.286$, $\Phi_0=105$ mV, 
    \item[Set 3] $Q_\text{W}=1.334$ nm$^{-2}$,  $N=1968$, $L_x=13.5$  nm \\
    \hspace*{0.3cm}$d\kappa_D=2.243$, $\Phi_0=72.19$ mV.
\end{itemize}

In determining the density profiles, we averaged particle positions over several snapshots at different times and 
in different simulations. For this purpose, we sampled the density profiles $\rho_i$ on the interval $[0,L_z]$ 
using an equidistant binning of 200 bins.  In each simulation set, we used snapshots after 100 time steps of 
step length $0.0001$ time units. The sampling time for Sets 1/2/3 was 290/515/610 time units after 
7.6/7.6/12.6 time units of equilibration, corresponding to averaging over around 0.87/1.57/1.84 million snapshots. Profiles are shown in Fig.~\ref{Fig:rho} where comparison is made with DFT results.

For the calculation of the pertinent decay lengths, we can choose to focus on the one-body density profiles, as discussed above, and shown in Fig.~\ref{Fig:rho}. Alternatively, as described in Sec.~\ref{sec:results}, we can choose to focus on pair-distribution functions for a bulk system. 
(Our choice of strategy will become clear in Sec.~\ref{sec:asymptoticDecayInBulk}, where we explain how the asymptotic decay of the bulk pair correlations $h_{ij}(r)$ connects directly to that of the one-body density profiles.) In bulk, the pair correlation functions are translationally invariant and for a prescribed computational effort their calculation leads to much better statistics than for an inhomogeneous system. For this reason, we performed \textit{bulk} MD simulations to calculate the decay lengths shown in Figs.~\ref{Fig:xi} and \ref{Fig:xi_lit}. We used a cubic simulation volume $L_x=L_y=L_z=30$ nm with periodic boundary conditions and sampled pair-distribution functions on the interval $[0,5]$ nm using 400 equidistant bins. The number of snapshots for taking the averages range from around $16500$ at $0.1$ M, reduced to around $2500$ at $6$ M.
 In order to calculate the decay lengths $\xi_N$ and $\xi_Z$ of, respectively, the total number and charge bulk  pair correlations $h_N(r)$ and $h_Z(r)$  we first choose to fit with  functions of the form of Eq.~\eqref{Eq:h_asymptotic} (see below), that assumes simple poles determine the asymptotic decay, over a range where our simulation data is sufficiently accurate. Typically we fit over a short range, approximately between $r=0.9\dots 1.1$ nm, limited by numerical noise and short decay lengths. Nevertheless, depending on the system parameters, this range could span down to $d$ and up to several nanometers. Furthermore, at these relatively short distances, often more than one pair $(A_n,\alpha_n)$ of amplitudes and poles (exponentials with different decay lengths) is required to fit the pair correlation functions obtained from simulations. Performing individual fits for each state point, we find two pairs are sufficient to fit the data. However, for number correlations at small concentrations we were guided by earlier literature \cite{Ulander_2001,Ennis} on asymptotic decay in bulk electrolytes where it was established that a branch point singularity dominates the decay except at extremely large $r$. The predicted decay for $h_N(r)$ is given in Eq.~\eqref{Eq:h_asymptotic_branch} (see below). For the narrow range of $r$ we have data available, a function of this form with $B=0$ provides an adequate fit. Although more advanced methods are available to extract the asymptotic decay lengths\cite{Ulander_2001,Gonzales_2013}, the relatively simple scheme we implement proved to be sufficient in the range of concentrations that we were most interested in.

\section{Results For Density Profiles and Decay Lengths}
\label{sec:results}

 DFT proves to be a valuable microscope in the near field, close to the electrode, where comparison with simulation is straightforward. It is also crucially important as a telescope in the far field, where the simulation results are limited by system size. 
 \subsection{Near Field}\label{sec:near_field}
\begin{figure*}
\centering
\includegraphics[width=0.49\textwidth]{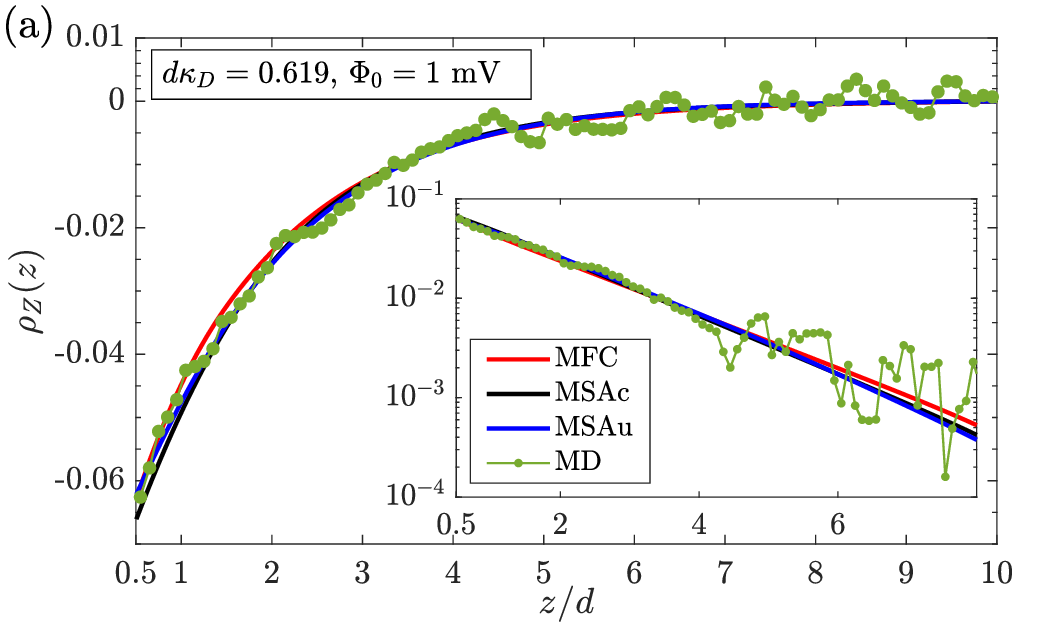} \includegraphics[width=0.49\textwidth]{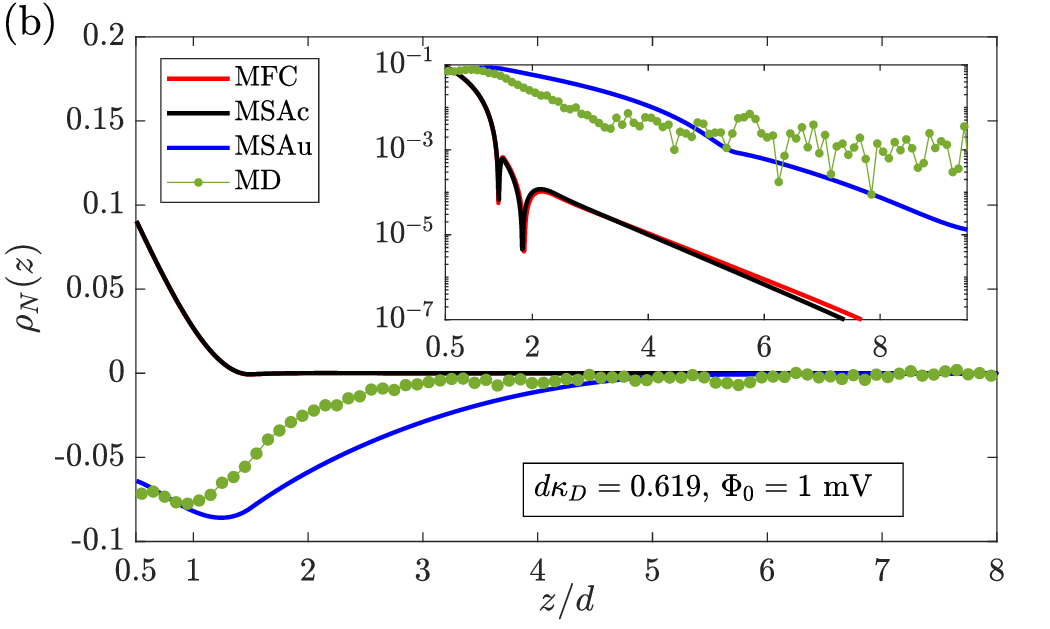} \\
\includegraphics[width=0.49\textwidth]{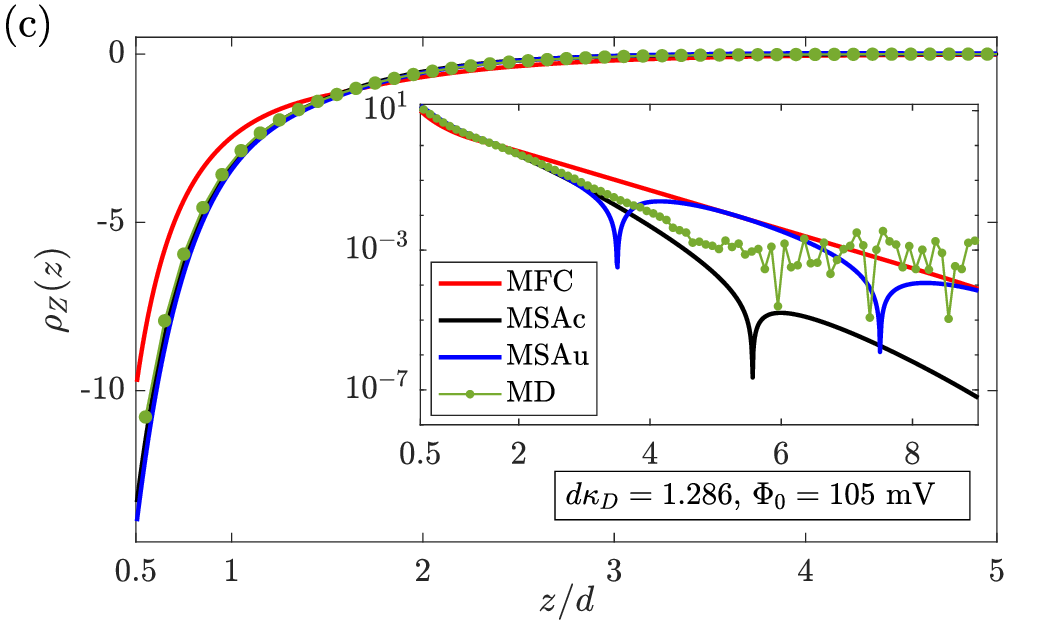} \includegraphics[width=0.49\textwidth]{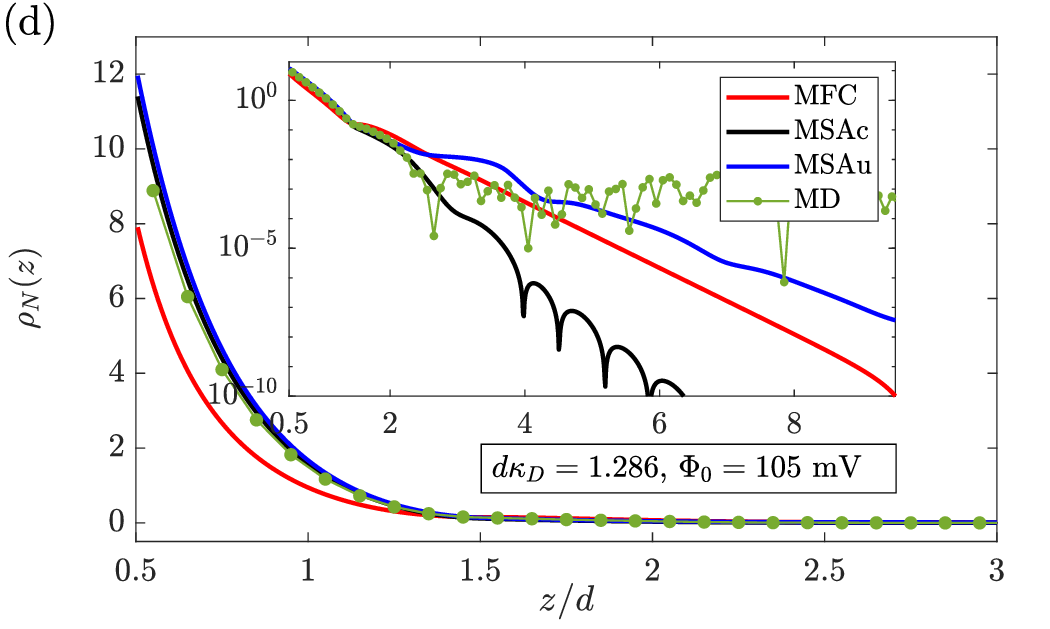} \\
\includegraphics[width=0.49\textwidth]{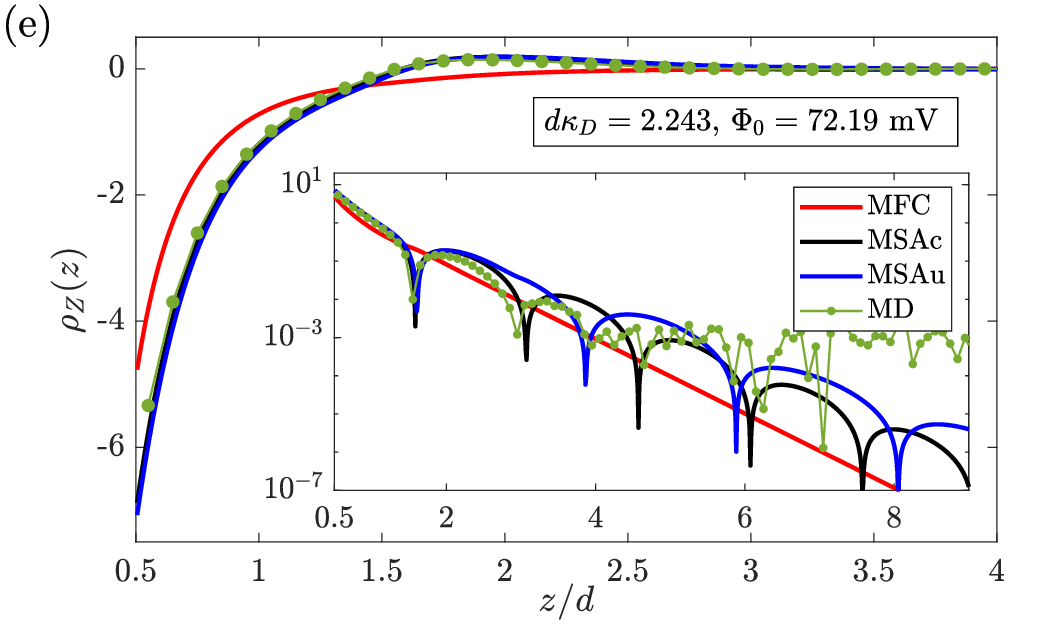} \includegraphics[width=0.49\textwidth]{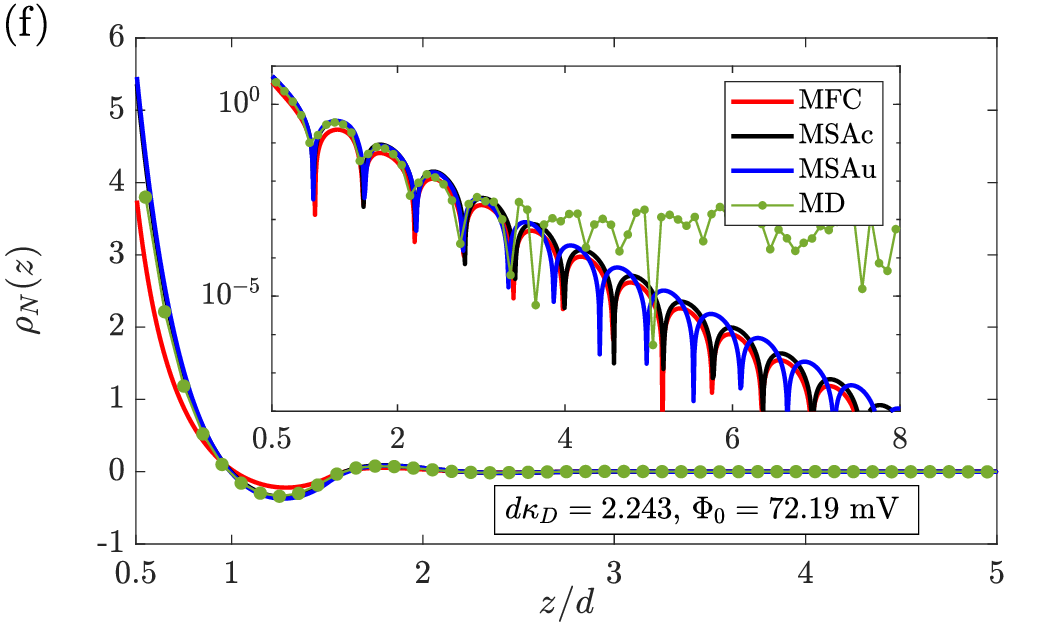} 
\caption{\label{Fig:rho} In (a), (c), and (e) the dimensionless charge density profile $\rho_Z(z)$ and in (b), (d), and (f) the dimensionless excess number density profile $\rho_N(z)$ for the RPM, modelling an aqueous 1:1 electrolyte, with ionic diameter $d=0.5$ nm in contact with a charged planar electrode located at $z=0$.  The surface potential $\Phi_0$ and the  ionic concentration, expressed as $d\kappa_D$, are given in each panel, while the number of ions and surface charge density for the simulations are given in Sec.~\ref{sec:DFT_MD}. Results obtained from MD simulation (green dots) and the three different DFT approximations (solid lines) are shown. The insets show the modulus of the density profiles plotted on a logarithmic scale. These plots, together with many others not shown here, are used to determine asymptotic decay lengths in the DFT studies. }
\end{figure*}

 In Fig.~\ref{Fig:rho} we plot the charge density profiles $\rho_Z(z)$ (left column) and excess number density profiles $\rho_N(z)$ (right column) obtained from the functionals MFC (red), MSAc (black) and MSAu (blue) and from MD simulations (green) for the sets of parameters given in the previous section.
 
In the near field, there is excellent agreement between the density profiles obtained from the MSAc and MSAu functionals and those from simulations, with the exception of $\rho_N(z)$  in Fig.~\ref{Fig:rho}(b). For this small concentration and low surface potential one observes depletion in the excess number density $\rho_N(z)$ near an electrode. This is caused by the negative electrostatic free energy (accounted for only by the MSAu functional) which dominates over the hard-sphere free energy; whereas the latter scales as $\rho_b^2$, the former scales as $\rho_b^{3/2}$. Apart from these special cases,  the difference between the MSAc and MSAu profiles in the near field is negligible. Hence, we deduce that adding the energy term to the functional is only important for the RPM at small concentrations and surface potentials.
 
 Note that the density profiles from the MFC functional are quite different from those of the MSA functionals. Clearly the details of near-field structure depend on the terms in the functional carrying the pair  direct correlation function (see Eq.~\eqref{Eq:FMSAc}), that are absent  in the MFC. 

Although in the near field the distinction in the density profiles between the MSAc and MSAu functionals is fairly small, one can easily distinguish those in the far field (see insets in Fig.~\ref{Fig:rho}). We focus on this important observation in the next subsection and in  Appendix~\ref{App:ES_cons}.


\subsection{Far Field}\label{sec:far_field}
We turn now to the interpretation of the far-field density profiles, i.e. results pertinent to large $z$  in Fig.~\ref{Fig:rho}. The  insets show that oscillations develop  in the asymptotic decay of both the charge and number density profiles as the concentration $d\kappa_D$  is increased. This is especially clear in the sequence  for the  MSAc number density profiles.

\subsubsection{Asymptotic Decay of Bulk Pair Correlations}
\label{sec:asymptoticDecayInBulk}

In the far field, we focus on the asymptotic decay of the one-body charge $\rho_Z(z)$  and excess number $\rho_N(z)$ densities far from the electrode(wall). The leading asymptotics for these densities are determined by the asymptotic decay of pair correlation functions in the corresponding uniform (bulk) fluid \cite{Evans_1993,Attard_1991,Attard_1992,Evans_1994,Kjellander_1992}. Appendix~\ref{App:decay_onebody} provides a simple argument. 
In the bulk, the asymptotic, large\textit{ r},  behavior of pair correlation functions can be obtained from the singularities that appear in the Fourier-transformed OZ equation (Eq.~\eqref{Eq:OZ}). For a single-component (neutral) system this takes the form
\begin{align}\label{Eq:h_k}
\hat{h}(k) = \frac{\hat{c}^{(2)}(k)}{1-\rho_b\hat{c}^{(2)}(k)}.
\end{align}
For models with short-ranged pair potentials (exponentially or faster decaying or of finite range) we expect the dominant singularities in Eq.\eqref{Eq:h_k} to be simple poles, at least for intermediate to high bulk concentrations. In this case the leading decay in three dimensions is given by 
\begin{align}\label{Eq:h_poles}
rh(r)\approx\sum_n \operatorname{Re}\left(A_n e^{i\alpha_n r}\right),
\end{align}
where $\operatorname{Re}$ denotes taking the real part.  $A_n$ is an amplitude and $\{\alpha\}$ denotes the set of poles $n$ with positive imaginary part in the complex $k$-plane, that satisfy the condition ${1-\rho_b\hat{c}^{(2)}(\alpha_n)=0}$.  When the poles are complex the asymptotic  behavior is determined by the pole ${\alpha=2\pi/\lambda+i\kappa}$, and its conjugate,  having the smallest imaginary part $\kappa$. The leading oscillatory decay of the total correlation function is then given in 3 dimensions by
\begin{align}\label{Eq:h_asymptotic}
rh(r)\underset{r\rightarrow\infty}{\approx} A\cos(2\pi r/\lambda+\varphi)e^{-\kappa r},
\end{align}
where the amplitude $A$ and phase $\varphi$  can be obtained from the residues\cite{Evans_1993,Evans_1994}. 
Generally, there are also  pure imaginary poles ${\alpha_n=i\kappa}$ giving rise to purely exponential decay of $rh(r)$. Whether the ultimate decay of $rh(r)$  is damped oscillatory or monotonic at a particular state point depends on whether the  lowest lying  pole, i.e. that with the smallest value of $\kappa$, is complex or pure imaginary. For model fluids exhibiting repulsive and attractive portions in the pair potential there is a crossover line in the phase diagram where the asymptotic decay of  $rh(r)$ changes from monotonic to damped oscillatory, termed the  Fisher-Widom (FW) line
\cite{Evans_1993,Evans_1994,Evans,FisherWidom,Dijkstra_2000}.
The procedure we employ for obtaining the asymptotic decay length, $\xi=1/\kappa$, in the bulk fluid is termed the IET route, since we usually invoke an integral equation closure or another explicit approximation, gleaned say from DFT,  for the bulk pair direct correlation function.
 
\begin{figure}
\includegraphics[width=\columnwidth]{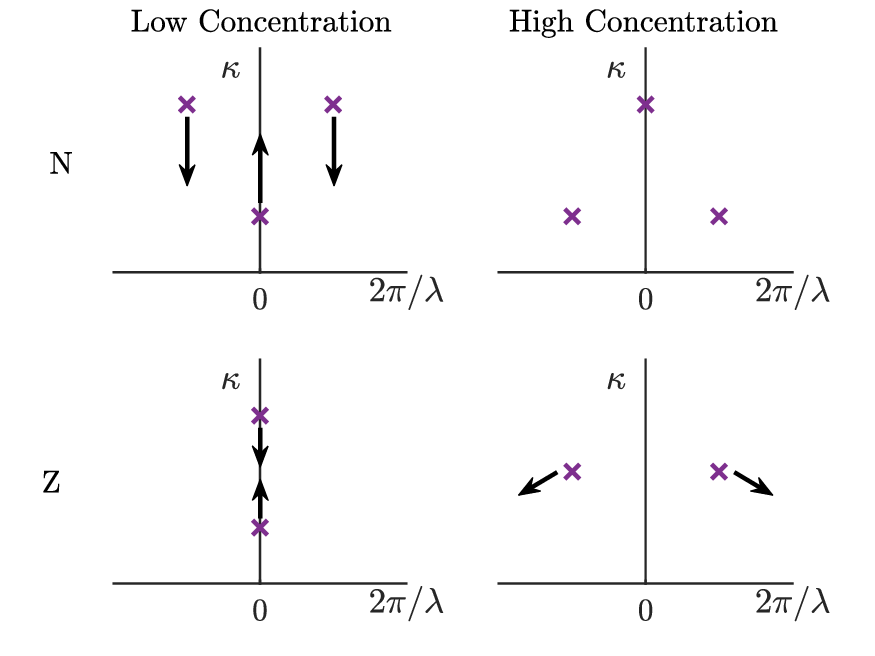}
\caption{\label{Fig:pole} The pole structure of the number-number (N) (top) and charge-charge (Z) (bottom) pair correlation function when increasing from low (dimensionless) concentrations $d\kappa_D$ (left) to high concentrations (right), as determined by IET.  The top describes  Fisher-Widom crossover and the bottom Kirkwood crossover. The scenario presented here makes no reference to other singularities, e.g. branch points; see text. }
\end{figure}

 For a system with two species, in our case cations (+) and anions (-), we must consider the total correlation matrix\begin{align}\label{Eq:Hmatrix}
H(r)=
\begin{bmatrix}
h_{++}(r)&h_{+-}(r)\\
h_{-+}(r)&h_{--}(r)
\end{bmatrix}
\end{align}
for which the Fourier-transformed OZ equation reads
\begin{align}
\hat{H}(k)=\left(\mathbb{1}-\hat{C}(k)\rho\right)^{-1}\hat{C}(k), \label{Eq:hMatrixFT}
\end{align}
where $\hat{C}$ has the same structure as in Eq.~\eqref{Eq:Hmatrix} and $\rho$ is a diagonal matrix whose elements are the bulk densities of each species.
Singularities on the r.h.s. of Eq.~\eqref{Eq:hMatrixFT} determine the asymptotic decay of the total correlation functions.
Within the RPM, $h_{++}=h_{--}$ and $h_{+-}=h_{-+}$ and it is  convenient to work with the combinations $h_N=h_{++}+h_{+-}$ and $h_Z=h_{++}-h_{+-}$, corresponding to the number-number $N$  and charge-charge $Z$ total correlation function, respectively. The combinations $h_N$ and $h_Z$  also follow naturally for the RPM from diagonalizing the matrix $H$. The special symmetry of the RPM suggests that these are decoupled and therefore the inverse decay lengths  $\kappa_N$ and $\kappa_Z$ are independent. Indeed within IET's that admit only simple poles this is the case \cite{Evans}.  Fig.~\ref{Fig:pole} summarises the pole structure of $h_N$ and $h_Z$  obtained from an approximate IET study (See Ref.~\onlinecite{Evans}).  The inverse decay length $\kappa$  is plotted on the vertical axis and the inverse wavelength $2\pi/\lambda$ on the  horizontal axis; crosses indicate a pole. The $N$ pole structure indicates that for small concentrations the pole with the smallest imaginary part is pure imaginary, and therefore the pair correlation function $h_N$ must decay monotonically. At larger concentrations, the conjugate pair of poles with the smallest imaginary part has a non-zero real part; $h_N$ will then exhibit oscillatory asymptotic decay. Hence, there should be crossover  from monotonic to oscillatory asymptotic $N$ decay, c.f. the FW crossover described above, as the concentration $d\kappa_D$ is increased. The $Z$ pole structure is different, as shown in the bottom panel of Fig.~\ref{Fig:pole}. Although one finds monotonic asymptotic decay of $rh_Z(r)$ at low concentrations and oscillatory decay at large concentrations, the crossover mechanism is that due to Kirkwood Ref.~\onlinecite{Kirkwood}. The key difference between the two types of crossover is:  at a FW point the real part of the pole with the smallest imaginary part jumps discontinuously  from zero to some non-zero  value, whereas  at a Kirkwood point the pole with the smallest imaginary part moves continuously away from the imaginary axis. Hence, the wavelength of oscillations diverges for the $Z$  decay at a Kirkwood point, but not for the $N$ decay at a FW point. (See Ref.~\onlinecite{Evans}).

Ionic systems bring additional subtleties. In particular, singularities other than simple poles are expected, reflecting `residual' coupling between number and charge correlations. This was recognized long ago by Kjellander and coworkers, e.g. Refs.~\onlinecite{Ulander_2001, Ennis,Kjellander_1992,Kjellander_1994}. Careful asymptotic analysis  for the bulk RPM reveals both a pole and a branch point singularity for number-number correlations implying
\begin{align}\label{Eq:h_asymptotic_branch}
h_N(r)\approx B \frac{e^{-\kappa_N r}}{r}+A \frac{e^{-\beta_N r}}{r^2}
\end{align}
should  provide an adequate description of the asymptotic decay. At moderately large values of $r$ and for low-intermediate ionic concentrations the second (branch point) term is expected to dominate. The same asymptotic analysis shows that the branch point  term gives the exponential decay length $\beta_N^{-1}=\kappa_Z^{-1}/2$, i.e. half that of the charge correlation length $\kappa_Z^{-1}$.  At low ionic concentrations $B\rightarrow0$ and we choose to fit $h_N(r)$ from simulations  according to Eq.~\eqref{Eq:h_asymptotic_branch}  with $B=0$, as indicated earlier. This procedure is, of course, empirical. In reality $B\neq 0$ and the pole contribution takes over as the concentration increases; see e.g. Eq.~(44) in Ennis et al. and Fig.~1 in Ulander and Kjellander \cite{Ulander_2001}. The decay length $\xi_N$ reported later, will be the larger of either $1/\kappa_N$ (pole) and $1/\beta_N$ (branch point). In Appendix~\ref{App:DFT_asymptotic} we discuss the origin of the term in the one-body number density profile at a planar wall that is analogous to the term corresponding to the branch point in Eq.~\eqref{Eq:h_asymptotic_branch}.



\subsubsection{Asymptotic Decay of One-Body Density Profiles at a Planar Electrode}

As mentioned above, there is a large body of work demonstrating that the asymptotic decay of the one-body density profiles of fluids adsorbed at planar walls is governed by the same physics that determines the decay of bulk pair correlation functions. Specifically, if we know the leading singularities from a calculation of the bulk pair direct correlation functions, in principle we know the decay lengths and the wavelength of any oscillations pertinent to the asymptotic decay of the density profiles at a planar electrode, see Appendix~\ref{App:decay_onebody}. This key observation  motivates our subsequent analysis. For example, as  $z\rightarrow \infty$, the charge and total density profiles in the RPM should take the form:
\begin{align}
\rho_{i}(z)\propto \cos(2\pi z/\lambda_{i})e^{-z/\xi_{i}}, \quad i\in\{Z,N\},
\end{align}
when an oscillatory contribution  dominates.
Then the asymptotic decay  lengths $\xi_Z$ and $\xi_N$  are identical to the corresponding decay length of the bulk fluid. In an oscillatory regime, the wavelengths $\lambda_Z$ and $\lambda_N$  are  identical to the corresponding bulk values. Moreover, any crossover that occurs in bulk must be reflected in the decay of the one-body profiles. Note that we have not indicated any amplitudes or phases in this equation. There is no simple way of determining these. Contrast this with the decay of bulk pair correlation functions where the amplitudes and phases are determined from the residues in the OZ analysis.

\begin{figure}
\includegraphics[width=0.48\textwidth]{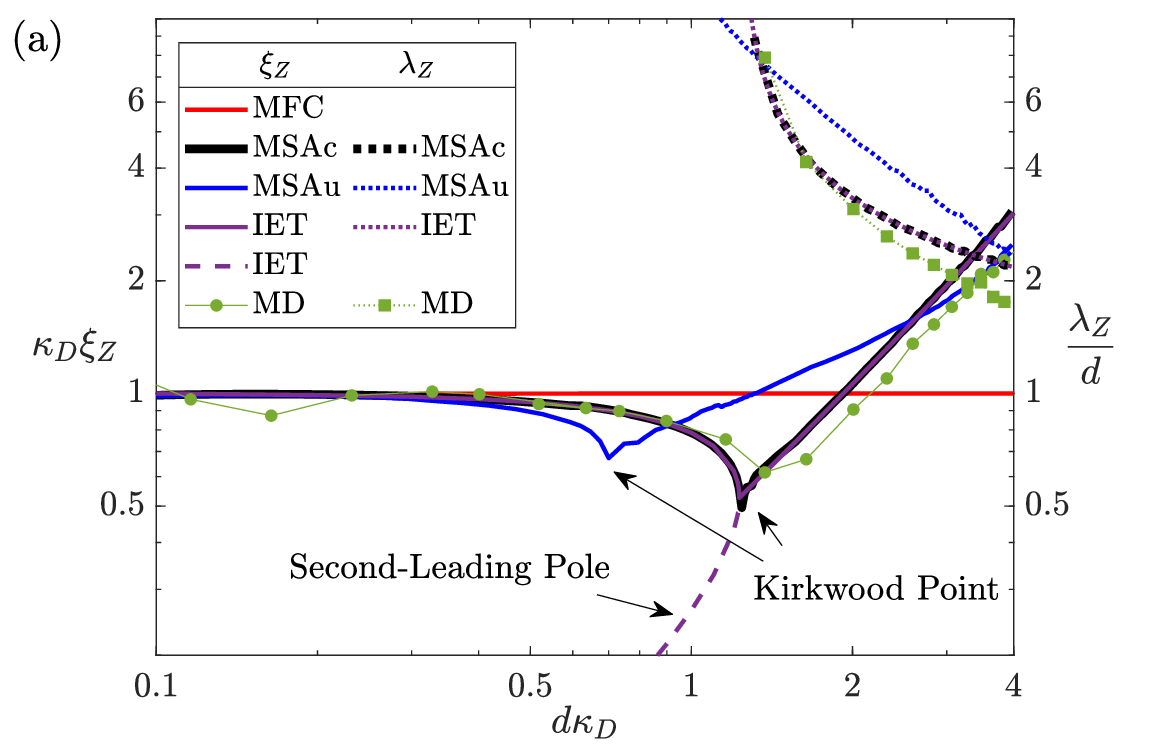}\\
\includegraphics[width=0.48\textwidth]{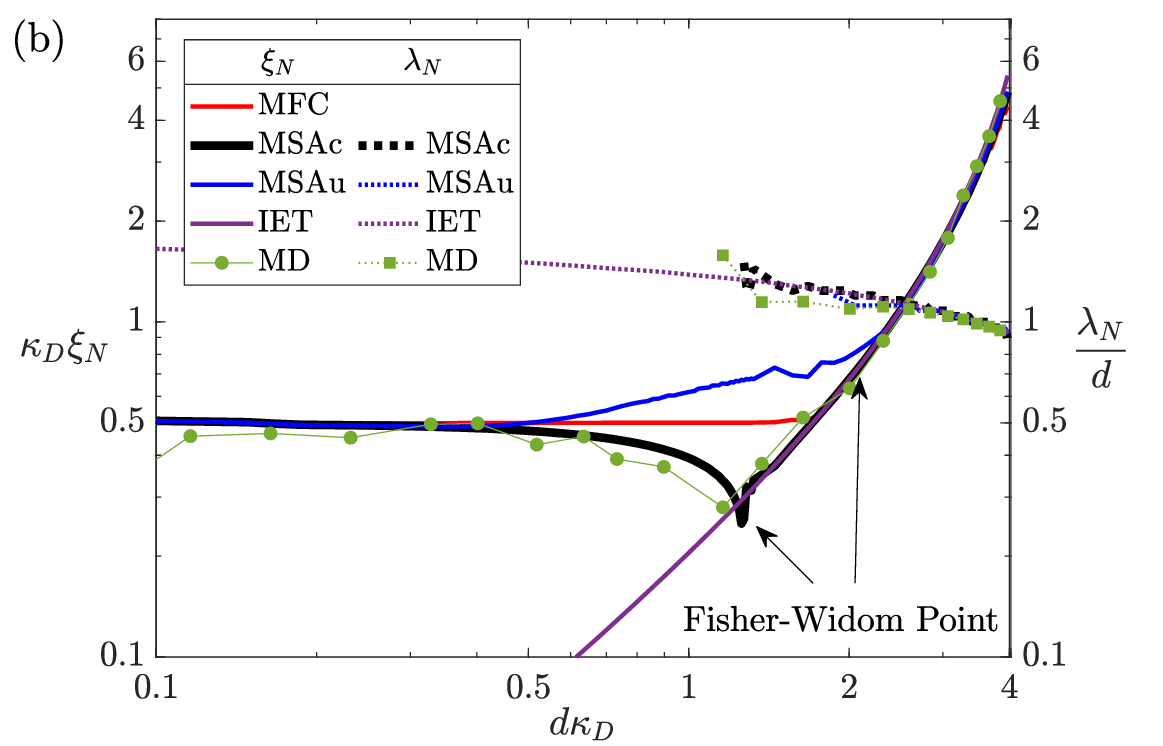}
\caption{\label{Fig:xi} (a) The charge decay length $\xi_Z$ (solid) and wavelength $\lambda_Z$ (dotted). The DFT results are for the one body density profiles obtained from the MFC (red), MSAc (black) or MSAu (blue) functional for the electrostatics. The results from IET (purple) correspond to the bulk IET MSA closure. These and the MD simulation results (green)  are for the decay of bulk pair correlation functions. The dashed-purple line in (a) represents the pole with the second-smallest imaginary part, i.e. the second-leading pole. Arrows point to the cusps where (Kirkwood) crossover from monotonic to oscillatory decay occurs. (b) The number decay length $\xi_N$ and wavelength $\lambda_N$  uses the same color coding as in (a). The IET route uses the FMT (HS) direct correlation function. The MD data for the decay length  $\xi_N$ below the Fisher-Widom point were fitted using the functional form of Eq.~\eqref{Eq:h_asymptotic_branch} with B=0.  Our numerical  results imply $\xi_N$ is close to the theoretical prediction $\xi_N=1/2\kappa_Z$. Results for $\xi_N$ calculated from the DFTs are plotted for the surface potential $\Phi_0=100$ mV; see text. Arrows point to  a (Fisher-Widom)  crossover from monotonic to oscillatory decay.}

\end{figure}

Guided by these observations, we can attempt to analyze the far-field results in Fig.~\ref{Fig:rho}. Extracting the asymptotic decay lengths from the one-body profiles calculated within DFT and simulations is non-trivial, since we must deal with  numerical limitations. Within DFT the asymptotic decay lengths $\xi_Z$ and $\xi_N$ and wavelengths $\lambda_Z$ and $\lambda_N$ are extracted from fits to the density profiles in Fig.~\ref{Fig:rho}. For MD we performed bulk simulations to achieve better statistics; see Sec.~\ref{sec:DFT_MD}. We confirmed that the results for the various decay lengths in DFT were independent of the surface potential $\Phi_0$. In  Fig.~\ref{Fig:xi}, we present the decay lengths, multiplied by the inverse  Debye length $\kappa_D$ (solid lines), and wavelengths divided by the HS diameter (dotted lines) obtained  by fitting the DFT results together with results from the IET route (purple).  
 For the latter we use the $ZZ$ pair direct correlation function from the MSA to determine the charge, $Z$, decay. This MSA result is well-known and the resulting poles are reported, e.g. in Refs.~\onlinecite{Evans,Outhwaite_1975}. The HS pair direct correlation function from FMT (WBII) is used for the $N$ decay. This treatment of number-number correlations  captures only the contributions from HS (steric) interactions. The  results from the MFC functional are plotted in red and those of the MSAc(u) functional in black(blue) while the results from MD simulations are plotted in green.

As predicted, $\xi_Z$ in Fig.~\ref{Fig:xi}(a) extracted from the MFC functional (red line) is given by the Debye length for all concentrations. At very low concentrations, $d\kappa_D\ll 1$,  the true decay length must converge to the Debye length for all theories, as dictated by the limiting law. Precisely how  $\xi_Z\kappa_D$  approaches unity at $d\kappa_D=0 $ is important and we return to this later. At intermediate concentrations ($d\kappa_D>0.5$), the limiting law is no longer valid and  $\xi_Z$  is found to be smaller than the Debye length. The  decay length obtained from the MSA IET  is universal as a function of $d\kappa_D$~\cite{Evans} and is given by the purple line. From its construction, the  MSAc functional should yield identical results and within our numerical accuracy it does; see black line. The MSAu functional (blue line), on the other hand, behaves quite differently. We argue this is due to the inconsistency inherent within this functional (Appendix~\ref{App:ES_cons}), which results in incorrect asymptotic behaviour.
The kinks that are observed for the DFT results indicate that the Kirkwood transition occurs at (using the notation $x=d\kappa_D$) $x_{K}^{MSAc}\approx 1.24$ and  $x_{K}^{MSAu}\approx 0.7004$  while the MSA IET value is $x_{K}^{IET}\approx 1.229$, As expected, the MSAc and IET Kirkwood points agree closely, i.e. to within 1 percent which is within the error of the fitting procedure used to calculate $\xi^{MSAc}_Z$. Strikingly, the MSAu Kirkwood point is smaller by almost a factor of two. The genesis of the kinks becomes clear when, within MSA IET, one plots the second smallest imaginary pole (purple-dashed line in Fig.~\ref{Fig:xi}(a)).  This plot indicates that the two smallest poles lie on the imaginary axis and move towards each other with increasing $d\kappa_D$, merging at the Kirkwood point. For larger concentrations the poles  move away from the imaginary axis, one to positive real values and the  other to equal but negative real values (as depicted in Fig.~\ref{Fig:pole}). The density profiles develop oscillatory decay for $d\kappa_D>x_{K}$ (see dotted lines), beginning with infinite wavelength at $d\kappa_D=x_K$. We find that the wavelengths from the MSAc and IET results are almost identical while the wavelength from the MSAu functional is very different. The MD results (in green) agree rather well with those from MSAc, and therefore with MSA IET. There is an indication within MD of a Kirkwood point at  around  $d\kappa_D\approx 1.37$ and for larger values of $d\kappa_D$ the MD results for $\xi_Z\kappa_D$ increase linearly with $d\kappa_D$ as found in MSA IET. Moreover, the wavelengths are close.\\

\begin{figure}
\includegraphics[width=0.48\textwidth]{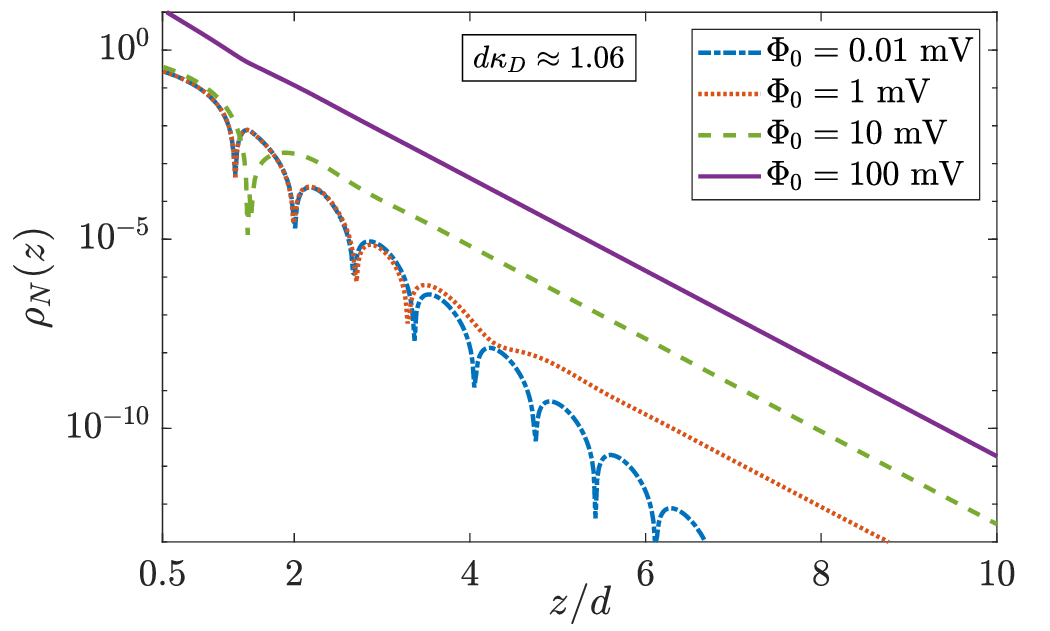}
\caption{\label{Fig:N_decay} The decay of the excess number density $\rho_N(z)$ obtained using the MSAc functional for several surface potentials $\Phi_0$ at fixed concentration $d\kappa_D=1.0589$ that is slightly below the Fisher-Widom point (see Fig.~\ref{Fig:xi}). For $\Phi_0=0.01$ mV we appear to observe only oscillatory asymptotic decay whereas for larger $\Phi_0$  we observe the true monotonic decay; see text.}
\end{figure}

\begin{figure}
\includegraphics[width=0.48\textwidth]{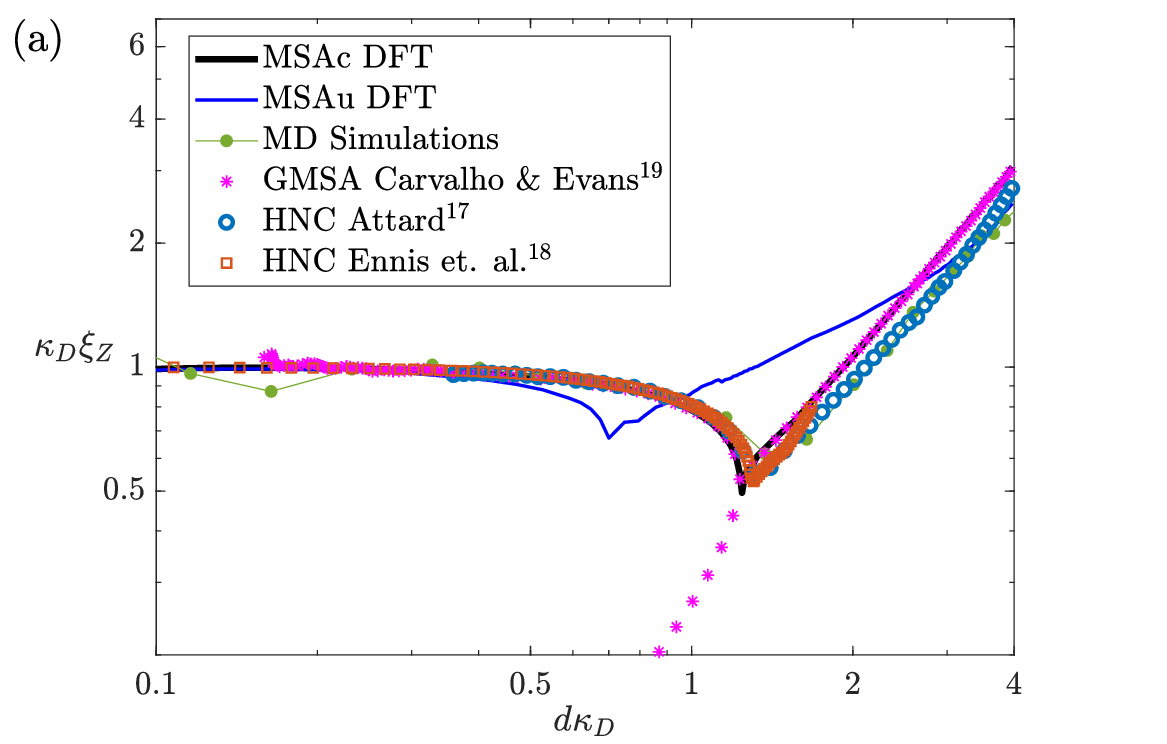}\\
\includegraphics[width=0.48\textwidth]{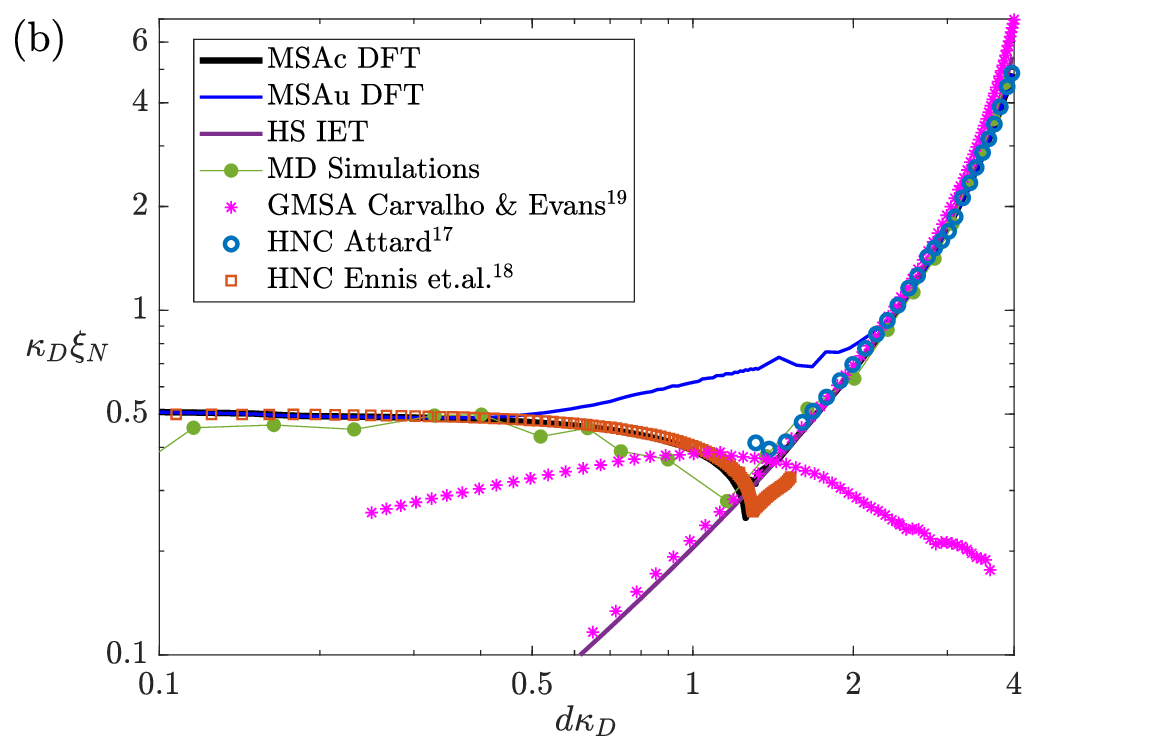}
\caption{\label{Fig:xi_lit}  The decay lengths obtained from different theories are compared with simulation results for the bulk decay lengths (green dots). In (a) the Z decay length obtained from our DFT calculations of one -body profiles (solid lines) are shown along with  the bulk decay lengths obtained using the GMSA from  Carvalho \& Evans Ref.~\onlinecite{Evans} (purple asterisks), the HNC from Attard \cite{Attard} (blue circles) and the HNC from Ennis et al. \cite{Ennis} (orange squares). In (b) the N decay length from the same sources, but also including the result from the HS IET (solid purple line).}
\end{figure}

In Fig.~\ref{Fig:xi} (b) we present the corresponding results for $\xi_N$ and $\lambda_N$; the color coding is the same as in (a). For the MFC functional  $\xi_N$ is exactly one half the Debye length  until pure HS correlations dominate at high concentrations. For the other two functionals,  $\xi_N\kappa_D $  is close to $1/2$ at small concentrations, $d\kappa_D<0.5$. This is expected and will be explained below. At high concentrations, $d\kappa_D>2$, the  $N$ decay lengths collapse onto a single curve and follow the result from IET, where, for all concentrations,  $\xi_N$ is obtained from the HS  pair direct correlation function given by  FMT. This collapse indicates that for sufficiently high concentrations the asymptotic N decay is  determined by  hard-sphere repulsion: electrostatic interactions hardly play a role. This is also reflected in the wavelengths (dotted lines). For the IET route $\lambda_N$ corresponds to the wavelength of the bulk (asymptotic) oscillations for the HS fluid and the MD simulation results (plotted in green)  agree closely. At intermediate concentrations, $0.5<d\kappa_D<2$, the three DFT  functionals show  very different results.  The decay lengths extracted from the MD simulations agree well with those from MSAc \sout{and IET}, for both the N and Z decay lengths. There are small differences in the Z decay length for concentrations beyond the Kirkwood point. However, as we will see in Fig.~\ref{Fig:xi_lit}, the differences are smaller when compared with the more accurate HNC IET results from Ref.~\onlinecite{Attard}.
From the number density profiles calculated in DFT  we were  able to determine the wavelength of oscillations for concentrations beyond the crossover from monotonic to oscillatory decay that we choose to   term the FW point, i.e. for $d\kappa_D>x_{FW}$, where $x_{FW}^{MSAc}\approx 1.26 $ and $x_{FW}^{MSAu}\approx 1.77$ for the MSAc and MSAu functionals, respectively. These values bracket the result $x_{FW} =1.41$   found in the Generalized MSA (GMSA) IET study ~\cite{Evans} of the bulk electrolyte; see below. Close to the FW point, ascertaining the concentration at which the oscillatory branch has the slower decay is not straightforward and, at first glance, appears to depend on the surface potential. This is illustrated in Fig.~\ref{Fig:N_decay}, where we plot $\rho_N(z)$ obtained from minimising the MSAc functional for surface potentials ranging from $\Phi_0=0.01$ mV (blue line) to $\Phi_0=100$ mV (purple line) at a fixed value of $d\kappa_D$, somewhat below the FW value. Although the true asymptotic decay must be monotonic, for the smallest surface potential we observe only oscillatory decay in the range of $z$  that we can access. For larger surface potentials we observe the correct monotonic decay at sufficiently large $z$. Such behaviour can be explained if we assume the decay of the number density profile has two competing portions:
\begin{align}
\rho_N(z)=A_1e^{-\alpha_1 z}+A_2e^{-\alpha_2 z}\cos(\omega z),
\end{align} 
For $d\kappa_D<x_{FW}$ we know $\alpha_1<\alpha_2$. However, if  $A_1<A_2$, then for a certain $z<z^*$,  the second  term dominates and we observe oscillatory decay. Only for $z>z^*$ will the first term dominate and then we observe the true asymptotic monotonic decay. 

Understanding how the number and charge decay lengths obtained from the three functionals vary with concentration and how their behaviour differs from MSA IET is non-trivial. It is necessary to consider the Euler-Lagrange equation obtained by minimizing the grand potential functional. In the far field, the number and charge densities can be expressed  (see Eqs.~\eqref{Eq:rho_Z_as_app} and \eqref{Eq:rho_N_as_app} in Appendix \ref{App:DFT_asymptotic}) as  
\begin{align}
       \rho_Z(z)&\approx2\Delta c_1(z;[\rho_Z,\rho_N]), \label{Eq:rho_Z_as}\\
    \rho_N(z)&\approx2\Delta c_2(z,[\rho_N,\rho_Z])+\Delta c_1(z;[\rho_Z,\rho_N])^2, \nonumber\\
    &=2\Delta c_2(z,[\rho_N,\rho_Z])+\frac{1}{4}\rho_Z(z)^2, \label{Eq:rho_N_as}
\end{align}
where $\Delta c_1(z)=c_1(z)-c_{1,b}$ denotes the deviation from bulk of the part of the one-body direct correlation function that is proportional to the valency of the species and $\Delta c_2(z)=c_2(z)-c_{2,b}$ is the part that is the same for the cations and anions; see Eqs.~\eqref{Eq:c1C} and~\eqref{Eq:c2C}.
For the simplest case, the MFC functional, these reduce to:
\begin{align}
\rho_Z(z)\approx & 2\Delta c^{(1),MFC}(z;[\rho_Z])=A \exp(-z/\xi_{MFC}),\\
\rho_N(z)\approx & 2\Delta c^{(1),HS}(z;[\rho_N])+\frac{1}{4}\rho_Z^2(z) \nonumber\\
= & B\cos(2\pi z/\lambda_{N,FMT})\exp(-z/\xi_{FMT})+\nonumber\\
&C\exp(-2z/\xi_{MFC}).
\end{align}
where we identified $\Delta c_1(z)=c^{(1),MFC}(z;[\rho_Z])$, $\Delta c_2=\Delta c^{(1),HS}(z;[\rho_N])$ and $A,$ $B$, $C$ are non-universal coefficients. Whilst the asymptotic decay of $\rho_Z(z)$ in the MFC is always given by the Debye length, i.e. $\xi_{MFC}=\kappa_D^{-1}$, for the number density $\rho_N(z)$ one finds a competition between terms decaying with the  FMT (HS) decay length $\xi_{FMT}$, dominating at high concentration, and those with  half the Debye length, dominating at low concentration. The competition results in the monotonic to oscillatory crossover (FW) point observed in Fig.~\ref{Fig:xi}. For the MSAc functional we find a similar result,
\begin{align}
\rho_Z(z)\approx & 2c^{(1),MFC}([\rho_Z];z)+2c^{(1),MSAc}([\rho_Z];z)\nonumber\\
= & A\exp(-z/\xi_{MSAc}),\\
\rho_N(z) \approx & 2\Delta c^{(1),HS}(z;[\rho_N])+\frac{1}{4}\rho_Z^2(z)\nonumber\\
= & B\cos(2\pi z/\lambda_{N,FMT})\exp(-z/\xi_{FMT})+\nonumber\\
&C\exp(-2z/\xi_{MSAc}),
\end{align}
 where we identified $\Delta c_1(z)=c^{(1),MFC}(z;[\rho_Z])+c^{(1),MSAc}(z;[\rho_Z])$ and $\Delta c_2(z)$ is the same as for the MFC. Hence, the decay has the same form as for the MFC functional except that $\xi_{MFC}=\kappa_D^{-1}$ is replaced with $\xi_{MSAc}$ (which is not $1/2\Gamma$), i.e. the value of the leading pole from the IET route. Note the presence of the Kirkwood point within MSAc  leads to oscillatory decay of $\rho_Z(z)$  for $d\kappa_D>x_K^{MSAc}$. 
From the results in Fig.~\ref{Fig:xi} it is clear that the MSAu functional exhibits very different behaviour from the other two functionals, regarding predictions for asymptotic decay of correlations. In Appendix \ref{App:ES_cons} we argue that MSAu has severe inconsistencies  that lead to erroneous predictions. The felon leading this inconsistency is the term $\eta(\{\tilde{n}(\mathbf{r})\})$ in Eq.~\eqref{Eq:MSA_phi_msa}, which of course, vanishes in the bulk RPM. Indeed one might argue that, given the symmetry of the RPM, the term should be omitted from the outset. If one adopts this strategy MSAu  returns the same asymptotic Z decay as found with the MSAc, while the number decay remains virtually unchanged. This is explained further in  Appendices~\ref{App:decay_onebody} and~\ref{App:DFT_asymptotic}. 

How do our far field results fare in the light of previous studies of asymptotic decay in the RPM?  Fig.~\ref{Fig:xi_lit}, attempts to address this question. We display the decay lengths calculated using different bulk IET, namely the GMSA \cite{Evans} and the hypernetted chain approximation (HNC) \cite{Attard,Ennis} and we present these in ranges for which we believe we can extract reliable numerical results from figures in the published papers.The \textit{Z} decay length obtained from our simulations follows the theoretical predictions quite well at small values of  $d\kappa_D$. Note that the HNC results from Ennis et al. focused on this regime where this closure is  expected to yield very accurate (bulk) decay lengths. Comparing  Kirkwood points, it is important to note that the crossover  value is universal within the MSA for the RPM: $x_{K}^{IET}\approx 1.229$. This is not the case within HNC where there is a very weak dependence on $d$. In the HNC results that we display in Fig.~\ref{Fig:xi_lit},  Attard \cite{Attard} used the same diameter as we used, $d=0.5$ nm, while Ennis et al. \cite{Ennis} report results for $d=0.46$ nm. The numerical values determined from HNC for Kirkwood crossover are very close to each other, i.e. Attard\cite{Attard} found $x_K\approx 1.3$ and Ennis et al.\cite{Ennis} found $x_{K}\approx 1.293$, which should be compared  to the MSA/GMSA value $x_K^{MSA}\approx 1.229$, and our simulation result $d\kappa_D\approx1.37$. Note that the original Kirkwood  theory gives a value $x_K$ =1.03 while the Modified Poisson Boltzmann theory\cite{Outhwaite_2019} yields $x_K$=1.241.

Locating the crossover for N decay is arguably more delicate as this depends on incorporating properly  hard-core correlations alongside any residual effects arising from the (net) electrostatics. GMSA and HNC theories attempt this. It is straightforward to show that the location of the FW point, as a function of $d\kappa_D$, is not universal.  Using the MSAc functional, we found crossover at $x_{FW}^{MSAc}\approx 1.26$; see Fig.~\ref{Fig:xi_lit}(b). Using the (bulk) HNC, Ennis et al.\cite{Ennis} (their Fig.~6)  found crossover to oscillatory decay at  $x_{FW}^{Ennis}\approx 1.52$.  From Fig.~5(a) of Attard\cite{Attard} we can deduce a value of $x_{FW}^{Attard}\approx 1.46$. In their pole analysis of the GMSA Carvalho and Evans found  $x_{FW}^{GMSA}\approx 1.41$. Note, however, the GMSA predicts values of $\xi_N\kappa_D  \ll 1/2$ for small values of $d\kappa_D$. This defect of the GMSA is elaborated upon in Appendix \ref{App:decay_onebody}.
Our  MD simulation results shown in Figs.~\ref{Fig:xi} and~\ref{Fig:xi_lit}  indicate crossover at a value of $d\kappa_D$  similar to that obtained from MSAc.

We remark that Attard, using HNC, and Carvalho and Evans, using GMSA, locate the point at which the N decay length becomes larger than the Z decay length; this occurs near  $d\kappa_D\approx3.0$ in both theories. The significance of this crossover will become clear in the next subsection.

\subsubsection{Asymptotic Decay of the Solvation Force}\label{sec:solv_force}

In light of the recent experimental surface force measurements \cite{Gebbie_etal_2015,Cheng_etal_2015,Espinosa_etal_2014,Smith_etal_2016}  that report long decay lengths, it is important to enquire what our DFT results predict for the decay length of the solvation force for a RPM electrolyte confined between two planar electrodes, separated by a distance $H$. The solvation force, see for instance Ref.~\onlinecite{Evans_1987},is defined formally by 
\begin{align}\label{Eq:solv_force}
f_s(H)=-\left.\frac{\partial\gamma(H)}{\partial H}\right|_{T,\mu,\Phi_0},
\end{align} 
 evaluated at fixed temperature $T$, chemical potential $\mu$ and surface potential $\Phi_0$. Here $\gamma=(\Omega+pV)/A$ is the surface tension, defined as the excess over bulk grand potential per unit area of the confined liquid. $A$ denotes the area of the electrodes, $V=AH$ is the accessible volume and $p$ is the bulk pressure, fixed by the reservoir chemical potential and temperature. $f_s(H)$, the excess pressure due to confinement, is related directly to the force measured in SFA experiments. It is not immediately obvious that the asymptotic, large $H$, decay of this thermodynamic quantity  should be given by the same singularities that determine the  asymptotic decay of the bulk pair correlation functions and of the one body density profiles. That this is the case, has been discussed by several authors, e.g. see Refs.~\onlinecite{Evans_1993,Attard_1991,Attard_1992,Evans_1994,Kjellander_1992}. The basic argument is that the potential of mean force between two big (spherical) solute particles immersed at infinite dilution in a reservoir of small `solvent' particles must, for large centre to centre separations $H$, decay with the same (exponential) decay length and period of oscillations (when the ultimate decay is oscillatory) as determined by the decay of the bulk pair correlation function in the small `solvent'. For the RPM  the ions constitute the small 'solvent' in this analysis. Allowing the radius of the big solute particle to become infinite we recover the case of two planar walls and then the potential of mean force yields the solvation force, or excess pressure. Since we have calculated the (bulk) charge and number decay lengths as a function of concentration, and examined the competition between these, we know the ultimate decay of the (thermodynamic) solvation force for each concentration. We denote the corresponding length scale as $\xi$, which represents the true correlation length in the liquid. The upshot is that the solvation force should decay as
\begin{align}\label{Eq:Om_dec}
    f_s(H)\propto\cos(2\pi H/\lambda+\varphi)e^{-H/\xi}, \quad H\rightarrow\infty,
\end{align}
where $\xi$ is the longest decay length in the system. In an oscillatory asymptotic regime, $\lambda$ is the wavelength of the slowest decaying (pole) contribution and $\varphi$ a non-universal phase shift.

\begin{figure}
\includegraphics[width=0.48\textwidth]{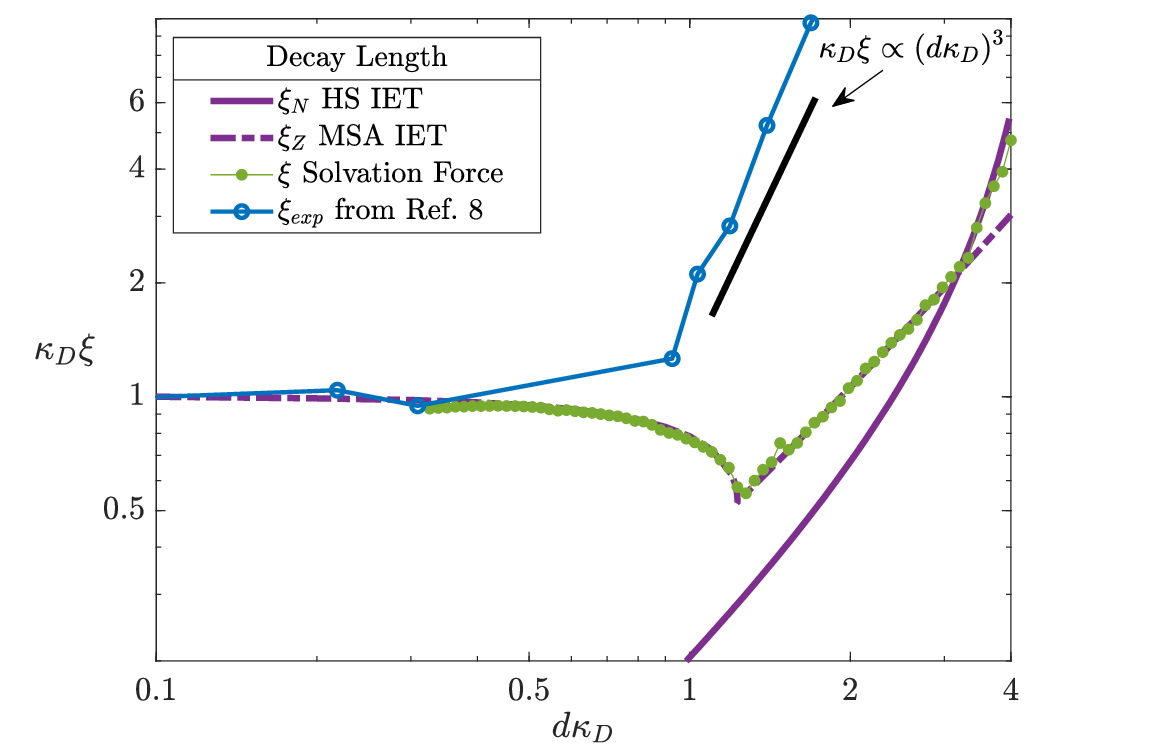}
\caption{\label{Fig:Om_dec} The decay length $\xi$ (green circles) of the solvation force obtained by measuring  the grand potential of the electrolyte,calculated from the  MSAc functional,  see Eqs.~\eqref{Eq:solv_force} and~\eqref{Eq:Om_dec}, as a function of the distance $H$ between two identical planar electrodes, for various bulk concentrations $d\kappa_D$.   $\xi$  is the larger of the charge (dashed-dot purple) and the total number (solid purple) decay lengths from IET. The latter predict a cross-over, near  $d\kappa_D \approx 3.1$, from longest-ranged decay governed by $\xi_Z$ to that governed by $\xi_N$. The experimental data (blue dots connected by a blue line) is for an aqueous NaCl electrolyte extracted from Ref.~\onlinecite{Smith_etal_2016}, and the black line indicates the often-cited power law $\xi\kappa_D\propto (d\kappa_D)^3$.}
\end{figure}

The decay length $\xi$ extracted from changing the planar distance $H$ from $2.5$ to $40$ nm at various (dimensionless) concentrations $d\kappa_D$  is presented  in Fig.~\ref{Fig:Om_dec}, where we used the MSAc functional to calculate the grand potential. For low concentrations $d\kappa_D<3.2$, $\xi$  is determined by the charge density decay $\xi_Z$, while for higher concentrations, $d\kappa_D>3.2$,  $\xi_N$ is longer. It is important to compare with results for bulk correlation lengths. From the HNC\cite{Attard} and from the GMSA\cite{Evans} calculations one finds this  crossover occurs  at $d\kappa_D\approx 3.0$, which is quite close to our DFT  value.  The decay length,  $\xi_{\text{exp}}$, that can be measured in an SFA experiment, at large plate separations $H$, should be the largest decay length in the confined liquid (the physical system), i.e. $\xi_{\text{exp}}(d\kappa_D)=\max_{a}\xi_a(d\kappa_D)$ where in our case $a \in \{Z,N \}$.

In Fig.~\ref{Fig:Om_dec} we also plotted in blue symbols the experimental results from Ref.~\onlinecite{Gebbie_2017} for NaCl dissolved in water. Clearly, the decay lengths extracted  from the SFA experiments are very different from those calculated for the RPM, except at very low concentrations. Interestingly, the decay length $\xi_{\text{exp}}$ measured for $d\kappa_D>1$ follows the power law $\kappa_D\xi\propto (d\kappa_D)^3$ as reported and emphasized in Refs.~\onlinecite{Lee_PRL,Smith_etal_2016,Lee_underscreen,Gebbie_2017,usfabian,Rotenberg_2018,uskorny}. This behavior  is depicted by the black line in the figure and is argued to be `universal', i.e. it describes a broad range of electrolytes and ionic liquids. This `universal' power law is not found within the RPM.

\section{Summary and discussion} 

In this paper we investigate the one- and two-body correlation functions of the RPM, a model electrolyte composed of equal-sized charged spheres of equal but opposite valency dissolved in a dielectric continuum with a Bjerrum length that is 1.46 times larger than the ionic diameter --these parameters are chosen to mimic an aqueous electrolyte  with monovalent ions, e.g. KCl, at room temperature. Using integral equation theory (IET), density functional theory (DFT), and molecular dynamics (MD) simulations, we focus on the asymptotic (far-field) decay of the correlation functions for a large range of ionic concentrations that extends from the very dilute regime where the Debye-H\"{u}ckel limiting law applies up to the regime where ionic hard-sphere packings dominate.  The one-body density profiles are calculated for the RPM in contact with a planar electrode at a fixed potential difference between the electrode and the bulk electrolyte, whereas the two-body correlations pertain to a homogeneous bulk system. Our DFT calculations make predictions for the asymptotic decay of the solvation force between two charged planar surfaces immersed in the RPM. The solvation force is the quantity that relates directly to the experimental SFA  measurements, where anomalously large decay lengths were observed at at high salt concentrations, and which stimulated this study

We distinguish between number-number and charge-charge correlations, which are decoupled in the bulk  RPM,where number-charge correlations are identically zero. The IET results for the decay of bulk pair correlations are based on two well-known direct correlation functions: the number-number combination uses the one obtained from fundamental measure theory (FMT) \cite{Roth_FMT} and the charge-charge correlation function uses the one that stems from the mean-spherical approximation (MSA) \cite{Waisman_1970,Waisman_1972_I,Waisman_1972_II}.
In our  DFT treatment we describe short-range repulsion in the RPM in terms of the FMT hard-sphere free-energy functional. We explore three different approximations to account for the electrostatic contributions to the free-energy functional, (i) the Poisson-Boltzmann-like mean-field expression (MFC) for the internal electrostatic energy, (ii) a correction to MFC that accounts for the finite-ion diameter by incorporating the MSA direct correlation function (MSAc) into the functional, and (iii) a further correction to MSAc that accounts for the correlation-induced MSA internal energy, termed MSAu. In the near-field, at distances of the order of the ion diameter, results from  IET and all three versions of DFT agree  reasonably well with those from MD although only the MSAu can account for the ionic depletion observed in the number density profile calculated  in simulations at low external electrode potentials; see Fig.~\ref{Fig:rho}. In contrast, in the far-field regime, i.e. for distances and separations much larger than the ion diameter, our IET and DFT results are mutually consistent apart from the MSAu implementation of DFT. It follows that employing a functional that yields optimal near-field performance does not guarantee the correct asymptotic behaviour, crucial for interpreting SFA measurements.

Two key results of our study are shown in Figs.~\ref{Fig:xi} and~\ref{Fig:xi_lit}. These concern the concentration dependence of the correlation lengths $\xi_Z$ and $\xi_N$ that dictate the asymptotic charge-charge and number-number decay lengths, respectively, and where we characterise the concentration in terms of the dimensionless quantity  $d\kappa_D$. At low concentrations, $d\kappa_D<0.5$, all the approaches we consider agree that $\xi_Z$  is close to $\kappa_D^{-1}$ and $\xi_N$ is close to $\frac{1}{2}\kappa_D^{-1}$, except for the GMSA results for $\xi_N$. At very high concentrations, $d\kappa_D>2$, we find good agreement between results from all the approaches apart from $\xi_Z$ in MFC which, of course, takes its dilute-limit value $\kappa_D^{-1}$ for all $d\kappa_D$.  In this high concentration regime  the structure of the RPM is dominated by steric repulsions rather than by Coulombic interactions. It follows that a  DFT must  incorporate properly hard-sphere correlations; this is not the case in MFC.  In the intermediate concentration regime, $0.5<d\kappa_D<2$, we find some substantial differences between several of our approaches, especially for $\xi_N$. This increases (MSAu), or decreases and then increases (MSAc), or stays constant (MFC) as the concentration increases. For $\xi_N$ the MSAc functional performs best, when compared to our MD simulations but also when compared to IET results from earlier studies \cite{Evans,Attard,Ennis}. For $\xi_Z$, the MSAc results agree quite well with those of the MD simulations, although at concentrations above the Kirkwood point it slightly overestimates the charge decay length. Overall, the MSAc results agree very well with those of HNC calculations in the range $0.5<d\kappa_D<1.5$ where we could extract reliable numbers from Ref.~\onlinecite{Ennis}. A very recent paper \cite{Kjellander_2020} introduced some new modifications/extensions of DH theory which make predictions for decay lengths. As far as we can tell, these are not significantly different from the results we present here.

Although our focus was on the RPM throughout, we also performed PM calculations (not reported here) with various ionic valency and diameter asymmetries. The resulting asymptotic decay properties are very similar to those of the RPM reported here. We find no long decay lengths, in line with what was reported in Ref.~\onlinecite{usfabian}.

The third key result, that connects with the SFA experiments, is presented in Fig.~\ref{Fig:Om_dec}. This shows the (true) decay length $\xi$ that characterises the decay of the solvation force as obtained from Eqs.~\eqref{Eq:solv_force} and~\eqref{Eq:Om_dec} using the excess grand potential determined from the MSAc functional. We find that for each concentration $\xi=\max(\xi_Z,\xi_N)$. There is excellent agreement between $\xi$ and $\xi_Z$ from MSA-based IET up to concentrations $d\kappa_D\approx 3$, and between  $\xi$ and $\xi_N$ from IET for hard spheres (FMT) at higher concentrations. Given the good agreement between MD simulations and DFT/IET  for $\xi_Z$ at low concentrations and $\xi_N$ at high concentrations, we are confident that our MSAc findings for $\xi$ in the RPM are reliable, at least for our parameter choice $T^*=d/\lambda_B=0.685$. Recall this choice describes a typical 1:1 aqueous electrolyte at room temperature. However, turning to the experimental data, the decay length in aqueous  NaCl as presented in Fig.~\ref{Fig:Om_dec} is vastly different from our theoretical predictions. The huge difference between the experimental results and those for the RPM is illustrated by comparing at 4.93 M NaCl concentration,  where $d_{exp}\kappa_D^{exp}\approx 2.2$ (using $d_{exp}=0.294$ nm and $\varepsilon_r=78$ instead of the concentration dependent $\varepsilon_r$ used in Ref.~\onlinecite{Smith_etal_2016}). One finds  $\kappa_D^{exp}\xi_{exp}\approx 24$ in the experiment\cite{Smith_etal_2016} whereas our RPM results predict $\kappa_D\xi\approx 2.3$. The difference is a about a factor 10, and is larger at higher concentrations. This cannot be explained easily  by some degree of arbitrariness in the exact definition of the ionic diameter, the slightly different size of sodium and chloride ions, or the small change of the dielectric constant at concentrations beyond say 2M NaCl from that of pure water; such considerations might allow at most a factor of 2 or so.  Moreover aqueous NaCl is not special. A great variety of ionic systems has been investigated experimentally in recent years using SFA or closely related techniques \cite{Lee_PRL,Smith_etal_2016,Lee_underscreen,Gebbie_2017}. These include aqueous LiCl, KCl, CsCl, but also several ionic liquid solutions in a particular solvent, as well as pure (room temperature)  ionic liquids.  As mentioned in Sec. IV B.3, the experimental correlation lengths (scaled as we scale Fig.\ref{Fig:Om_dec}) appear to fall on top of the result for NaCl. The empirical `universal' scaling relation $\kappa_D\xi\propto(d \kappa_D)^3$ for $d\kappa_D>1$ actually extends way beyond the scale of Fig.\ref{Fig:Om_dec}, up to data points for ionic liquids at $d \kappa_D\approx7$  where $\kappa_D\xi\approx 120$.  From the SFA measurements one might conclude that the measured correlation length in concentrated electrolytes and ionic liquids  is at least an order of magnitude larger than our RPM predictions.   
It is important to recognize that  the  large correlation lengths were measured at separations of several nanometers in the SFA experiments \cite{Smith_etal_2016}. At shorter separations, an additional \textit{structural} decay length was measured \cite{Smith_PRL118}, which is much shorter. Although our DFT calculations find no indication of a long decay length, and we measure across 8 decades, we cannot rule out the possibility of a large decay length, buried in the noise that sets in beyond  about three decades in our MD results.

The full story is more subtle. There is good reason to reconsider earlier work on molten alkali halides, conventionally regarded as archetypal ionic liquids. Of course, these salts have high melting temperatures, making experiments difficult. Nevertheless, it is well known that the RPM accounts well for the main features of the partial structure factors of molten salts such as KCl or NaCl where  cations and anions have similar size\cite{HansenMC}. Careful neutron (isotopic substitution) diffraction  experiments\cite{Zeidler} for molten NaCl at 1093 K extracted the three partial structure factors from which the total pair correlation functions $h_{ij}(r)$ can be obtained by Fourier Transforming. Fitting the resulting data to formulae equivalent to the mixture generalization of  Eq.~\eqref{Eq:h_asymptotic}, decay lengths for the partial (and thus the total  number and charge correlation functions) were determined, along with accompanying wavelengths. At this temperature, not far above the melting point,  the longest decay length  observed is about 0.46 nm,  i.e. $< 2$  ionic diameters. There is no indication of a long decay length.
Also pertinent are MD simulations for NaCl from Keblinski et al.\cite{Keblinsky_2000} These employ a symmetrized version of the standard Born-Mayer-Huggins potentials for alkali halides, i.e. the anion-anion  and the cation-cation potentials are identical, mimicking the symmetry of the RPM. Data were analyzed using a mean diameter of $d$ = 0.276 nm. Key observations from this far-reaching study are: i) for fixed, very high temperature well above the critical temperature, which is slightly below 3000K in their model, the authors find (Kirkwood) crossover between monotonic and oscillatory decay of charge correlations at $d\kappa_D\approx 1.4 $, a value that does not depend much on their choice of (high) temperature. This scenario is predicted within the MSA for the RPM, where the Kirkwood line is universal, and  is almost vertical in  the $\rho^*-T^*$ plane; see Fig.~1 of  Ref.~\onlinecite{Evans}. ii) Keblinsky et al.  find crossover between monotonic and oscillatory decay of the total number correlations at fixed $T$  = 3000K. Although they do not locate the crossover density precisely, the broad range identified brackets the  FW  crossover density predicted by the GMSA;  see Fig.~1 of  Ref.~\onlinecite{Evans}. iii) Most importantly, at all state points away from the critical point, the decay lengths reported in\cite{Keblinsky_2000} are short, i.e. < 2$d$.

\section{Conclusion}

 Our main conclusion, which, of course, also relies upon significant previous literature on bulk decay lengths, is that the (R)PM in equilibrium cannot explain the experimental (SFA) measurements reporting an anomalously large decay length of the solvation force  in concentrated electrolytes and certain ionic liquids.  This is in line with findings reported in less idealized models, e.g. Refs.~\onlinecite{Rotenberg_2018,usfabian,Rotenberg,Holm}. Perhaps this is not too surprising when addressing room temperature ionic liquids with non-spherical ions that contain organic rings and tails etc. It is more discomforting in the case of aqueous alkali halide solutions. We  distinguish here between the model (the RPM in thermodynamic equilibrium) and the method used to analyse it (DFT, IET, MD). Given that the various theories and simulation  methods mutually agree on their predictions for the longest correlation length $\xi$, the source of the discrepancy must lie in the model. The RPM seems to lack a crucial ingredient to explain the experimental findings. Assuming the experiments pertain to equilibrium, the key question is `Which piece of physics is missing?' Before addressing this question, we emphasize once again that careful experimental (neutron diffraction)\cite{Zeidler}  and simulation studies \cite{Keblinsky_2000} of the bulk pair correlation functions in molten NaCl, an archetypal ionic liquid, find no evidence for a long decay length. More specifically, we find that the MD results of Keblinsky et al. for a Born-Mayer-Huggins model of molten NaCl at $T=10000$ K and $T=50000$ K agree qualitatively with our RPM results for all concentrations and quantitatively ( within 15\%) for concentrations exceeding the Kirkwood point. For example, for $T=10000$ K these authors report  at $d\kappa_D=3.1$ (using their $d=0.276$ nm) a decay length $\xi\kappa_D\approx1.7$, see their Fig.~8, whereas we report $\xi\kappa_D\approx2.$

Noting that we have already pointed out that asymmetries of the ionic valencies and diameters yield decay lengths very similar to those of the RPM, the first possible candidate to explain the discrepancy between predictions from the (R)PM and the SFA measurements is the description of the solvent as a (uniform) dielectric continuum. However, recent computer simulations and theories for several electrolyte models that include the solvent explicitly also find decay lengths, measured  at high ionic concentrations, of the order of the particle diameter. A broad range of models is considered: the solvent is either modeled as a hard-sphere fluid \cite{usfabian,Rotenberg_2018}, or as the SPC/E model for water\cite{Rotenberg, Holm} in the case of aqueous alkali halides. For the ionic liquids, models of organic solvents such as dimethoxyethane-dioxolane\cite{Rotenberg} or racemic propylene-carbonate\cite{Holm} are considered. Although these explicit-solvent models show an increase of the longest correlation length at high concentrations, the  observed increase is very similar to the one we find here for the RPM. We conclude that current treatments of solvent effects changes little the primitive model predictions of  decay/correlation lengths that are about an order of magnitude smaller than measured in SFA experiments.

Another obvious candidate is the omission of polarizability; this is absent completely in the RPM and is, at best, included approximately in some of the explicit-solvent models. An interesting approach was put forward by Kjellander\cite{Kjellander_2019,Kjellander_2016,Kjellander_2020}, who shows that electrostatic screening and the static dielectric function $\epsilon(k)$, with wave-number $k$, are intimately coupled, such that the long-wavelength limit $\epsilon(0)$ equals the static dielectric constant only in the absence of ions, e.g in dipolar fluids or non-electrolytes, but not in their presence. The upshot is that in an electrolyte, the screening and the dielectric response cannot be disentangled\cite{Kjellander_2019}. To best of our knowledge, there are no specific predictions for decay lengths that might be tested quantitatively against experimental results. Polarizability also leads to fluid-fluid and fluid-wall dispersion forces, giving rise to a power-law decay of the solvation force  \cite{Maciolek_2004}. Although dispersion forces are long-ranged, we expect these to be relatively weak such that they become manifest in the solvation force only beyond separations of many particle diameters, for instance beyond 15$d$ for the (reasonable) parameters of Ref.~\onlinecite{Maciolek_2004}.

Of course, there are  other factors that could account for the disagreement between results of theory and simulation on the one hand and SFA experiments on the other. Strictly speaking, there is a possibility that the measured long decay length could just be buried in the noise of all our calculations and all simulation studies \cite{Rotenberg,Holm}, although for instance the simulations of Ref.~\onlinecite{Holm} show statistics that allow observation of decay over as many as five decades before the signal disappears in the noise, and over eight decades in our DFT calculations.
  Significantly, the variety and  number of experimental systems studied, along with the apparent success of the empirical power law scaling mentioned earlier,  suggest there should be a\textit{ general}, rather than a materials specific, explanation of the difference. Our present contribution, which establishes the consistency of results from DFT, IET and MD for the RPM, makes very clear why it is important to understand  the origin of the difference. We conclude by re-emphasizing:  the large decay length measured in SFA experiments,  for a variety of concentrated electrolytes and several room temperature ionic liquids, cannot be accounted for by primitive electrolyte models. New physical ingredients and/or new interpretations of the experiments are required  in order to understand the recent SFA results.

\begin{acknowledgments}
We thank C. Holm for sharing a preprint of Ref. \onlinecite{Holm} with us and  P.S. Salmon for sending Ref.~\onlinecite{Zeidler}. We are grateful to R. Roth for insightful comments and to a referee who pointed us to pertinent literature which improved our analysis and discussion.  This work is part of the D-ITP consortium, a program of the Netherlands Organisation for Scientific Research (NWO) that is funded by the Dutch Ministry of Education, Culture and Science (OCW).
It forms part of the NWO programme `Data-driven science for smart and sustainable energy research', with project number 16DDS014. AH acknowledges support by the 
state of Baden-Württemberg through bwHPC and the 
German Research Foundation (DFG) through grant no
INST 39/963-1 FUGG (bwForCluster NEMO) and through project number 406121234. RE was supported by the Leverhulme Trust  through EM 2020-029/4.
\end{acknowledgments}


\bibliography{biblio}

\begin{thebibliography}{76}%
\makeatletter
\providecommand \@ifxundefined [1]{%
 \@ifx{#1\undefined}
}%
\providecommand \@ifnum [1]{%
 \ifnum #1\expandafter \@firstoftwo
 \else \expandafter \@secondoftwo
 \fi
}%
\providecommand \@ifx [1]{%
 \ifx #1\expandafter \@firstoftwo
 \else \expandafter \@secondoftwo
 \fi
}%
\providecommand \natexlab [1]{#1}%
\providecommand \enquote  [1]{``#1''}%
\providecommand \bibnamefont  [1]{#1}%
\providecommand \bibfnamefont [1]{#1}%
\providecommand \citenamefont [1]{#1}%
\providecommand \href@noop [0]{\@secondoftwo}%
\providecommand \href [0]{\begingroup \@sanitize@url \@href}%
\providecommand \@href[1]{\@@startlink{#1}\@@href}%
\providecommand \@@href[1]{\endgroup#1\@@endlink}%
\providecommand \@sanitize@url [0]{\catcode `\\12\catcode `\$12\catcode
  `\&12\catcode `\#12\catcode `\^12\catcode `\_12\catcode `\%12\relax}%
\providecommand \@@startlink[1]{}%
\providecommand \@@endlink[0]{}%
\providecommand \url  [0]{\begingroup\@sanitize@url \@url }%
\providecommand \@url [1]{\endgroup\@href {#1}{\urlprefix }}%
\providecommand \urlprefix  [0]{URL }%
\providecommand \Eprint [0]{\href }%
\providecommand \doibase [0]{http://dx.doi.org/}%
\providecommand \selectlanguage [0]{\@gobble}%
\providecommand \bibinfo  [0]{\@secondoftwo}%
\providecommand \bibfield  [0]{\@secondoftwo}%
\providecommand \translation [1]{[#1]}%
\providecommand \BibitemOpen [0]{}%
\providecommand \bibitemStop [0]{}%
\providecommand \bibitemNoStop [0]{.\EOS\space}%
\providecommand \EOS [0]{\spacefactor3000\relax}%
\providecommand \BibitemShut  [1]{\csname bibitem#1\endcsname}%
\let\auto@bib@innerbib\@empty
\bibitem [{\citenamefont {Helmholtz}(1853)}]{Helmholtz}%
  \BibitemOpen
  \bibfield  {author} {\bibinfo {author} {\bibfnamefont {H.}~\bibnamefont
  {Helmholtz}},\ }\bibfield  {title} {\enquote {\bibinfo {title} {Ueber einige
  gesetze der vertheilung elektrischer ströme in körperlichen leitern mit
  anwendung auf die thierisch-elektrischen versuche},}\ }\href {\doibase
  10.1002/andp.18531650603} {\bibfield  {journal} {\bibinfo  {journal} {Annalen
  der Physik}\ }\textbf {\bibinfo {volume} {165}},\ \bibinfo {pages} {211--233}
  (\bibinfo {year} {1853})}\BibitemShut {NoStop}%
\bibitem [{\citenamefont {{Gouy, M.}}(1910)}]{Gouy}%
  \BibitemOpen
  \bibfield  {author} {\bibinfo {author} {\bibnamefont {{Gouy, M.}}},\
  }\bibfield  {title} {\enquote {\bibinfo {title} {Sur la constitution de la
  charge \'electrique \`a la surface d'un \'electrolyte},}\ }\href {\doibase
  10.1051/jphystap:019100090045700} {\bibfield  {journal} {\bibinfo  {journal}
  {J. Phys. Theor. Appl.}\ }\textbf {\bibinfo {volume} {9}},\ \bibinfo {pages}
  {457--468} (\bibinfo {year} {1910})}\BibitemShut {NoStop}%
\bibitem [{\citenamefont {Chapman}(1913)}]{Chapman}%
  \BibitemOpen
  \bibfield  {author} {\bibinfo {author} {\bibfnamefont {D.~L.}\ \bibnamefont
  {Chapman}},\ }\bibfield  {title} {\enquote {\bibinfo {title} {Li. a
  contribution to the theory of electrocapillarity},}\ }\href {\doibase
  10.1080/14786440408634187} {\bibfield  {journal} {\bibinfo  {journal} {The
  London, Edinburgh, and Dublin Philosophical Magazine and Journal of Science}\
  }\textbf {\bibinfo {volume} {25}},\ \bibinfo {pages} {475--481} (\bibinfo
  {year} {1913})}\BibitemShut {NoStop}%
\bibitem [{\citenamefont {Debye}\ and\ \citenamefont {Hückel}(1923)}]{DH}%
  \BibitemOpen
  \bibfield  {author} {\bibinfo {author} {\bibfnamefont {P.}~\bibnamefont
  {Debye}}\ and\ \bibinfo {author} {\bibfnamefont {E.}~\bibnamefont
  {Hückel}},\ }\bibfield  {title} {\enquote {\bibinfo {title} {Zur theorie der
  elektrolyte. i. gefrierpunktserniedrigung und verwandte erscheinungen},}\
  }\href@noop {} {\bibfield  {journal} {\bibinfo  {journal} {Physikalische
  Zeitschrift}\ }\textbf {\bibinfo {volume} {24}},\ \bibinfo {pages} {305}
  (\bibinfo {year} {1923})}\BibitemShut {NoStop}%
\bibitem [{\citenamefont {Gebbie}\ \emph {et~al.}(2015)\citenamefont {Gebbie},
  \citenamefont {Dobbs}, \citenamefont {Valtiner},\ and\ \citenamefont
  {Israelachvili}}]{Gebbie_etal_2015}%
  \BibitemOpen
  \bibfield  {author} {\bibinfo {author} {\bibfnamefont {M.~A.}\ \bibnamefont
  {Gebbie}}, \bibinfo {author} {\bibfnamefont {H.~A.}\ \bibnamefont {Dobbs}},
  \bibinfo {author} {\bibfnamefont {M.}~\bibnamefont {Valtiner}}, \ and\
  \bibinfo {author} {\bibfnamefont {J.~N.}\ \bibnamefont {Israelachvili}},\
  }\bibfield  {title} {\enquote {\bibinfo {title} {Long-range electrostatic
  screening in ionic liquids},}\ }\href {\doibase 10.1073/pnas.1508366112}
  {\bibfield  {journal} {\bibinfo  {journal} {Proc. Natl. Acad. Sci. U. S. A.}\
  }\textbf {\bibinfo {volume} {112}},\ \bibinfo {pages} {7432--7437} (\bibinfo
  {year} {2015})}\BibitemShut {NoStop}%
\bibitem [{\citenamefont {Cheng}\ \emph {et~al.}(2015)\citenamefont {Cheng},
  \citenamefont {Stock}, \citenamefont {Moeremans}, \citenamefont {Baimpos},
  \citenamefont {Banquy}, \citenamefont {Renner},\ and\ \citenamefont
  {Valtiner}}]{Cheng_etal_2015}%
  \BibitemOpen
  \bibfield  {author} {\bibinfo {author} {\bibfnamefont {H.-W.}\ \bibnamefont
  {Cheng}}, \bibinfo {author} {\bibfnamefont {P.}~\bibnamefont {Stock}},
  \bibinfo {author} {\bibfnamefont {B.}~\bibnamefont {Moeremans}}, \bibinfo
  {author} {\bibfnamefont {T.}~\bibnamefont {Baimpos}}, \bibinfo {author}
  {\bibfnamefont {X.}~\bibnamefont {Banquy}}, \bibinfo {author} {\bibfnamefont
  {F.~U.}\ \bibnamefont {Renner}}, \ and\ \bibinfo {author} {\bibfnamefont
  {M.}~\bibnamefont {Valtiner}},\ }\bibfield  {title} {\enquote {\bibinfo
  {title} {Characterizing the influence of water on charging and layering at
  electrified ionic-liquid/solid interfaces},}\ }\href {\doibase
  10.1002/admi.201500159} {\bibfield  {journal} {\bibinfo  {journal} {Advanced
  Materials Interfaces}\ }\textbf {\bibinfo {volume} {2}},\ \bibinfo {pages}
  {1500159} (\bibinfo {year} {2015})}\BibitemShut {NoStop}%
\bibitem [{\citenamefont {Espinosa-Marzal}\ \emph {et~al.}(2014)\citenamefont
  {Espinosa-Marzal}, \citenamefont {Arcifa}, \citenamefont {Rossi},\ and\
  \citenamefont {Spencer}}]{Espinosa_etal_2014}%
  \BibitemOpen
  \bibfield  {author} {\bibinfo {author} {\bibfnamefont {R.~M.}\ \bibnamefont
  {Espinosa-Marzal}}, \bibinfo {author} {\bibfnamefont {A.}~\bibnamefont
  {Arcifa}}, \bibinfo {author} {\bibfnamefont {A.}~\bibnamefont {Rossi}}, \
  and\ \bibinfo {author} {\bibfnamefont {N.~D.}\ \bibnamefont {Spencer}},\
  }\bibfield  {title} {\enquote {\bibinfo {title} {Microslips to
  “avalanches” in confined, molecular layers of ionic liquids},}\ }\href
  {\doibase 10.1021/jz402451v} {\bibfield  {journal} {\bibinfo  {journal} {The
  Journal of Physical Chemistry Letters}\ }\textbf {\bibinfo {volume} {5}},\
  \bibinfo {pages} {179--184} (\bibinfo {year} {2014})}\BibitemShut {NoStop}%
\bibitem [{\citenamefont {Smith}, \citenamefont {Lee},\ and\ \citenamefont
  {Perkin}(2016)}]{Smith_etal_2016}%
  \BibitemOpen
  \bibfield  {author} {\bibinfo {author} {\bibfnamefont {A.~M.}\ \bibnamefont
  {Smith}}, \bibinfo {author} {\bibfnamefont {A.~A.}\ \bibnamefont {Lee}}, \
  and\ \bibinfo {author} {\bibfnamefont {S.}~\bibnamefont {Perkin}},\
  }\bibfield  {title} {\enquote {\bibinfo {title} {The electrostatic screening
  length in concentrated electrolytes increases with concentration},}\ }\href
  {\doibase 10.1021/acs.jpclett.6b00867} {\bibfield  {journal} {\bibinfo
  {journal} {The journal of physical chemistry letters}\ }\textbf {\bibinfo
  {volume} {7}},\ \bibinfo {pages} {2157—2163} (\bibinfo {year}
  {2016})}\BibitemShut {NoStop}%
\bibitem [{\citenamefont {Gebbie}\ \emph {et~al.}(2017)\citenamefont {Gebbie},
  \citenamefont {Smith}, \citenamefont {Dobbs}, \citenamefont {Lee},
  \citenamefont {Warr}, \citenamefont {Banquy}, \citenamefont {Valtiner},
  \citenamefont {Rutland}, \citenamefont {Israelachvili}, \citenamefont
  {Perkin},\ and\ \citenamefont {Atkin}}]{Gebbie_2017}%
  \BibitemOpen
  \bibfield  {author} {\bibinfo {author} {\bibfnamefont {M.~A.}\ \bibnamefont
  {Gebbie}}, \bibinfo {author} {\bibfnamefont {A.~M.}\ \bibnamefont {Smith}},
  \bibinfo {author} {\bibfnamefont {H.~A.}\ \bibnamefont {Dobbs}}, \bibinfo
  {author} {\bibfnamefont {A.~A.}\ \bibnamefont {Lee}}, \bibinfo {author}
  {\bibfnamefont {G.~G.}\ \bibnamefont {Warr}}, \bibinfo {author}
  {\bibfnamefont {X.}~\bibnamefont {Banquy}}, \bibinfo {author} {\bibfnamefont
  {M.}~\bibnamefont {Valtiner}}, \bibinfo {author} {\bibfnamefont {M.~W.}\
  \bibnamefont {Rutland}}, \bibinfo {author} {\bibfnamefont {J.~N.}\
  \bibnamefont {Israelachvili}}, \bibinfo {author} {\bibfnamefont
  {S.}~\bibnamefont {Perkin}}, \ and\ \bibinfo {author} {\bibfnamefont
  {R.}~\bibnamefont {Atkin}},\ }\bibfield  {title} {\enquote {\bibinfo {title}
  {Long range electrostatic forces in ionic liquids},}\ }\href {\doibase
  10.1039/C6CC08820A} {\bibfield  {journal} {\bibinfo  {journal} {Chem.
  Commun.}\ }\textbf {\bibinfo {volume} {53}},\ \bibinfo {pages} {1214--1224}
  (\bibinfo {year} {2017})}\BibitemShut {NoStop}%
\bibitem [{\citenamefont {Simon}\ and\ \citenamefont {Y.}(2008)}]{Simon_2008}%
  \BibitemOpen
  \bibfield  {author} {\bibinfo {author} {\bibfnamefont {P.}~\bibnamefont
  {Simon}}\ and\ \bibinfo {author} {\bibfnamefont {G.}~\bibnamefont {Y.}},\
  }\bibfield  {title} {\enquote {\bibinfo {title} {Materials for
  electrochemical capacitors},}\ }\href {\doibase 10.1038/nmat2297} {\bibfield
  {journal} {\bibinfo  {journal} {Nat. Mater.}\ }\textbf {\bibinfo {volume}
  {7}},\ \bibinfo {pages} {845--854} (\bibinfo {year} {2008})}\BibitemShut
  {NoStop}%
\bibitem [{\citenamefont {C.}\ \emph {et~al.}(2012)\citenamefont {C.},
  \citenamefont {B.}, \citenamefont {P.A.}, \citenamefont {PL.}, \citenamefont
  {P.}, \citenamefont {Y.},\ and\ \citenamefont {M.}}]{Merlet_2012}%
  \BibitemOpen
  \bibfield  {author} {\bibinfo {author} {\bibfnamefont {M.}~\bibnamefont
  {C.}}, \bibinfo {author} {\bibfnamefont {R.}~\bibnamefont {B.}}, \bibinfo
  {author} {\bibfnamefont {M.}~\bibnamefont {P.A.}}, \bibinfo {author}
  {\bibfnamefont {T.}~\bibnamefont {PL.}}, \bibinfo {author} {\bibfnamefont
  {S.}~\bibnamefont {P.}}, \bibinfo {author} {\bibfnamefont {G.}~\bibnamefont
  {Y.}}, \ and\ \bibinfo {author} {\bibfnamefont {S.}~\bibnamefont {M.}},\
  }\bibfield  {title} {\enquote {\bibinfo {title} {On the molecular origin of
  supercapacitance in nanoporous carbon electrodes},}\ }\href {\doibase
  10.1038/nmat3260} {\bibfield  {journal} {\bibinfo  {journal} {Nat. Mater.}\
  }\textbf {\bibinfo {volume} {11}},\ \bibinfo {pages} {306--310} (\bibinfo
  {year} {2012})}\BibitemShut {NoStop}%
\bibitem [{\citenamefont {Limmer}\ \emph {et~al.}(2013)\citenamefont {Limmer},
  \citenamefont {Merlet}, \citenamefont {Salanne}, \citenamefont {Chandler},
  \citenamefont {Madden}, \citenamefont {van Roij},\ and\ \citenamefont
  {Rotenberg}}]{Limmer_2013}%
  \BibitemOpen
  \bibfield  {author} {\bibinfo {author} {\bibfnamefont {D.~T.}\ \bibnamefont
  {Limmer}}, \bibinfo {author} {\bibfnamefont {C.}~\bibnamefont {Merlet}},
  \bibinfo {author} {\bibfnamefont {M.}~\bibnamefont {Salanne}}, \bibinfo
  {author} {\bibfnamefont {D.}~\bibnamefont {Chandler}}, \bibinfo {author}
  {\bibfnamefont {P.~A.}\ \bibnamefont {Madden}}, \bibinfo {author}
  {\bibfnamefont {R.}~\bibnamefont {van Roij}}, \ and\ \bibinfo {author}
  {\bibfnamefont {B.}~\bibnamefont {Rotenberg}},\ }\bibfield  {title} {\enquote
  {\bibinfo {title} {Charge fluctuations in nanoscale capacitors},}\ }\href
  {\doibase 10.1103/PhysRevLett.111.106102} {\bibfield  {journal} {\bibinfo
  {journal} {Phys. Rev. Lett.}\ }\textbf {\bibinfo {volume} {111}},\ \bibinfo
  {pages} {106102} (\bibinfo {year} {2013})}\BibitemShut {NoStop}%
\bibitem [{\citenamefont {Härtel}\ \emph {et~al.}(2015)\citenamefont
  {Härtel}, \citenamefont {Janssen}, \citenamefont {Weingarth}, \citenamefont
  {Presser},\ and\ \citenamefont {van Roij}}]{Hartel_2015}%
  \BibitemOpen
  \bibfield  {author} {\bibinfo {author} {\bibfnamefont {A.}~\bibnamefont
  {Härtel}}, \bibinfo {author} {\bibfnamefont {M.}~\bibnamefont {Janssen}},
  \bibinfo {author} {\bibfnamefont {D.}~\bibnamefont {Weingarth}}, \bibinfo
  {author} {\bibfnamefont {V.}~\bibnamefont {Presser}}, \ and\ \bibinfo
  {author} {\bibfnamefont {R.}~\bibnamefont {van Roij}},\ }\bibfield  {title}
  {\enquote {\bibinfo {title} {Heat-to-current conversion of low-grade heat
  from a thermocapacitive cycle by supercapacitors},}\ }\href {\doibase
  10.1039/C5EE01192B} {\bibfield  {journal} {\bibinfo  {journal} {Energy
  Environ. Sci.}\ }\textbf {\bibinfo {volume} {8}},\ \bibinfo {pages}
  {2396--2401} (\bibinfo {year} {2015})}\BibitemShut {NoStop}%
\bibitem [{\citenamefont {Brogioli}(2009)}]{Brogioli_2009}%
  \BibitemOpen
  \bibfield  {author} {\bibinfo {author} {\bibfnamefont {D.}~\bibnamefont
  {Brogioli}},\ }\bibfield  {title} {\enquote {\bibinfo {title} {Extracting
  renewable energy from a salinity difference using a capacitor},}\ }\href
  {\doibase 10.1103/PhysRevLett.103.058501} {\bibfield  {journal} {\bibinfo
  {journal} {Phys. Rev. Lett.}\ }\textbf {\bibinfo {volume} {103}},\ \bibinfo
  {pages} {058501} (\bibinfo {year} {2009})}\BibitemShut {NoStop}%
\bibitem [{\citenamefont {Janssen}\ and\ \citenamefont {van
  Roij}(2017)}]{Janssen_2017}%
  \BibitemOpen
  \bibfield  {author} {\bibinfo {author} {\bibfnamefont {M.}~\bibnamefont
  {Janssen}}\ and\ \bibinfo {author} {\bibfnamefont {R.}~\bibnamefont {van
  Roij}},\ }\bibfield  {title} {\enquote {\bibinfo {title} {Reversible heating
  in electric double layer capacitors},}\ }\href {\doibase
  10.1103/PhysRevLett.118.096001} {\bibfield  {journal} {\bibinfo  {journal}
  {Phys. Rev. Lett.}\ }\textbf {\bibinfo {volume} {118}},\ \bibinfo {pages}
  {096001} (\bibinfo {year} {2017})}\BibitemShut {NoStop}%
\bibitem [{\citenamefont {Porada}\ \emph {et~al.}(2013)\citenamefont {Porada},
  \citenamefont {Zhao}, \citenamefont {{van der Wal}}, \citenamefont
  {Presser},\ and\ \citenamefont {Biesheuvel}}]{Poroda_etal_2013}%
  \BibitemOpen
  \bibfield  {author} {\bibinfo {author} {\bibfnamefont {S.}~\bibnamefont
  {Porada}}, \bibinfo {author} {\bibfnamefont {R.}~\bibnamefont {Zhao}},
  \bibinfo {author} {\bibfnamefont {A.}~\bibnamefont {{van der Wal}}}, \bibinfo
  {author} {\bibfnamefont {V.}~\bibnamefont {Presser}}, \ and\ \bibinfo
  {author} {\bibfnamefont {P.}~\bibnamefont {Biesheuvel}},\ }\bibfield  {title}
  {\enquote {\bibinfo {title} {Review on the science and technology of water
  desalination by capacitive deionization},}\ }\href {\doibase
  https://doi.org/10.1016/j.pmatsci.2013.03.005} {\bibfield  {journal}
  {\bibinfo  {journal} {Progress in Materials Science}\ }\textbf {\bibinfo
  {volume} {58}},\ \bibinfo {pages} {1388 -- 1442} (\bibinfo {year}
  {2013})}\BibitemShut {NoStop}%
\bibitem [{\citenamefont {Attard}(1993)}]{Attard}%
  \BibitemOpen
  \bibfield  {author} {\bibinfo {author} {\bibfnamefont {P.}~\bibnamefont
  {Attard}},\ }\bibfield  {title} {\enquote {\bibinfo {title} {Asymptotic
  analysis of primitive model electrolytes and the electrical double layer},}\
  }\href {\doibase 10.1103/PhysRevE.48.3604} {\bibfield  {journal} {\bibinfo
  {journal} {Phys. Rev. E}\ }\textbf {\bibinfo {volume} {48}},\ \bibinfo
  {pages} {3604--3621} (\bibinfo {year} {1993})}\BibitemShut {NoStop}%
\bibitem [{\citenamefont {Ennis}, \citenamefont {Kjellander},\ and\
  \citenamefont {Mitchell}(1995)}]{Ennis}%
  \BibitemOpen
  \bibfield  {author} {\bibinfo {author} {\bibfnamefont {J.}~\bibnamefont
  {Ennis}}, \bibinfo {author} {\bibfnamefont {R.}~\bibnamefont {Kjellander}}, \
  and\ \bibinfo {author} {\bibfnamefont {D.~J.}\ \bibnamefont {Mitchell}},\
  }\bibfield  {title} {\enquote {\bibinfo {title} {Dressed ion theory for bulk
  symmetric electrolytes in the restricted primitive model},}\ }\href {\doibase
  10.1063/1.469166} {\bibfield  {journal} {\bibinfo  {journal} {The Journal of
  Chemical Physics}\ }\textbf {\bibinfo {volume} {102}},\ \bibinfo {pages}
  {975--991} (\bibinfo {year} {1995})}\BibitemShut {NoStop}%
\bibitem [{\citenamefont {de~Carvalho}\ and\ \citenamefont
  {Evans}(1994)}]{Evans}%
  \BibitemOpen
  \bibfield  {author} {\bibinfo {author} {\bibfnamefont {R.~L.}\ \bibnamefont
  {de~Carvalho}}\ and\ \bibinfo {author} {\bibfnamefont {R.}~\bibnamefont
  {Evans}},\ }\bibfield  {title} {\enquote {\bibinfo {title} {The decay of
  correlations in ionic fluids},}\ }\href {\doibase 10.1080/00268979400101491}
  {\bibfield  {journal} {\bibinfo  {journal} {Molecular Physics}\ }\textbf
  {\bibinfo {volume} {83}},\ \bibinfo {pages} {619--654} (\bibinfo {year}
  {1994})}\BibitemShut {NoStop}%
\bibitem [{\citenamefont {Mermin}(1965)}]{Mermin}%
  \BibitemOpen
  \bibfield  {author} {\bibinfo {author} {\bibfnamefont {N.~D.}\ \bibnamefont
  {Mermin}},\ }\bibfield  {title} {\enquote {\bibinfo {title} {Thermal
  properties of the inhomogeneous electron gas},}\ }\href {\doibase
  10.1103/PhysRev.137.A1441} {\bibfield  {journal} {\bibinfo  {journal} {Phys.
  Rev.}\ }\textbf {\bibinfo {volume} {137}},\ \bibinfo {pages} {A1441--A1443}
  (\bibinfo {year} {1965})}\BibitemShut {NoStop}%
\bibitem [{\citenamefont {Evans}(1979)}]{Evans1979}%
  \BibitemOpen
  \bibfield  {author} {\bibinfo {author} {\bibfnamefont {R.}~\bibnamefont
  {Evans}},\ }\bibfield  {title} {\enquote {\bibinfo {title} {The nature of the
  liquid-vapour interface and other topics in the statistical mechanics of
  non-uniform, classical fluids},}\ }\href {\doibase 10.1080/00018737900101365}
  {\bibfield  {journal} {\bibinfo  {journal} {Advances in Physics}\ }\textbf
  {\bibinfo {volume} {28}},\ \bibinfo {pages} {143--200} (\bibinfo {year}
  {1979})}\BibitemShut {NoStop}%
\bibitem [{\citenamefont {Hansen}\ and\ \citenamefont
  {McDonald}(2013)}]{HansenMC}%
  \BibitemOpen
  \bibfield  {author} {\bibinfo {author} {\bibfnamefont {J.-P.}\ \bibnamefont
  {Hansen}}\ and\ \bibinfo {author} {\bibfnamefont {I.}~\bibnamefont
  {McDonald}},\ }\href@noop {} {\emph {\bibinfo {title} {{Theory of simple
  liquids; 4th ed.}}}}\ (\bibinfo  {publisher} {Academic Press},\ \bibinfo
  {address} {New York, NY},\ \bibinfo {year} {2013})\BibitemShut {NoStop}%
\bibitem [{\citenamefont {Ornstein}\ and\ \citenamefont {Zernike}(1914)}]{OZ}%
  \BibitemOpen
  \bibfield  {author} {\bibinfo {author} {\bibfnamefont {L.}~\bibnamefont
  {Ornstein}}\ and\ \bibinfo {author} {\bibfnamefont {F.}~\bibnamefont
  {Zernike}},\ }\bibfield  {title} {\enquote {\bibinfo {title} {Accidental
  deviations of density and opalescence at the critical point of a single
  substance},}\ }\href@noop {} {\bibfield  {journal} {\bibinfo  {journal}
  {KNAW}\ }\textbf {\bibinfo {volume} {17}},\ \bibinfo {pages} {793--806}
  (\bibinfo {year} {1914})}\BibitemShut {NoStop}%
\bibitem [{\citenamefont {Evans}\ \emph {et~al.}(2016)\citenamefont {Evans},
  \citenamefont {Oettel}, \citenamefont {Roth},\ and\ \citenamefont
  {Kahl}}]{EvansJPCM}%
  \BibitemOpen
  \bibfield  {author} {\bibinfo {author} {\bibfnamefont {R.}~\bibnamefont
  {Evans}}, \bibinfo {author} {\bibfnamefont {M.}~\bibnamefont {Oettel}},
  \bibinfo {author} {\bibfnamefont {R.}~\bibnamefont {Roth}}, \ and\ \bibinfo
  {author} {\bibfnamefont {G.}~\bibnamefont {Kahl}},\ }\bibfield  {title}
  {\enquote {\bibinfo {title} {New developments in classical density functional
  theory},}\ }\href {\doibase 10.1088/0953-8984/28/24/240401} {\bibfield
  {journal} {\bibinfo  {journal} {Journal of Physics: Condensed Matter}\
  }\textbf {\bibinfo {volume} {28}},\ \bibinfo {pages} {240401} (\bibinfo
  {year} {2016})}\BibitemShut {NoStop}%
\bibitem [{\citenamefont {Härtel}(2017)}]{Andreas_2017}%
  \BibitemOpen
  \bibfield  {author} {\bibinfo {author} {\bibfnamefont {A.}~\bibnamefont
  {Härtel}},\ }\bibfield  {title} {\enquote {\bibinfo {title} {Structure of
  electric double layers in capacitive systems and to what extent (classical)
  density functional theory describes it},}\ }\href {\doibase
  10.1088/1361-648x/aa8342} {\bibfield  {journal} {\bibinfo  {journal} {Journal
  of Physics: Condensed Matter}\ }\textbf {\bibinfo {volume} {29}},\ \bibinfo
  {pages} {423002} (\bibinfo {year} {2017})}\BibitemShut {NoStop}%
\bibitem [{\citenamefont {Blum}(1974)}]{Blum_1}%
  \BibitemOpen
  \bibfield  {author} {\bibinfo {author} {\bibfnamefont {L.}~\bibnamefont
  {Blum}},\ }\bibfield  {title} {\enquote {\bibinfo {title} {Solution of a
  model for the solvent‐electrolyte interactions in the mean spherical
  approximation},}\ }\href {\doibase 10.1063/1.1682224} {\bibfield  {journal}
  {\bibinfo  {journal} {The Journal of Chemical Physics}\ }\textbf {\bibinfo
  {volume} {61}},\ \bibinfo {pages} {2129--2133} (\bibinfo {year}
  {1974})}\BibitemShut {NoStop}%
\bibitem [{\citenamefont {Blum}(1975)}]{Blum_2}%
  \BibitemOpen
  \bibfield  {author} {\bibinfo {author} {\bibfnamefont {L.}~\bibnamefont
  {Blum}},\ }\bibfield  {title} {\enquote {\bibinfo {title} {Mean spherical
  model for asymmetric electrolytes},}\ }\href {\doibase
  10.1080/00268977500103051} {\bibfield  {journal} {\bibinfo  {journal}
  {Molecular Physics}\ }\textbf {\bibinfo {volume} {30}},\ \bibinfo {pages}
  {1529--1535} (\bibinfo {year} {1975})}\BibitemShut {NoStop}%
\bibitem [{\citenamefont {Blum}\ and\ \citenamefont {Hoeye}(1977)}]{Blum_3}%
  \BibitemOpen
  \bibfield  {author} {\bibinfo {author} {\bibfnamefont {L.}~\bibnamefont
  {Blum}}\ and\ \bibinfo {author} {\bibfnamefont {J.~S.}\ \bibnamefont
  {Hoeye}},\ }\bibfield  {title} {\enquote {\bibinfo {title} {Mean spherical
  model for asymmetric electrolytes. 2. thermodynamic properties and the pair
  correlation function},}\ }\href {\doibase 10.1021/j100528a019} {\bibfield
  {journal} {\bibinfo  {journal} {The Journal of Physical Chemistry}\ }\textbf
  {\bibinfo {volume} {81}},\ \bibinfo {pages} {1311--1316} (\bibinfo {year}
  {1977})}\BibitemShut {NoStop}%
\bibitem [{\citenamefont {Hiroike}(1977)}]{Hiroike}%
  \BibitemOpen
  \bibfield  {author} {\bibinfo {author} {\bibfnamefont {K.}~\bibnamefont
  {Hiroike}},\ }\bibfield  {title} {\enquote {\bibinfo {title} {Supplement to
  blum's theory for asymmetric electrolytes},}\ }\href {\doibase
  10.1080/00268977700101011} {\bibfield  {journal} {\bibinfo  {journal}
  {Molecular Physics}\ }\textbf {\bibinfo {volume} {33}},\ \bibinfo {pages}
  {1195--1198} (\bibinfo {year} {1977})}\BibitemShut {NoStop}%
\bibitem [{\citenamefont {Waisman}\ and\ \citenamefont
  {Lebowitz}(1970)}]{Waisman_1970}%
  \BibitemOpen
  \bibfield  {author} {\bibinfo {author} {\bibfnamefont {E.}~\bibnamefont
  {Waisman}}\ and\ \bibinfo {author} {\bibfnamefont {J.~L.}\ \bibnamefont
  {Lebowitz}},\ }\bibfield  {title} {\enquote {\bibinfo {title} {Exact solution
  of an integral equation for the structure of a primitive model of
  electrolytes},}\ }\href {\doibase 10.1063/1.1673642} {\bibfield  {journal}
  {\bibinfo  {journal} {The Journal of Chemical Physics}\ }\textbf {\bibinfo
  {volume} {52}},\ \bibinfo {pages} {4307--4309} (\bibinfo {year}
  {1970})}\BibitemShut {NoStop}%
\bibitem [{\citenamefont {Waisman}\ and\ \citenamefont
  {Lebowitz}(1972{\natexlab{a}})}]{Waisman_1972_I}%
  \BibitemOpen
  \bibfield  {author} {\bibinfo {author} {\bibfnamefont {E.}~\bibnamefont
  {Waisman}}\ and\ \bibinfo {author} {\bibfnamefont {J.~L.}\ \bibnamefont
  {Lebowitz}},\ }\bibfield  {title} {\enquote {\bibinfo {title} {Mean spherical
  model integral equation for charged hard spheres i. method of solution},}\
  }\href {\doibase 10.1063/1.1677644} {\bibfield  {journal} {\bibinfo
  {journal} {The Journal of Chemical Physics}\ }\textbf {\bibinfo {volume}
  {56}},\ \bibinfo {pages} {3086--3093} (\bibinfo {year}
  {1972}{\natexlab{a}})}\BibitemShut {NoStop}%
\bibitem [{\citenamefont {Waisman}\ and\ \citenamefont
  {Lebowitz}(1972{\natexlab{b}})}]{Waisman_1972_II}%
  \BibitemOpen
  \bibfield  {author} {\bibinfo {author} {\bibfnamefont {E.}~\bibnamefont
  {Waisman}}\ and\ \bibinfo {author} {\bibfnamefont {J.~L.}\ \bibnamefont
  {Lebowitz}},\ }\bibfield  {title} {\enquote {\bibinfo {title} {Mean spherical
  model integral equation for charged hard spheres. ii. results},}\ }\href
  {\doibase 10.1063/1.1677645} {\bibfield  {journal} {\bibinfo  {journal} {The
  Journal of Chemical Physics}\ }\textbf {\bibinfo {volume} {56}},\ \bibinfo
  {pages} {3093--3099} (\bibinfo {year} {1972}{\natexlab{b}})}\BibitemShut
  {NoStop}%
\bibitem [{\citenamefont {Roth}\ and\ \citenamefont
  {Gillespie}(2016)}]{Roth_shells}%
  \BibitemOpen
  \bibfield  {author} {\bibinfo {author} {\bibfnamefont {R.}~\bibnamefont
  {Roth}}\ and\ \bibinfo {author} {\bibfnamefont {D.}~\bibnamefont
  {Gillespie}},\ }\bibfield  {title} {\enquote {\bibinfo {title} {Shells of
  charge: a density functional theory for charged hard spheres},}\ }\href
  {\doibase 10.1088/0953-8984/28/24/244006} {\bibfield  {journal} {\bibinfo
  {journal} {Journal of Physics: Condensed Matter}\ }\textbf {\bibinfo {volume}
  {28}},\ \bibinfo {pages} {244006} (\bibinfo {year} {2016})}\BibitemShut
  {NoStop}%
\bibitem [{\citenamefont {Roth}(2010)}]{Roth_FMT}%
  \BibitemOpen
  \bibfield  {author} {\bibinfo {author} {\bibfnamefont {R.}~\bibnamefont
  {Roth}},\ }\bibfield  {title} {\enquote {\bibinfo {title} {Fundamental
  measure theory for hard-sphere mixtures: a review},}\ }\href {\doibase
  10.1088/0953-8984/22/6/063102} {\bibfield  {journal} {\bibinfo  {journal}
  {Journal of Physics: Condensed Matter}\ }\textbf {\bibinfo {volume} {22}},\
  \bibinfo {pages} {063102} (\bibinfo {year} {2010})}\BibitemShut {NoStop}%
\bibitem [{\citenamefont {Mier‐y‐Teran}\ \emph {et~al.}(1990)\citenamefont
  {Mier‐y‐Teran}, \citenamefont {Suh}, \citenamefont {White},\ and\
  \citenamefont {Davis}}]{MSAc}%
  \BibitemOpen
  \bibfield  {author} {\bibinfo {author} {\bibfnamefont {L.}~\bibnamefont
  {Mier‐y‐Teran}}, \bibinfo {author} {\bibfnamefont {S.~H.}\ \bibnamefont
  {Suh}}, \bibinfo {author} {\bibfnamefont {H.~S.}\ \bibnamefont {White}}, \
  and\ \bibinfo {author} {\bibfnamefont {H.~T.}\ \bibnamefont {Davis}},\
  }\bibfield  {title} {\enquote {\bibinfo {title} {A nonlocal free‐energy
  density‐functional approximation for the electrical double layer},}\ }\href
  {\doibase 10.1063/1.458542} {\bibfield  {journal} {\bibinfo  {journal} {The
  Journal of Chemical Physics}\ }\textbf {\bibinfo {volume} {92}},\ \bibinfo
  {pages} {5087--5098} (\bibinfo {year} {1990})}\BibitemShut {NoStop}%
\bibitem [{\citenamefont {Voukadinova}, \citenamefont {Valisk\'o},\ and\
  \citenamefont {Gillespie}(2018)}]{Gillespie}%
  \BibitemOpen
  \bibfield  {author} {\bibinfo {author} {\bibfnamefont {A.}~\bibnamefont
  {Voukadinova}}, \bibinfo {author} {\bibfnamefont {M.}~\bibnamefont
  {Valisk\'o}}, \ and\ \bibinfo {author} {\bibfnamefont {D.}~\bibnamefont
  {Gillespie}},\ }\bibfield  {title} {\enquote {\bibinfo {title} {Assessing the
  accuracy of three classical density functional theories of the electrical
  double layer},}\ }\href {\doibase 10.1103/PhysRevE.98.012116} {\bibfield
  {journal} {\bibinfo  {journal} {Phys. Rev. E}\ }\textbf {\bibinfo {volume}
  {98}},\ \bibinfo {pages} {012116} (\bibinfo {year} {2018})}\BibitemShut
  {NoStop}%
\bibitem [{\citenamefont {Blum}\ and\ \citenamefont
  {Rosenfeld}(1991)}]{Blum_c}%
  \BibitemOpen
  \bibfield  {author} {\bibinfo {author} {\bibfnamefont {L.}~\bibnamefont
  {Blum}}\ and\ \bibinfo {author} {\bibfnamefont {Y.}~\bibnamefont
  {Rosenfeld}},\ }\bibfield  {title} {\enquote {\bibinfo {title} {Relation
  between the free energy and the direct correlation function in the mean
  spherical approximation},}\ }\href {\doibase 10.1007/BF01030005} {\bibfield
  {journal} {\bibinfo  {journal} {J. Stat. Phys.}\ }\textbf {\bibinfo {volume}
  {63}},\ \bibinfo {pages} {1177–1190} (\bibinfo {year} {1991})}\BibitemShut
  {NoStop}%
\bibitem [{\citenamefont {Orkoulas}\ and\ \citenamefont
  {Panagiotopoulos}(1994)}]{RPMCP}%
  \BibitemOpen
  \bibfield  {author} {\bibinfo {author} {\bibfnamefont {G.}~\bibnamefont
  {Orkoulas}}\ and\ \bibinfo {author} {\bibfnamefont {A.~Z.}\ \bibnamefont
  {Panagiotopoulos}},\ }\bibfield  {title} {\enquote {\bibinfo {title} {Free
  energy and phase equilibria for the restricted primitive model of ionic
  fluids from monte carlo simulations},}\ }\href {\doibase 10.1063/1.467770}
  {\bibfield  {journal} {\bibinfo  {journal} {The Journal of Chemical Physics}\
  }\textbf {\bibinfo {volume} {101}},\ \bibinfo {pages} {1452--1459} (\bibinfo
  {year} {1994})}\BibitemShut {NoStop}%
\bibitem [{\citenamefont {Weik}\ \emph {et~al.}(2019)\citenamefont {Weik},
  \citenamefont {Weeber}, \citenamefont {Szuttor}, \citenamefont
  {Breitsprecher}, \citenamefont {de~Graaf}, \citenamefont {Kuron},
  \citenamefont {J.}, \citenamefont {H.}, \citenamefont {D.},\ and\
  \citenamefont {C.}}]{weik_espresso_2019}%
  \BibitemOpen
  \bibfield  {author} {\bibinfo {author} {\bibfnamefont {F.}~\bibnamefont
  {Weik}}, \bibinfo {author} {\bibfnamefont {R.}~\bibnamefont {Weeber}},
  \bibinfo {author} {\bibnamefont {Szuttor}}, \bibinfo {author} {\bibfnamefont
  {K.}~\bibnamefont {Breitsprecher}}, \bibinfo {author} {\bibfnamefont
  {J.}~\bibnamefont {de~Graaf}}, \bibinfo {author} {\bibfnamefont
  {M.}~\bibnamefont {Kuron}}, \bibinfo {author} {\bibfnamefont
  {L.}~\bibnamefont {J.}}, \bibinfo {author} {\bibfnamefont {M.}~\bibnamefont
  {H.}}, \bibinfo {author} {\bibfnamefont {S.}~\bibnamefont {D.}}, \ and\
  \bibinfo {author} {\bibfnamefont {H.}~\bibnamefont {C.}},\ }\bibfield
  {title} {\enquote {\bibinfo {title} {Particle methods in natural science and
  engineering},}\ }\href {\doibase 10.1140/epjst/e2019-900008-2} {\bibfield
  {journal} {\bibinfo  {journal} {Eur. Phys. J. Special Topics}\ }\textbf
  {\bibinfo {volume} {227}},\ \bibinfo {pages} {1493–1499} (\bibinfo {year}
  {2019})}\BibitemShut {NoStop}%
\bibitem [{\citenamefont {Hockney}\ and\ \citenamefont
  {Eastwood}(1988)}]{hockney_computer_1988}%
  \BibitemOpen
  \bibfield  {author} {\bibinfo {author} {\bibfnamefont {R.}~\bibnamefont
  {Hockney}}\ and\ \bibinfo {author} {\bibfnamefont {J.}~\bibnamefont
  {Eastwood}},\ }\href@noop {} {\emph {\bibinfo {title} {Computer Simulation
  Using Particles}}}\ (\bibinfo  {publisher} {CRC Press},\ \bibinfo {year}
  {1988})\BibitemShut {NoStop}%
\bibitem [{\citenamefont {Andersen}, \citenamefont {Weeks},\ and\ \citenamefont
  {Chandler}(1971)}]{andersen_relationship_1971}%
  \BibitemOpen
  \bibfield  {author} {\bibinfo {author} {\bibfnamefont {H.~C.}\ \bibnamefont
  {Andersen}}, \bibinfo {author} {\bibfnamefont {J.~D.}\ \bibnamefont {Weeks}},
  \ and\ \bibinfo {author} {\bibfnamefont {D.}~\bibnamefont {Chandler}},\
  }\bibfield  {title} {\enquote {\bibinfo {title} {Relationship between the
  hard-sphere fluid and fluids with realistic repulsive forces},}\ }\href
  {\doibase 10.1103/PhysRevA.4.1597} {\bibfield  {journal} {\bibinfo  {journal}
  {Phys. Rev. A}\ }\textbf {\bibinfo {volume} {4}},\ \bibinfo {pages}
  {1597--1607} (\bibinfo {year} {1971})}\BibitemShut {NoStop}%
\bibitem [{\citenamefont {Weeks}, \citenamefont {Chandler},\ and\ \citenamefont
  {Andersen}(1971)}]{weeks_role_1971}%
  \BibitemOpen
  \bibfield  {author} {\bibinfo {author} {\bibfnamefont {J.~D.}\ \bibnamefont
  {Weeks}}, \bibinfo {author} {\bibfnamefont {D.}~\bibnamefont {Chandler}}, \
  and\ \bibinfo {author} {\bibfnamefont {H.~C.}\ \bibnamefont {Andersen}},\
  }\bibfield  {title} {\enquote {\bibinfo {title} {Role of repulsive forces in
  determining the equilibrium structure of simple liquids},}\ }\href {\doibase
  10.1063/1.1674820} {\bibfield  {journal} {\bibinfo  {journal} {The Journal of
  Chemical Physics}\ }\textbf {\bibinfo {volume} {54}},\ \bibinfo {pages}
  {5237--5247} (\bibinfo {year} {1971})}\BibitemShut {NoStop}%
\bibitem [{\citenamefont {Arnold}, \citenamefont {de~Joannis},\ and\
  \citenamefont {Holm}(2002)}]{elc-method-part1}%
  \BibitemOpen
  \bibfield  {author} {\bibinfo {author} {\bibfnamefont {A.}~\bibnamefont
  {Arnold}}, \bibinfo {author} {\bibfnamefont {J.}~\bibnamefont {de~Joannis}},
  \ and\ \bibinfo {author} {\bibfnamefont {C.}~\bibnamefont {Holm}},\
  }\bibfield  {title} {\enquote {\bibinfo {title} {Electrostatics in periodic
  slab geometries. i},}\ }\href {\doibase 10.1063/1.1491955} {\bibfield
  {journal} {\bibinfo  {journal} {J. Chem. Phys.}\ }\textbf {\bibinfo {volume}
  {117}},\ \bibinfo {pages} {2496--2502} (\bibinfo {year} {2002})}\BibitemShut
  {NoStop}%
\bibitem [{\citenamefont {de~Joannis}, \citenamefont {Arnold},\ and\
  \citenamefont {Holm}(2002)}]{elc-method-part2}%
  \BibitemOpen
  \bibfield  {author} {\bibinfo {author} {\bibfnamefont {J.}~\bibnamefont
  {de~Joannis}}, \bibinfo {author} {\bibfnamefont {A.}~\bibnamefont {Arnold}},
  \ and\ \bibinfo {author} {\bibfnamefont {C.}~\bibnamefont {Holm}},\
  }\bibfield  {title} {\enquote {\bibinfo {title} {Electrostatics in periodic
  slab geometries. ii},}\ }\href {\doibase 10.1063/1.1491954} {\bibfield
  {journal} {\bibinfo  {journal} {J. Chem. Phys.}\ }\textbf {\bibinfo {volume}
  {117}},\ \bibinfo {pages} {2503--2512} (\bibinfo {year} {2002})}\BibitemShut
  {NoStop}%
\bibitem [{\citenamefont {Ulander}\ and\ \citenamefont
  {Kjellander}(2001)}]{Ulander_2001}%
  \BibitemOpen
  \bibfield  {author} {\bibinfo {author} {\bibfnamefont {J.}~\bibnamefont
  {Ulander}}\ and\ \bibinfo {author} {\bibfnamefont {R.}~\bibnamefont
  {Kjellander}},\ }\bibfield  {title} {\enquote {\bibinfo {title} {The decay of
  pair correlation functions in ionic fluids: A dressed ion theory analysis of
  monte carlo simulations},}\ }\href {\doibase 10.1063/1.1350449} {\bibfield
  {journal} {\bibinfo  {journal} {The Journal of Chemical Physics}\ }\textbf
  {\bibinfo {volume} {114}},\ \bibinfo {pages} {4893--4904} (\bibinfo {year}
  {2001})}\BibitemShut {NoStop}%
\bibitem [{\citenamefont {González-Mozuelos}, \citenamefont
  {Guerrero-García},\ and\ \citenamefont {Olvera de~la
  Cruz}(2013)}]{Gonzales_2013}%
  \BibitemOpen
  \bibfield  {author} {\bibinfo {author} {\bibfnamefont {P.}~\bibnamefont
  {González-Mozuelos}}, \bibinfo {author} {\bibfnamefont {G.~I.}\ \bibnamefont
  {Guerrero-García}}, \ and\ \bibinfo {author} {\bibfnamefont
  {M.}~\bibnamefont {Olvera de~la Cruz}},\ }\bibfield  {title} {\enquote
  {\bibinfo {title} {An exact method to obtain effective electrostatic
  interactions from computer simulations: The case of effective charge
  amplification},}\ }\href {\doibase 10.1063/1.4817776} {\bibfield  {journal}
  {\bibinfo  {journal} {The Journal of Chemical Physics}\ }\textbf {\bibinfo
  {volume} {139}},\ \bibinfo {pages} {064709} (\bibinfo {year}
  {2013})}\BibitemShut {NoStop}%
\bibitem [{\citenamefont {Evans}\ \emph {et~al.}(1993)\citenamefont {Evans},
  \citenamefont {Henderson}, \citenamefont {Hoyle}, \citenamefont {Parry},\
  and\ \citenamefont {Sabeur}}]{Evans_1993}%
  \BibitemOpen
  \bibfield  {author} {\bibinfo {author} {\bibfnamefont {R.}~\bibnamefont
  {Evans}}, \bibinfo {author} {\bibfnamefont {J.}~\bibnamefont {Henderson}},
  \bibinfo {author} {\bibfnamefont {D.}~\bibnamefont {Hoyle}}, \bibinfo
  {author} {\bibfnamefont {A.}~\bibnamefont {Parry}}, \ and\ \bibinfo {author}
  {\bibfnamefont {Z.}~\bibnamefont {Sabeur}},\ }\bibfield  {title} {\enquote
  {\bibinfo {title} {Asymptotic decay of liquid structure: Oscillatory
  liquid-vapour density profiles and the fisher-widom line},}\ }\href@noop {}
  {\bibfield  {journal} {\bibinfo  {journal} {Molecular Physics}\ }\textbf
  {\bibinfo {volume} {80}},\ \bibinfo {pages} {755 -- 775} (\bibinfo {year}
  {1993})}\BibitemShut {NoStop}%
\bibitem [{\citenamefont {Attard}\ \emph {et~al.}(1991)\citenamefont {Attard},
  \citenamefont {B\'erard}, \citenamefont {Ursenbach},\ and\ \citenamefont
  {Patey}}]{Attard_1991}%
  \BibitemOpen
  \bibfield  {author} {\bibinfo {author} {\bibfnamefont {P.}~\bibnamefont
  {Attard}}, \bibinfo {author} {\bibfnamefont {D.~R.}\ \bibnamefont
  {B\'erard}}, \bibinfo {author} {\bibfnamefont {C.~P.}\ \bibnamefont
  {Ursenbach}}, \ and\ \bibinfo {author} {\bibfnamefont {G.~N.}\ \bibnamefont
  {Patey}},\ }\bibfield  {title} {\enquote {\bibinfo {title} {Interaction free
  energy between planar walls in dense fluids: An ornstein-zernike approach
  with results for hard-sphere, lennard-jones, and dipolar systems},}\ }\href
  {\doibase 10.1103/PhysRevA.44.8224} {\bibfield  {journal} {\bibinfo
  {journal} {Phys. Rev. A}\ }\textbf {\bibinfo {volume} {44}},\ \bibinfo
  {pages} {8224--8234} (\bibinfo {year} {1991})}\BibitemShut {NoStop}%
\bibitem [{\citenamefont {Attard}, \citenamefont {Ursenbach},\ and\
  \citenamefont {Patey}(1992)}]{Attard_1992}%
  \BibitemOpen
  \bibfield  {author} {\bibinfo {author} {\bibfnamefont {P.}~\bibnamefont
  {Attard}}, \bibinfo {author} {\bibfnamefont {C.~P.}\ \bibnamefont
  {Ursenbach}}, \ and\ \bibinfo {author} {\bibfnamefont {G.~N.}\ \bibnamefont
  {Patey}},\ }\bibfield  {title} {\enquote {\bibinfo {title} {Long-range
  attractions between solutes in near-critical fluids},}\ }\href {\doibase
  10.1103/PhysRevA.45.7621} {\bibfield  {journal} {\bibinfo  {journal} {Phys.
  Rev. A}\ }\textbf {\bibinfo {volume} {45}},\ \bibinfo {pages} {7621--7623}
  (\bibinfo {year} {1992})}\BibitemShut {NoStop}%
\bibitem [{\citenamefont {Evans}\ \emph {et~al.}(1994)\citenamefont {Evans},
  \citenamefont {Leote~de Carvalho}, \citenamefont {Henderson},\ and\
  \citenamefont {Hoyle}}]{Evans_1994}%
  \BibitemOpen
  \bibfield  {author} {\bibinfo {author} {\bibfnamefont {R.}~\bibnamefont
  {Evans}}, \bibinfo {author} {\bibfnamefont {R.~J.~F.}\ \bibnamefont {Leote~de
  Carvalho}}, \bibinfo {author} {\bibfnamefont {J.~R.}\ \bibnamefont
  {Henderson}}, \ and\ \bibinfo {author} {\bibfnamefont {D.~C.}\ \bibnamefont
  {Hoyle}},\ }\bibfield  {title} {\enquote {\bibinfo {title} {Asymptotic decay
  of correlations in liquids and their mixtures},}\ }\href {\doibase
  10.1063/1.466920} {\bibfield  {journal} {\bibinfo  {journal} {The Journal of
  Chemical Physics}\ }\textbf {\bibinfo {volume} {100}},\ \bibinfo {pages}
  {591--603} (\bibinfo {year} {1994})}\BibitemShut {NoStop}%
\bibitem [{\citenamefont {Kjellander}\ and\ \citenamefont
  {Mitchell}(1992)}]{Kjellander_1992}%
  \BibitemOpen
  \bibfield  {author} {\bibinfo {author} {\bibfnamefont {R.}~\bibnamefont
  {Kjellander}}\ and\ \bibinfo {author} {\bibfnamefont {D.}~\bibnamefont
  {Mitchell}},\ }\bibfield  {title} {\enquote {\bibinfo {title} {An exact but
  linear and poisson—boltzmann-like theory for electrolytes and colloid
  dispersions in the primitive model},}\ }\href {\doibase
  https://doi.org/10.1016/0009-2614(92)87048-T} {\bibfield  {journal} {\bibinfo
   {journal} {Chemical Physics Letters}\ }\textbf {\bibinfo {volume} {200}},\
  \bibinfo {pages} {76 -- 82} (\bibinfo {year} {1992})}\BibitemShut {NoStop}%
\bibitem [{\citenamefont {Fisher}\ and\ \citenamefont
  {Widom}(1969)}]{FisherWidom}%
  \BibitemOpen
  \bibfield  {author} {\bibinfo {author} {\bibfnamefont {M.~E.}\ \bibnamefont
  {Fisher}}\ and\ \bibinfo {author} {\bibfnamefont {B.}~\bibnamefont {Widom}},\
  }\bibfield  {title} {\enquote {\bibinfo {title} {Decay of correlations in
  linear systems},}\ }\href {\doibase 10.1063/1.1671624} {\bibfield  {journal}
  {\bibinfo  {journal} {The Journal of Chemical Physics}\ }\textbf {\bibinfo
  {volume} {50}},\ \bibinfo {pages} {3756--3772} (\bibinfo {year}
  {1969})}\BibitemShut {NoStop}%
\bibitem [{\citenamefont {Dijkstra}\ and\ \citenamefont
  {Evans}(2000)}]{Dijkstra_2000}%
  \BibitemOpen
  \bibfield  {author} {\bibinfo {author} {\bibfnamefont {M.}~\bibnamefont
  {Dijkstra}}\ and\ \bibinfo {author} {\bibfnamefont {R.}~\bibnamefont
  {Evans}},\ }\bibfield  {title} {\enquote {\bibinfo {title} {A simulation
  study of the decay of the pair correlation function in simple fluids},}\
  }\href {\doibase 10.1063/1.480598} {\bibfield  {journal} {\bibinfo  {journal}
  {The Journal of Chemical Physics}\ }\textbf {\bibinfo {volume} {112}},\
  \bibinfo {pages} {1449--1456} (\bibinfo {year} {2000})}\BibitemShut {NoStop}%
\bibitem [{\citenamefont {Kirkwood}(1936)}]{Kirkwood}%
  \BibitemOpen
  \bibfield  {author} {\bibinfo {author} {\bibfnamefont {J.~G.}\ \bibnamefont
  {Kirkwood}},\ }\bibfield  {title} {\enquote {\bibinfo {title} {Statistical
  mechanics of liquid solutions.}}\ }\href {\doibase 10.1021/cr60064a007}
  {\bibfield  {journal} {\bibinfo  {journal} {Chemical Reviews}\ }\textbf
  {\bibinfo {volume} {19}},\ \bibinfo {pages} {275--307} (\bibinfo {year}
  {1936})}\BibitemShut {NoStop}%
\bibitem [{\citenamefont {Kjellander}\ and\ \citenamefont
  {Mitchell}(1994)}]{Kjellander_1994}%
  \BibitemOpen
  \bibfield  {author} {\bibinfo {author} {\bibfnamefont {R.}~\bibnamefont
  {Kjellander}}\ and\ \bibinfo {author} {\bibfnamefont {D.~J.}\ \bibnamefont
  {Mitchell}},\ }\bibfield  {title} {\enquote {\bibinfo {title} {Dressed‐ion
  theory for electrolyte solutions: A debye–hückel‐like reformulation of
  the exact theory for the primitive model},}\ }\href {\doibase
  10.1063/1.468116} {\bibfield  {journal} {\bibinfo  {journal} {The Journal of
  Chemical Physics}\ }\textbf {\bibinfo {volume} {101}},\ \bibinfo {pages}
  {603--626} (\bibinfo {year} {1994})}\BibitemShut {NoStop}%
\bibitem [{\citenamefont {Outhwaite}\ and\ \citenamefont
  {Hutson}(1975)}]{Outhwaite_1975}%
  \BibitemOpen
  \bibfield  {author} {\bibinfo {author} {\bibfnamefont {C.}~\bibnamefont
  {Outhwaite}}\ and\ \bibinfo {author} {\bibfnamefont {V.}~\bibnamefont
  {Hutson}},\ }\bibfield  {title} {\enquote {\bibinfo {title} {The mean
  spherical model for charged hard spheres},}\ }\href {\doibase
  10.1080/00268977500101331} {\bibfield  {journal} {\bibinfo  {journal}
  {Molecular Physics}\ }\textbf {\bibinfo {volume} {29}},\ \bibinfo {pages}
  {1521--1531} (\bibinfo {year} {1975})}\BibitemShut {NoStop}%
\bibitem [{\citenamefont {Outhwaite}\ and\ \citenamefont
  {Bhuiyan}(2019)}]{Outhwaite_2019}%
  \BibitemOpen
  \bibfield  {author} {\bibinfo {author} {\bibfnamefont {C.~W.}\ \bibnamefont
  {Outhwaite}}\ and\ \bibinfo {author} {\bibfnamefont {L.}~\bibnamefont
  {Bhuiyan}},\ }\bibfield  {title} {\enquote {\bibinfo {title} {Comments on the
  linear modified poisson-boltzmann equation in electrolyte solution theory},}\
  }\href {\doibase 10.5488/CMP.22.23801} {\bibfield  {journal} {\bibinfo
  {journal} {Condensed Matter Physics}\ }\textbf {\bibinfo {volume} {22}},\
  \bibinfo {pages} {23801} (\bibinfo {year} {2019})}\BibitemShut {NoStop}%
\bibitem [{\citenamefont {Evans}\ and\ \citenamefont {Marini
  Bettolo~Marconi}(1987)}]{Evans_1987}%
  \BibitemOpen
  \bibfield  {author} {\bibinfo {author} {\bibfnamefont {R.}~\bibnamefont
  {Evans}}\ and\ \bibinfo {author} {\bibfnamefont {U.}~\bibnamefont {Marini
  Bettolo~Marconi}},\ }\bibfield  {title} {\enquote {\bibinfo {title} {Phase
  equilibria and solvation forces for fluids confined between parallel
  walls},}\ }\href {\doibase 10.1063/1.452363} {\bibfield  {journal} {\bibinfo
  {journal} {The Journal of Chemical Physics}\ }\textbf {\bibinfo {volume}
  {86}},\ \bibinfo {pages} {7138--7148} (\bibinfo {year} {1987})}\BibitemShut
  {NoStop}%
\bibitem [{\citenamefont {Lee}\ \emph {et~al.}(2017{\natexlab{a}})\citenamefont
  {Lee}, \citenamefont {Perez-Martinez}, \citenamefont {Smith},\ and\
  \citenamefont {Perkin}}]{Lee_PRL}%
  \BibitemOpen
  \bibfield  {author} {\bibinfo {author} {\bibfnamefont {A.~A.}\ \bibnamefont
  {Lee}}, \bibinfo {author} {\bibfnamefont {C.~S.}\ \bibnamefont
  {Perez-Martinez}}, \bibinfo {author} {\bibfnamefont {A.~M.}\ \bibnamefont
  {Smith}}, \ and\ \bibinfo {author} {\bibfnamefont {S.}~\bibnamefont
  {Perkin}},\ }\bibfield  {title} {\enquote {\bibinfo {title} {Scaling analysis
  of the screening length in concentrated electrolytes},}\ }\href {\doibase
  10.1103/PhysRevLett.119.026002} {\bibfield  {journal} {\bibinfo  {journal}
  {Phys. Rev. Lett.}\ }\textbf {\bibinfo {volume} {119}},\ \bibinfo {pages}
  {026002} (\bibinfo {year} {2017}{\natexlab{a}})}\BibitemShut {NoStop}%
\bibitem [{\citenamefont {Lee}\ \emph {et~al.}(2017{\natexlab{b}})\citenamefont
  {Lee}, \citenamefont {Perez-Martinez}, \citenamefont {Smith},\ and\
  \citenamefont {Perkin}}]{Lee_underscreen}%
  \BibitemOpen
  \bibfield  {author} {\bibinfo {author} {\bibfnamefont {A.~A.}\ \bibnamefont
  {Lee}}, \bibinfo {author} {\bibfnamefont {C.~S.}\ \bibnamefont
  {Perez-Martinez}}, \bibinfo {author} {\bibfnamefont {A.~M.}\ \bibnamefont
  {Smith}}, \ and\ \bibinfo {author} {\bibfnamefont {S.}~\bibnamefont
  {Perkin}},\ }\bibfield  {title} {\enquote {\bibinfo {title} {Underscreening
  in concentrated electrolytes},}\ }\href {\doibase 10.1039/C6FD00250A}
  {\bibfield  {journal} {\bibinfo  {journal} {Faraday Discuss.}\ }\textbf
  {\bibinfo {volume} {199}},\ \bibinfo {pages} {239--259} (\bibinfo {year}
  {2017}{\natexlab{b}})}\BibitemShut {NoStop}%
\bibitem [{\citenamefont {Coupette}, \citenamefont {Lee},\ and\ \citenamefont
  {H\"artel}(2018)}]{usfabian}%
  \BibitemOpen
  \bibfield  {author} {\bibinfo {author} {\bibfnamefont {F.}~\bibnamefont
  {Coupette}}, \bibinfo {author} {\bibfnamefont {A.~A.}\ \bibnamefont {Lee}}, \
  and\ \bibinfo {author} {\bibfnamefont {A.}~\bibnamefont {H\"artel}},\
  }\bibfield  {title} {\enquote {\bibinfo {title} {Screening lengths in ionic
  fluids},}\ }\href {\doibase 10.1103/PhysRevLett.121.075501} {\bibfield
  {journal} {\bibinfo  {journal} {Phys. Rev. Lett.}\ }\textbf {\bibinfo
  {volume} {121}},\ \bibinfo {pages} {075501} (\bibinfo {year}
  {2018})}\BibitemShut {NoStop}%
\bibitem [{\citenamefont {Rotenberg}, \citenamefont {Bernard},\ and\
  \citenamefont {Hansen}(2018)}]{Rotenberg_2018}%
  \BibitemOpen
  \bibfield  {author} {\bibinfo {author} {\bibfnamefont {B.}~\bibnamefont
  {Rotenberg}}, \bibinfo {author} {\bibfnamefont {O.}~\bibnamefont {Bernard}},
  \ and\ \bibinfo {author} {\bibfnamefont {J.-P.}\ \bibnamefont {Hansen}},\
  }\bibfield  {title} {\enquote {\bibinfo {title} {Underscreening in ionic
  liquids: a first principles analysis},}\ }\href {\doibase
  10.1088/1361-648X/aaa3ac} {\bibfield  {journal} {\bibinfo  {journal} {J Phys
  Condens Matter}\ }\textbf {\bibinfo {volume} {30}},\ \bibinfo {pages}
  {054005} (\bibinfo {year} {2018})}\BibitemShut {NoStop}%
\bibitem [{\citenamefont {Feng}\ \emph {et~al.}(2019)\citenamefont {Feng},
  \citenamefont {Chen}, \citenamefont {Bi}, \citenamefont {Goodwin},
  \citenamefont {Postnikov}, \citenamefont {Brilliantov}, \citenamefont
  {Urbakh},\ and\ \citenamefont {Kornyshev}}]{uskorny}%
  \BibitemOpen
  \bibfield  {author} {\bibinfo {author} {\bibfnamefont {G.}~\bibnamefont
  {Feng}}, \bibinfo {author} {\bibfnamefont {M.}~\bibnamefont {Chen}}, \bibinfo
  {author} {\bibfnamefont {S.}~\bibnamefont {Bi}}, \bibinfo {author}
  {\bibfnamefont {Z.~A.~H.}\ \bibnamefont {Goodwin}}, \bibinfo {author}
  {\bibfnamefont {E.~B.}\ \bibnamefont {Postnikov}}, \bibinfo {author}
  {\bibfnamefont {N.}~\bibnamefont {Brilliantov}}, \bibinfo {author}
  {\bibfnamefont {M.}~\bibnamefont {Urbakh}}, \ and\ \bibinfo {author}
  {\bibfnamefont {A.~A.}\ \bibnamefont {Kornyshev}},\ }\bibfield  {title}
  {\enquote {\bibinfo {title} {Free and bound states of ions in ionic liquids,
  conductivity, and underscreening paradox},}\ }\href {\doibase
  10.1103/PhysRevX.9.021024} {\bibfield  {journal} {\bibinfo  {journal} {Phys.
  Rev. X}\ }\textbf {\bibinfo {volume} {9}},\ \bibinfo {pages} {021024}
  (\bibinfo {year} {2019})}\BibitemShut {NoStop}%
\bibitem [{\citenamefont {Kjellander}(2020)}]{Kjellander_2020}%
  \BibitemOpen
  \bibfield  {author} {\bibinfo {author} {\bibfnamefont {R.}~\bibnamefont
  {Kjellander}},\ }\bibfield  {title} {\enquote {\bibinfo {title} {A multiple
  decay-length extension of the debye–hückel theory: to achieve high
  accuracy also for concentrated solutions and explain under-screening in
  dilute symmetric electrolytes},}\ }\href {\doibase 10.1039/D0CP02742A}
  {\bibfield  {journal} {\bibinfo  {journal} {Phys. Chem. Chem. Phys.}\
  }\textbf {\bibinfo {volume} {22}},\ \bibinfo {pages} {23952--23985} (\bibinfo
  {year} {2020})}\BibitemShut {NoStop}%
\bibitem [{\citenamefont {Smith}, \citenamefont {Lee},\ and\ \citenamefont
  {Perkin}(2017)}]{Smith_PRL118}%
  \BibitemOpen
  \bibfield  {author} {\bibinfo {author} {\bibfnamefont {A.~M.}\ \bibnamefont
  {Smith}}, \bibinfo {author} {\bibfnamefont {A.~A.}\ \bibnamefont {Lee}}, \
  and\ \bibinfo {author} {\bibfnamefont {S.}~\bibnamefont {Perkin}},\
  }\bibfield  {title} {\enquote {\bibinfo {title} {Switching the structural
  force in ionic liquid-solvent mixtures by varying composition},}\ }\href
  {\doibase 10.1103/PhysRevLett.118.096002} {\bibfield  {journal} {\bibinfo
  {journal} {Phys. Rev. Lett.}\ }\textbf {\bibinfo {volume} {118}},\ \bibinfo
  {pages} {096002} (\bibinfo {year} {2017})}\BibitemShut {NoStop}%
\bibitem [{\citenamefont {Zeidler}(2009)}]{Zeidler}%
  \BibitemOpen
  \bibfield  {author} {\bibinfo {author} {\bibfnamefont {A.}~\bibnamefont
  {Zeidler}},\ }\emph {\bibinfo {title} {Ordering in Amorphous Binary
  Systems}},\ \href@noop {} {Ph.D. thesis},\ \bibinfo  {school} {University of
  Bath} (\bibinfo {year} {2009}),\ \bibinfo {note} {chapter 3: The Structure of
  Molten Sodium Chloride}\BibitemShut {NoStop}%
\bibitem [{\citenamefont {Keblinski}\ \emph {et~al.}(2000)\citenamefont
  {Keblinski}, \citenamefont {Eggebrecht}, \citenamefont {Wolf},\ and\
  \citenamefont {Phillpot}}]{Keblinsky_2000}%
  \BibitemOpen
  \bibfield  {author} {\bibinfo {author} {\bibfnamefont {P.}~\bibnamefont
  {Keblinski}}, \bibinfo {author} {\bibfnamefont {J.}~\bibnamefont
  {Eggebrecht}}, \bibinfo {author} {\bibfnamefont {D.}~\bibnamefont {Wolf}}, \
  and\ \bibinfo {author} {\bibfnamefont {S.~R.}\ \bibnamefont {Phillpot}},\
  }\bibfield  {title} {\enquote {\bibinfo {title} {Molecular dynamics study of
  screening in ionic fluids},}\ }\href {\doibase 10.1063/1.481819} {\bibfield
  {journal} {\bibinfo  {journal} {The Journal of Chemical Physics}\ }\textbf
  {\bibinfo {volume} {113}},\ \bibinfo {pages} {282--291} (\bibinfo {year}
  {2000})}\BibitemShut {NoStop}%
\bibitem [{\citenamefont {Coles}\ \emph {et~al.}(2020)\citenamefont {Coles},
  \citenamefont {Park}, \citenamefont {Nikam}, \citenamefont {Kanduč},
  \citenamefont {Dzubiella},\ and\ \citenamefont {Rotenberg}}]{Rotenberg}%
  \BibitemOpen
  \bibfield  {author} {\bibinfo {author} {\bibfnamefont {S.~W.}\ \bibnamefont
  {Coles}}, \bibinfo {author} {\bibfnamefont {C.}~\bibnamefont {Park}},
  \bibinfo {author} {\bibfnamefont {R.}~\bibnamefont {Nikam}}, \bibinfo
  {author} {\bibfnamefont {M.}~\bibnamefont {Kanduč}}, \bibinfo {author}
  {\bibfnamefont {J.}~\bibnamefont {Dzubiella}}, \ and\ \bibinfo {author}
  {\bibfnamefont {B.}~\bibnamefont {Rotenberg}},\ }\bibfield  {title} {\enquote
  {\bibinfo {title} {Correlation length in concentrated electrolytes: Insights
  from all-atom molecular dynamics simulations},}\ }\href {\doibase
  10.1021/acs.jpcb.9b10542} {\bibfield  {journal} {\bibinfo  {journal} {The
  Journal of Physical Chemistry B}\ }\textbf {\bibinfo {volume} {124}},\
  \bibinfo {pages} {1778--1786} (\bibinfo {year} {2020})}\BibitemShut {NoStop}%
\bibitem [{\citenamefont {Zeman}, \citenamefont {Kondrat},\ and\ \citenamefont
  {Holm}(2020)}]{Holm}%
  \BibitemOpen
  \bibfield  {author} {\bibinfo {author} {\bibfnamefont {J.}~\bibnamefont
  {Zeman}}, \bibinfo {author} {\bibfnamefont {S.}~\bibnamefont {Kondrat}}, \
  and\ \bibinfo {author} {\bibfnamefont {C.}~\bibnamefont {Holm}},\ }\bibfield
  {title} {\enquote {\bibinfo {title} {Bulk ionic screening lengths from
  extremely large-scale molecular dynamics simulations},}\ }\href {\doibase
  10.1039/D0CC05023G} {\bibfield  {journal} {\bibinfo  {journal} {Chem.
  Commun.}\ }\textbf {\bibinfo {volume} {56}},\ \bibinfo {pages} {15635--15638}
  (\bibinfo {year} {2020})}\BibitemShut {NoStop}%
\bibitem [{\citenamefont {Kjellander}(2019)}]{Kjellander_2019}%
  \BibitemOpen
  \bibfield  {author} {\bibinfo {author} {\bibfnamefont {R.}~\bibnamefont
  {Kjellander}},\ }\bibfield  {title} {\enquote {\bibinfo {title} {The intimate
  relationship between the dielectric response and the decay of intermolecular
  correlations and surface forces in electrolytes},}\ }\href {\doibase
  10.1039/C9SM00712A} {\bibfield  {journal} {\bibinfo  {journal} {Soft Matter}\
  }\textbf {\bibinfo {volume} {15}},\ \bibinfo {pages} {5866--5895} (\bibinfo
  {year} {2019})}\BibitemShut {NoStop}%
\bibitem [{\citenamefont {Kjellander}(2016)}]{Kjellander_2016}%
  \BibitemOpen
  \bibfield  {author} {\bibinfo {author} {\bibfnamefont {R.}~\bibnamefont
  {Kjellander}},\ }\bibfield  {title} {\enquote {\bibinfo {title} {Nonlocal
  electrostatics in ionic liquids: The key to an understanding of the screening
  decay length and screened interactions},}\ }\href {\doibase
  10.1063/1.4962756} {\bibfield  {journal} {\bibinfo  {journal} {The Journal of
  Chemical Physics}\ }\textbf {\bibinfo {volume} {145}},\ \bibinfo {pages}
  {124503} (\bibinfo {year} {2016})}\BibitemShut {NoStop}%
\bibitem [{\citenamefont {Macio\l{}ek}, \citenamefont {Drzewi\'{n}ski},\ and\
  \citenamefont {Bryk}(2004)}]{Maciolek_2004}%
  \BibitemOpen
  \bibfield  {author} {\bibinfo {author} {\bibfnamefont {A.}~\bibnamefont
  {Macio\l{}ek}}, \bibinfo {author} {\bibfnamefont {A.}~\bibnamefont
  {Drzewi\'{n}ski}}, \ and\ \bibinfo {author} {\bibfnamefont {P.}~\bibnamefont
  {Bryk}},\ }\bibfield  {title} {\enquote {\bibinfo {title} {Solvation force
  for long-ranged wall-fluid potentials},}\ }\href {\doibase 10.1063/1.1635807}
  {\bibfield  {journal} {\bibinfo  {journal} {J. Chem. Phys.}\ }\textbf
  {\bibinfo {volume} {120}},\ \bibinfo {pages} {1921--1934} (\bibinfo {year}
  {2004})}\BibitemShut {NoStop}%
\bibitem [{\citenamefont {Percus}(1962)}]{Percus_1962}%
  \BibitemOpen
  \bibfield  {author} {\bibinfo {author} {\bibfnamefont {J.~K.}\ \bibnamefont
  {Percus}},\ }\bibfield  {title} {\enquote {\bibinfo {title} {Approximation
  methods in classical statistical mechanics},}\ }\href {\doibase
  10.1103/PhysRevLett.8.462} {\bibfield  {journal} {\bibinfo  {journal} {Phys.
  Rev. Lett.}\ }\textbf {\bibinfo {volume} {8}},\ \bibinfo {pages} {462--463}
  (\bibinfo {year} {1962})}\BibitemShut {NoStop}%
\bibitem [{\citenamefont {Walters}\ \emph {et~al.}(2018)\citenamefont
  {Walters}, \citenamefont {Subramanian}, \citenamefont {Archer},\ and\
  \citenamefont {Evans}}]{Walters_2018}%
  \BibitemOpen
  \bibfield  {author} {\bibinfo {author} {\bibfnamefont {M.~C.}\ \bibnamefont
  {Walters}}, \bibinfo {author} {\bibfnamefont {P.}~\bibnamefont
  {Subramanian}}, \bibinfo {author} {\bibfnamefont {A.~J.}\ \bibnamefont
  {Archer}}, \ and\ \bibinfo {author} {\bibfnamefont {R.}~\bibnamefont
  {Evans}},\ }\bibfield  {title} {\enquote {\bibinfo {title} {Structural
  crossover in a model fluid exhibiting two length scales: Repercussions for
  quasicrystal formation},}\ }\href {\doibase 10.1103/PhysRevE.98.012606}
  {\bibfield  {journal} {\bibinfo  {journal} {Phys. Rev. E}\ }\textbf {\bibinfo
  {volume} {98}},\ \bibinfo {pages} {012606} (\bibinfo {year}
  {2018})}\BibitemShut {NoStop}%
\bibitem [{\citenamefont {Stopper}\ \emph {et~al.}(2019)\citenamefont
  {Stopper}, \citenamefont {Hansen-Goos}, \citenamefont {Roth},\ and\
  \citenamefont {Evans}}]{Stopper_2019}%
  \BibitemOpen
  \bibfield  {author} {\bibinfo {author} {\bibfnamefont {D.}~\bibnamefont
  {Stopper}}, \bibinfo {author} {\bibfnamefont {H.}~\bibnamefont
  {Hansen-Goos}}, \bibinfo {author} {\bibfnamefont {R.}~\bibnamefont {Roth}}, \
  and\ \bibinfo {author} {\bibfnamefont {R.}~\bibnamefont {Evans}},\ }\bibfield
   {title} {\enquote {\bibinfo {title} {On the decay of the pair correlation
  function and the line of vanishing excess isothermal compressibility in
  simple fluids},}\ }\href {\doibase 10.1063/1.5110044} {\bibfield  {journal}
  {\bibinfo  {journal} {The Journal of Chemical Physics}\ }\textbf {\bibinfo
  {volume} {151}},\ \bibinfo {pages} {014501} (\bibinfo {year}
  {2019})}\BibitemShut {NoStop}%
\bibitem [{\citenamefont {Evans}\ and\ \citenamefont
  {de~Carvalho}(1996)}]{Evans_book}%
  \BibitemOpen
  \bibfield  {author} {\bibinfo {author} {\bibfnamefont {R.}~\bibnamefont
  {Evans}}\ and\ \bibinfo {author} {\bibfnamefont {R.~J. F.~L.}\ \bibnamefont
  {de~Carvalho}},\ }\enquote {\bibinfo {title} {Decay of correlations in bulk
  fluids and at interfaces: A density-functional perspective},}\ in\ \href
  {\doibase 10.1021/bk-1996-0629.ch012} {\emph {\bibinfo {booktitle} {Chemical
  Applications of Density-Functional Theory}}}\ (\bibinfo  {publisher}
  {American Chemical Society},\ \bibinfo {year} {1996})\ Chap.~\bibinfo
  {chapter} {12}, pp.\ \bibinfo {pages} {166--184}\BibitemShut {NoStop}%
\end{thebibliography}%

\appendix


\section{Adding Electrostatic Energy terms to the functional and the inconsistency of the MSAu}\label{App:ES_cons}

In this section we focus on the electrostatics part of the functional, especially on (the lack of) consistency between several routes one can take from a free-energy functional to thermodynamic and structural properties of the electrolyte.  The pair direct correlation functions obtained from the crudest MF/Poisson-Boltzmann treatment and from the MSA closure for the bulk RPM are such that the ionic valencies can be factored out so that $c^{(2),ES}_{ij}(r)=z_iz_jc^{ES}(r)$. Here $c^{ES}(r)$ depends on the closure but vanishes for $r\rightarrow\infty$. This structural form of the direct correlations can be reconstructed by considering second functional derivatives of an (electrostatic) excess free-energy functional of the form
\begin{align}\label{Eq:FESc}
\beta\mathcal{F}_{ex}^{ES,c}[\{\rho\}]=-\frac{1}{2}\int\mathrm{d}\mathbf{r}\int\mathrm{d}\mathbf{r}'\rho_Z(\mathbf{r})c^{ES}(|\mathbf{r}-\mathbf{r}'|;\{\rho_b\})\rho_Z(\mathbf{r}'),
\end{align}
where $\rho_Z=\sum_j z_j\rho_j=\rho_+-\rho_-$. (Note that $\rho_Z$ introduced in the main text was normalized w.r.t. the bulk density $\rho_b$). In a bulk system, where the profiles are constant and charge neutrality holds such that $\rho_Z=0$, a functional of the form of Eq.~\eqref{Eq:FESc} gives a vanishing electrostatic free energy. However, if we calculate the bulk pair direct correlation function  $c_Z(r)=c_{++}^{(2),ES}(r)-c_{+-}^{(2),ES}(r)=2c^{ES}(r)$ from Eq.~\eqref{Eq:FESc}, and use the OZ equation to find $h_Z(r)$, the electrostatic internal energy $U^{ES}$  follows via the energy route as
\begin{align}\label{Eq:energy_route}
\beta\frac{U^{ES}}{V}=4\pi\lambda_B \rho_b^2\int_{0}^{\infty}\mathrm{d}r r h_Z(r).
\end{align}
The internal energy can then be used to obtain the Helmholtz free energy $F^{ES}$ via the standard temperature or charging integration
\begin{align}\label{Eq:UF}
\beta F^{ES}(\beta)=\int_0^{\beta} \mathrm{d}\beta' U^{ES}(\beta').
\end{align}
In the dilute limit the resulting reduced free energy density $\Phi^{ES}=\beta F^{ES}/V$ should reduce to the exactly known limiting law
\begin{align}\label{Eq:F_MF}
\lim_{\{\rho_b\}\rightarrow 0}\Phi^{ES}(\{\rho_b\})=-\frac{\kappa_D^3(\{\rho_b\})}{12\pi},
\end{align}
which is manifestly non-zero. Hence, using the same $c^{ES}(r)$ in Eq.~\eqref{Eq:FESc} we find, depending on the chosen route, either a vanishing or the correct (physical) non-zero bulk electrostatic free energy. Revisiting the original DH paper\cite{DH}, or considering the exact expression given by Eq.~(\ref{Eq:energy_route}), shows that one can interpret the internal energy $U^{ES}$ as the sum of Coulomb energies of each ion with its surrounding screening cloud of ions of opposite sign. This is overall a  negative energy contribution arising from the cohesive energy due to the Coulombic attraction between (relatively) nearby opposite charges that dominates over positive energy contributions from the repulsions between like charges at (relatively) large distances.  In order to remedy the inconsistency between these two routes, one might consider including an additional term in the free-energy functional that takes the cohesive Coulomb energy into account explicitly, e.g.
\begin{align}\label{Eq:ESu}
    \beta\mathcal{F}^{ESu}[\{\rho\}]=\int\mathrm{d}\mathbf{r}\Phi^{ES}(\{\tilde{\rho}(\mathbf{r})\}),
\end{align}
where $\tilde{\rho}$ denotes a weighted  density  with an arbitrary weight function $\omega$ that does \textit{not} depend on the bulk density.
However,  in order to retain consistency between the two routes discussed above, such an additional term as written in  Eq.~\eqref{Eq:ESu} must be chosen so that it does not affect the bulk direct correlation function $c_Z$ that enters the calculation of the charge correlation function $h_Z$ from which the internal energy of Eq.~\eqref{Eq:energy_route} follows. This implies that the contribution to $c_Z(r)$ due to Eq.~\eqref{Eq:ESu} must vanish, i.e.
\begin{align}\label{Eq:cZ}
-{\beta}\frac{\delta^2\mathcal{F}_{ex}^{ESu}[\{\rho\}]}{\delta\rho_Z(\mathbf{r})\delta\rho_Z(\mathbf{r}')}=0,
\end{align}
Hence, $\Phi^{ES}(\{\tilde{\rho}\})$ may be at most linear in $\rho_Z$. If this is the case, then the sum of Eq.~\eqref{Eq:FESc} and the additional Helmholtz free-energy contribution of Eq.~\eqref{Eq:ESu}  is consistent, when comparing the free energy that results from evaluating the resulting functional in bulk, with the one from the energy  route leading to Eqs.~\eqref{Eq:UF} and~\eqref{Eq:F_MF}. Moreover, if $c_Z$ does not change by adding Eq.~\eqref{Eq:ESu} to the free-energy functional, then the asymptotic decay of the charge-charge correlations also remains unchanged. 

If we consider Eq.~\eqref{Eq:F_MF}, and recognize that $\kappa_D$ depends only on the total density $\rho_N=\rho_++\rho_-$ in the RPM, it is clear that including a contribution to the free-energy functional such as
\begin{align}\label{Eq:FES_MF}
\beta\mathcal{F}_{ex}^{ES,F}[\{\rho\}]=-\int\mathrm{d}\mathbf{r}\frac{\kappa^3_D(\{\tilde{\rho}(\mathbf{r})\})}{12\pi}
\end{align}
does not breach electrostatic consistency. We found that minimizing the resulting functional actually accounts for the depletion observed in the number density profile in Fig.~\ref{Fig:rho}(a).
Importantly, however, the MSAu functional that we used in the main text has a different structure as it depends not only on the number density but \textit{also} on the square of the  charge density. This can be ascertained by examining the MSAu contribution to the functional and write Eq.~\eqref{Eq:MSA_phi_msa} as
\begin{align}\label{Eq:Phi_MSA_RPM}
\Phi^{MSA}_{RPM}(\tilde{n}_Z(\mathbf{r}),\tilde{n}_N(\mathbf{r}))=\vartheta_1(\tilde{n}_N(\mathbf{r}))+\vartheta_2(\tilde{n}_N(\mathbf{r}))\tilde{n}^2_Z(\mathbf{r}),
\end{align}
where  $\vartheta_1(\tilde{n}_N)$ and $\vartheta_2(\tilde{n}_N)$ are functions of $\tilde{n}_N$  and not of $\tilde{n}_Z$. The final term in this equation originates  in the term proportional to $\eta$ in Eq.~\eqref{Eq:MSA_phi_msa}. It follows that  $\Phi^{ES}$ is quadratic in $\rho_Z$ and will therefore breach consistency. This suggests that the MSAu functional, as implemented by Roth and  Gillespie Ref.~\onlinecite{Roth_shells}, appears to perform well for density profiles close to charged surfaces\cite{Gillespie}, but suffers from inconsistencies that dictate its (poor) performance in the far-field regime. In order to investigate further, we examined  the MSAu functional for the case where $\eta(\{\tilde{n}(\mathbf{r})\})=0$ for all positions. Then  $\vartheta_2$ vanishes and we retrieve precisely the same asymptotic Z decay as from MSAc and IET. We note, however, any infinitesimal asymmetry in the ion sizes will cause the $\eta$ term to reappear, creating the same issues. For these reasons we choose to work with a non-vanishing $\eta$, in line with Ref.~\onlinecite{Roth_shells}, and our figures display results for this choice. 


\section{Relating the Decay of the one-body Density Profile to that of the two-body Bulk Correlation Function}\label{App:decay_onebody}

We show that the asymptotic decay of the one-body density profile is determined by the asymptotic decay of the bulk pair correlation function. The Euler-Lagrange equation for a single species system is:
\begin{align}
\rho(\mathbf{r})&=\rho_b\exp\left(-\beta V_{ext}(\mathbf{r})+c^{(1)}(\mathbf{r};[\rho])-c^{(1)}_b\right),
\end{align}
with the one-body direct correlation function $c_b^{(1)}=c^{(1)}(\rho_b)$. We choose the external potential $V_{ext}(\mathbf{r})$ to correspond to a planar wall or to a big (spherical) solute, such that $V_{ext}(\mathbf{r})\rightarrow 0$ as $\mathbf{r}\rightarrow \infty$, and $\rho(\mathbf{r})\rightarrow\rho_b$, the bulk reservoir value.
Asymptotically ($\mathbf{r}\rightarrow\infty$) the term in the exponential is small, which allows us to write
\begin{align}
\rho(\mathbf{r})=\rho_b\left(1-\beta V_{ext}(\mathbf{r})+c^{(1)}(\mathbf{r};[\rho])-c^{(1)}_b\right).
\end{align}
Noting that $c^{(1)}(\mathbf{r};[\rho])$ is both a functional of $\rho$ and a function of $\mathbf{r}$, we can expand this term around the bulk density to lowest order:
\begin{align}\label{Eq:c1_expansion}
c^{(1)}(\mathbf{r};[\rho])=&c^{(1)}_b+\int \mathrm{d}\mathbf{r}' \left.\frac{\delta c^{(1)}(\mathbf{r};[\rho])}{\delta \rho(\mathbf{r}')}\right|_{\rho=\rho_b}(\rho(\mathbf{r}')-\rho_b)
\end{align}
and write, 
\begin{align}
c^{(1)}(\mathbf{r};[\rho])-c^{(1)}_b=\int \mathrm{d}\mathbf{r}{'} c^{(2)}(|\mathbf{r}-\mathbf{r}'|;\rho_b)\Delta\rho(\mathbf{r}'),
\end{align}
with density deviation  $\Delta\rho(\mathbf{r})=\rho(\mathbf{r})-\rho_b$.
The density deviation can then be expressed, in lowest order, as
\begin{align}
\frac{\Delta\rho(\mathbf{r})}{\rho_b}=-\beta V_{ext}(\mathbf{r})+\int \mathrm{d}\mathbf{r}' c^{(2)}(|\mathbf{r}-\mathbf{r}'|;\rho_b)\Delta\rho(\mathbf{r}').
\end{align}
We now suppose that $V_{ext}(\mathbf{r})$ is sufficiently smooth that its Fourier transform exists. Then
\begin{align}
\frac{\Delta\hat{\rho}(k)}{\rho_b}=-\beta \hat{V}_{ext}(k)+\hat{c}^{(2)}(k;\rho_b)\Delta\hat{\rho}(k).
\end{align}
and the Fourier transform of the density deviation takes the simple form:
\begin{align}\label{Eq:Drho_k}
\Delta\hat{\rho}(k)=\frac{-\beta \hat{V}_{ext}(k)\rho_b}{1-\rho_b\hat{c}^{(2)}(k;\rho_b)}.
\end{align}
Note that the Fourier transform, denoted by \^{}, is three-dimensional. For the spherical solute, the structure of Eq.~\eqref{Eq:Drho_k} is equivalent to that of the bulk OZ Eq.~\eqref{Eq:h_k}. In that case poles of the total pair correlation function $\hat{h}(k)$ are determined by the zeroes of $1-\hat{c}^{(2)}(k;\rho_b)\rho_b$. Similarly the poles of the density deviation $\Delta\hat{\rho}(k)$ for a (large) solute are determined by the same zeroes. In practice, this means that we should consider model fluids where the pair potential $u(r)$ is short-ranged, i.e. it should decay faster than power law and the potential $V_{ext}(r)$ should decay faster than power-law \underline{and}, if this is exponentially decaying, should have a decay length that is shorter than the bulk correlation length of the liquid. This simple argument focuses on the poles. We do not address explicitly the case of branch point singularities.

An important limiting case is when the solute is made identical to a solvent particle. Then the one-body density profile $\rho(r)=\rho_bg(r)=\rho_b(1+h(r))$, which is the famous Percus test particle result\cite{Percus_1962}.
It follows that employing the test particle route within the framework of DFT must yield the same poles, where these dictate the decay, and therefore the same asymptotic decay length and wavelength, where pertinent, as those determined from the bulk OZ equation~\eqref{Eq:h_k}, with $\hat{c}^{(2)}(k;\rho_b)$ obtained from Eq.~\eqref{Eq:c2F} in the homogeneous limit, see Refs.~\onlinecite{Walters_2018,Stopper_2019}.  

Taking the limit of the radius of the spherical solute particle to infinity is fairly straightforward. Alternatively, one can impose planar geometry from the outset  and perform appropriate Fourier transforms. The upshot is that one finds:
\begin{align}
    \rho(z)-\rho_b\propto A_w e^{-\kappa z}\cos(2\pi z/\lambda+\phi_w), \quad z\rightarrow\infty,
\end{align}
where $\kappa$ and $\lambda$ refer to the leading pole identified in Eq.~\eqref{Eq:h_asymptotic}, for $rh(r)$. The amplitude $A_w$  and phase $\phi_w$  are not related to the corresponding quantities in Eq.~\eqref{Eq:h_asymptotic}. This argument is, of course, based on a linear response treatment of the asymptotics; it is close to that presented  in  Ref.~\onlinecite{Evans_book}.

The argument laid out above is deceptively simple. It implies that knowledge of the exact $c^{(2)}(r;\rho_b)$ is sufficient to determine the exact asymptotic decay of the one-body density profiles at a planar wall, and, indeed, of the solvation force for the liquid confined between two planar walls, provided the external potential is sufficiently short-ranged. In practice, one never has the exact $c^{(2)}(r;\rho_b)$ and it is not always clear what physics is omitted in employing an approximate  $c^{(2)}(r;\rho_b)$. In one-component neutral fluids this is not a major issue, apart from state points very close to the (bulk) critical point where  $c^{(2)}(r;\rho_b)$  develops power-law decay.

The situation is very different in ionic liquids where approximate IET's might omit crucial physics. For example, the MSA IET for the RPM decouples completely charge and number density correlations. The resulting $c^{(2)}_N(r;\rho_b)$, the pair direct correlation function for number-number correlations, is equal to  $c^{(2)}_{HS}(r;\rho_b)$, the HS pair direct correlation function. This is why in the MSA IET the decay length $\xi_N$ always takes the HS value and does not approach the correct value $\kappa_D\xi_N=0.5$, appropriate in the dilute limit. The MSA is termed a linear approximation, because it neglects the coupling. Its generalization, the GMSA, is a different beast. The corresponding $c^{(2)}_N(r;\rho_b)$ is not simply equal to  $c^{(2)}_{HS}(r;\rho_b)$, as there is some feedback from the charge correlations. This is why the GMSA results for $\xi_N$ exhibit FW crossover, as illustrated in Ref.~\onlinecite{Evans} and discussed in Sec.~\ref{sec:far_field}. However, the GMSA does not capture the correct low concentration limiting behaviour. In contrast, the Hypernetted Chain  (HNC) IET is regarded as a fully non-linear theory. Here one finds  $c^{(2)}_N(r;\rho_b)$ has a term in $h_Z^2(r)$ (see e.g. Ref.~\onlinecite{Ennis}), not present in the MSA or  GMSA, that leads to the correct behaviour: $\kappa_D\xi_N\rightarrow 0.5$,  in the dilute  limit. Recall the HNC IET is known to be very accurate for the (bulk) RPM, across a variety of regimes \cite{HansenMC}. 

Our DFT treatment of the asymptotics, see Appendix~\ref{App:DFT_asymptotic}, works at the level of one-body direct correlation functions. The three treatments we implement are explicitly non-linear and capture crucial couplings. As one sees in Eq.~\eqref{Eq:rho_N_as}, the number density profile $\rho_N(z)$ includes the term $\rho_Z^2(z)$, which is analogous to the  HNC closure for the bulk.


\section{Analysing the Euler Lagrange Equations in the Far Field}\label{App:DFT_asymptotic}

 We consider again the one-body profiles for the RPM in a planar geometry, as given by the self-consistency relations
\begin{align}\label{Eq:EL_app}
\rho_{\pm}(z)=\rho_b\exp\left[-\beta V_{ext}(z) +c^{(1)}_\pm(z)-c^{(1)}_{\pm,b} \right],
\end{align}
where subscript $b$ denotes bulk. We assume a short-ranged external potential $V_{ext}$ that is the same for both species. The one-body direct correlation functions $c_\pm^{(1)}(z)$ depend on the underlying functional. For the RPM, this will be of the form
\begin{align}\label{Eq:cES_split}
    c_\pm^{(1)}(z)=\pm c_1(z)+c_2(z),
\end{align}
where distinction has been made between terms that are proportional to the valency and those that are not. Typically, $c_1(z)$  has the form
\begin{align}\label{Eq:c1C}
 c_1(z;[\rho_Z,\rho_N])=\int\mathrm{d}z'\rho_Z(z')c^{ES}(|z-z'|;[\rho_N]),
\end{align}
where $\rho_Z=(\rho_+-\rho_-)/\rho_b$ and $\rho_N=(\rho_++\rho_-)/\rho_b-2$ are the charge and (excess) number densities as defined in the main text, and $c^{ES}(z)$ has its origin in electrostatics. For the MFC and MSAc functionals  $c^{ES}(z)$ is independent of $\rho_N(z)$. However, this is not the case for the MSAu functional due to the final term in Eq.~\eqref{Eq:Phi_MSA_RPM}. Nevertheless, the predominant contribution to $c_1(z)$ comes from $\rho_Z$. The other term $c_2(z)$ in Eq.~\eqref{Eq:cES_split}  has its origin solely in the steric (HS) repulsions for both the MFC and MSAc. In the case of the MSAu there is an additional term from the derivative of $\Phi^{MSA}$ w.r.t density. Generally we can write
\begin{align}\label{Eq:c2C}
    c_2(z;[\rho_N,\rho_Z])&=-\int\mathrm{d}z'\sum_\alpha\frac{\partial \Phi(\{\tilde{n}_\alpha\})}{\partial \tilde{n}_\alpha}(z')w_\alpha(|z-z'|),
\end{align}
where the free energy density $\Phi$ contains both the FMT contribution \cite{Roth_FMT} as well as possible others, such as the MSAu term Eq.~\eqref{Eq:MSA_phi_msa} or  that from $\Phi^{ES}$ introduced in the previous Appendix Eq.~\eqref{Eq:F_MF}, and the sum is over all weighted densities $\alpha$ with corresponding weight function $w_\alpha$.

Using Eqs.~\eqref{Eq:EL_app} and~\eqref{Eq:cES_split}, the Euler Lagrange equations for $\rho_Z$ and $\rho_N$ read
\begin{align}
    \rho_Z(z)=2\exp&\left[-\beta V_{ext}(z)+\Delta c_2(z;[\rho_N,\rho_Z])\right]\times\nonumber\\
     &\sinh(\Delta c_1(z;[\rho_Z,\rho_N])),\label{Eq:rhoZ_app}\\
    \rho_N(z)=2\exp&\left[-\beta V_{ext}(z)+\Delta c_2(z;[\rho_N,\rho_Z])\right]\times\nonumber\\
    &\cosh(\Delta c_1(z;[\rho_Z,\rho_N]))-2,\label{Eq:rhoN_app}
\end{align}
where $\Delta c_i(z)=c_i(z)-c_{i,b}$, $i\in \{1,2\}$.  In the far field, $ z\rightarrow\infty$, the quantities $\Delta c_i(z)$ 
are small and $\beta V_{ext}$ vanishes, which allows us to expand Eqs.~\eqref{Eq:rhoZ_app} and~\eqref{Eq:rhoN_app} as 
\begin{align}
       \rho_Z(z)&\approx2\Delta c_1(z;[\rho_Z,\rho_N]), \label{Eq:rho_Z_as_app}\\
    \rho_N(z)&\approx2\Delta c_2(z,[\rho_N,\rho_Z])+\Delta c_1(z;[\rho_Z,\rho_N])^2, \nonumber\\
    &=2\Delta c_2(z,[\rho_N,\rho_Z])+\frac{1}{4}\rho_Z(z)^2, \label{Eq:rho_N_as_app}
\end{align}
where in the last line we substituted $\rho_Z$ for $2\Delta c_1$. This asymptotic representation of the charge and number density profiles contains important information. First we consider the case where $c_1(z;[\rho_Z])$ depends on the charge density only, which holds for the MFC and MSAc functionals. Then, to lowest order,  $\rho_Z$ does not depend on $\rho_N$, consistent with the notion that charge is (essentially) decoupled from the number density in the RPM. However, the expression for $\rho_N(z)$ contains the $\rho_Z(z)^2$ term. We consider first the situation where  $c_2(z;[\rho_N])$ does not depend on the charge density, as is the case for the MFC and MSAc functionals. Then from Eqs.~\eqref{Eq:rho_Z_as_app} and~\eqref{Eq:rho_N_as_app} we find the leading asymptotic behaviour to be
\begin{align}
\rho_Z(z)=&Ae^{-z/\xi_Z}\\
    \rho_N(z)=&B\cos\left(\frac{2\pi}{\lambda_{FMT}}z\right)e^{-z/\xi_{FMT}}+Ce^{-2z/\xi_Z},
\end{align}
as $z\rightarrow\infty$. Here $\xi_{FMT}$ and $\lambda_{FMT}$ denote, respectively, the decay length and the wavelength for a pure hard-sphere system treated by FMT, $\xi_Z$ is the charge decay length and $A,B,C$ are constants. These results pertain to concentrations below any Kirkwood point, where $\rho_Z(z)$ is monotonically decreasing. In this case, the ultimate decay length $\xi_N$ of $\rho_N(Z)$ depends on whether $\xi_Z/2$ is larger or smaller than $\xi_{FMT}$. Fig.~\ref{Fig:xi} illustrates the variation of $\xi_{FMT}$ and $\xi_{Z}$ with concentration $d\kappa_D$. At small concentrations, $\xi_Z/2$ is the larger length scale, and in the dilute limit $\kappa_D\xi_N=0.5$. For concentrations above the Kirkwood point, where $\rho_Z(z)$ acquires an oscillatory factor; there is intricate competition between the two length scales. At sufficiently high concentrations $\xi_{FMT}$ will dictate the asymptotics. 

In the second situation, where $c_1$ and $c_2$ depend both on the charge and number density profiles, as is the case for the MSAu functional, a stronger coupling between the charge and number density profiles emerges. The resulting  MSAu decay lengths differ markedly from those from MSAc, IET and simulation results for concentrations between $0.5<d\kappa_D<2$ (see Fig.~\ref{Fig:xi}), suggesting the coupling is  not treated correctly in the MSAu. In particular, $\kappa_D\xi_N$  from the MSAu increases with $d\kappa_D$ in this range, see Fig.~\ref{Fig:xi}(b).  Note that in the limit of low concentration and of high concentration, $\kappa_D\xi_N$  from the MSAu does approach the correct behaviour. 
As mentioned previously in the text and  in Appendix~\ref{App:ES_cons}, one could consider an  MSAu functional that sets  $\eta(\{\tilde{n}(\mathbf{r})\})=0$. For this choice we recover the same asymptotic Z decay as with MSAc. With this choice $c_1(z)$ is the same for both MSAc and MSAu. In sharp contrast, we found that the N decay length changes very slightly by setting $\eta(\{\tilde{n}(\mathbf{r})\})=0$. This is because the $\vartheta_2$ term in Eq.~\eqref{Eq:Phi_MSA_RPM}, which is proportional to $\eta$, is subordinate to $\vartheta_1$. Hence, omitting $\vartheta_2$ barely influences $\Delta c_2(z)$ and therefore barely influences Eq.~\eqref{Eq:rho_N_as_app}, thereby  leaving almost no mark on the N decay length. To conclude, setting $\eta(\{\tilde{n}(\mathbf{r})\})=0$ in Eq.~\eqref{Eq:MSA_phi_msa}  influences only the charge density profiles, leaving the number density profiles almost unchanged. However, we decided not to follow this choice in presenting our results as the procedure cannot be applied generically:  breaking any symmetry in the ionic system (changing either their size or valency) will generate a non-vanishing $\eta$ in the bulk. We follow Ref.~\onlinecite{Roth_shells} and retain $\eta(\{\tilde{n}(\mathbf{r})\}$.

\section*{DATA AVAILABILITY}

The data that support the findings of this study are available
from the corresponding authors upon reasonable request.

\end{document}